\documentclass[twocolumn,superscriptaddress,showpacs,nofootinbib]{revtex4}

\usepackage{bm,color}

\usepackage{graphicx}
\usepackage{amsmath}
\usepackage{amssymb}
\usepackage{amsfonts}
\usepackage{bm}
\usepackage{color}

\def\be{\begin{equation}}
\def\ee{\end{equation}}
\def\bea{\begin{eqnarray}}
\def\eea{\end{eqnarray}}

\begin{document}

\title{Models of dark matter halos based on statistical mechanics: \\
II. The fermionic King model}

\author{Pierre-Henri Chavanis}
\affiliation{Laboratoire de Physique Th\'eorique, Universit\'e Paul
Sabatier, 118 route de Narbonne 31062 Toulouse, France}
\author{Mohammed Lemou}
\affiliation{CNRS and IRMAR, Universit\'e de Rennes 1 and INRIA-Rennes Bretagne Atlantique, France}
\author{Florian M\'ehats}
\affiliation{CNRS and IRMAR, Universit\'e de Rennes 1 and INRIA-Rennes Bretagne Atlantique, France}

\begin{abstract}

We discuss the nature of phase transitions in the fermionic King model which
describes tidally truncated quantum self-gravitating systems. This distribution
function takes into account the escape of high energy particles and has a finite
mass. On the other hand, the Pauli exclusion principle puts an upper bound on
the phase space density of the system and stabilizes it against gravitational
collapse. As a result, there exists a statistical equilibrium state for
any accessible values of energy and temperature. We
plot the caloric curves  and
investigate the nature of phase transitions as a function of the degeneracy parameter in both microcanonical and canonical
ensembles. We consider stable and metastable states and emphasize the importance
of the latter for systems with long-range interactions. Phase transitions can take place between a
``gaseous'' phase  unaffected by quantum mechanics and  a ``condensed''
phase dominated by quantum mechanics. The phase diagram exhibits two critical points, one in each ensemble, beyond which
the phase transitions disappear. There also exist a
region of negative specific heats and a situation of ensemble inequivalence for
sufficiently large systems.
In the microcanonical ensemble, gravitational  collapse (gravothermal catastrophe) results in the formation a small degenerate object  containing a small mass. This is accompanied by the expulsion of a hot envelope containing a large mass. In the canonical ensemble, gravitational collapse (isothermal collapse) leads to a small degenerate object  containing  almost all the mass. It is surrounded by a tenuous envelope. We apply the fermionic King model to the case of dark matter halos made of massive
neutrinos. The gaseous phase describes large halos and the condensed phase
describes dwarf
halos. Partially degenerate configurations describe intermediate size halos.
We argue that large dark matter halos cannot harbor a fermion ball because these
nucleus-halo
configurations are thermodynamically unstable (saddle points of entropy).
Large dark matter halos may rather contain a central black hole resulting from
a dynamical instability of relativistic origin occuring during the gravothermal
catastrophe. We relate the existence of black holes to the microcanonical
critical point and determine the minimum halo mass above which black holes
can
form. We also compare fermionic and bosonic models of dark matter and discuss
the value of the mass of the dark matter particle in each case.

\end{abstract}



\maketitle


\section{Introduction}
\label{sec_introduction}

Self-gravitating systems have a very particular thermodynamics first investigated by Antonov \cite{antonov} and Lynden-Bell \& Wood \cite{lbw} in relation to stellar systems such as globular clusters made of classical point
mass stars. A first curiosity is the existence of negative specific heats leading to  the notion of ensemble inequivalence. It is well-known in astrophysics that self-gravitating systems have negative specific heats \cite{eddington}. However, when considered from the viewpoint of statistical mechanics, this property leads to an apparent paradox since the specific heat must be positive in the canonical ensemble as it measures the variance of the fluctuations of energy.  As first understood by Thirring \cite{thirring}, this paradox is solved by realizing that the statistical ensembles are inequivalent. Negative specific heats are allowed in the microcanonical ensemble (MCE) while they are forbidden in the canonical ensemble (CE).\footnote{MCE describes an isolated system evolving at fixed energy while CE describes a dissipative system coupled to a thermal bath fixing its temperature. For the sake of completeness, we shall consider the two ensembles in this paper even if MCE is usually the most relevant to describe astrophysical systems.}  The inequivalence of statistical ensembles  is not restricted to self-gravitating systems.  It may arise in other systems with long-range interactions due to the non-additivity of the energy \cite{cdr}. However, the statistical mechanics of self-gravitating systems presents specific difficulties \cite{paddy,katzrevue,ijmpb} that are absent in other systems with long-range interactions.

First, there is no statistical equilibrium state in a strict sense because
a self-gravitating system in an infinite domain has no maximum of entropy or free energy.\footnote{We can always increase the entropy at fixed mass and energy in MCE and we can always increase the free energy  at fixed mass in CE by spreading the system to infinity  (see Appendices A and B of \cite{sc}). The absence of statistical equilibrium state in an unbounded domain can also be directly inferred from the fact that  the integrals defining the density of states in MCE and the partition function in CE diverge  at large distances \cite{paddy}.} There are not even critical points of entropy or free energy because the isothermal self-gravitating sphere, corresponding to the Boltzmann distribution coupled to the Poisson equation, has infinite
mass \cite{chandra}. Therefore, the statistical mechanics
of self-gravitating systems is essentially an out-of-equilibrium problem \cite{aakin}. The
absence of statistical equilibrium state in an unbounded domain
is related to the fact that self-gravitating systems like globular clusters have the tendency to evaporate \cite{bt}.
However, evaporation is a slow process so that, on an intermediate timescale, self-gravitating systems appear to be self-confined. Furthermore, stellar
systems like globular clusters are never totally isolated from the surrounding.
In practice, they feel the tides of a nearby galaxy.  As a result, the stars escape when
they reach sufficiently high energies. This implies that the density profile of the cluster
vanishes at a finite radius $R$ interpreted as a tidal radius.

There are two possibilities to solve the infinite
mass problem of the self-gravitating isothermal sphere. A first possibility, introduced by Antonov \cite{antonov}, is to enclose the system within  a ``box'' so as to artificially prevent  its evaporation. The box radius mimics  the tidal radius of more realistic systems.
This procedure is appreciated by theorists first because it is simple, and secondly because it allows one to develop a
rigorous statistical mechanics of self-gravitating systems based on the ordinary Boltzmann distribution. However, for astrophysical
applications, this model is too much idealized because self-gravitating systems in nature are not enclosed in boxes! Another possibility is
to take  the evaporation of high energy stars into account and use the  King model \cite{king}. This is  a truncated
Boltzmann distribution obtained from the usual Boltzmann distribution by
subtracting a constant term 
so that the distribution vanishes at the escape energy. This distribution has a
finite mass. It can be derived  from a kinetic theory based on  the classical
Landau equation \cite{king}.

\begin{figure}
\begin{center}
\includegraphics[clip,scale=0.3]{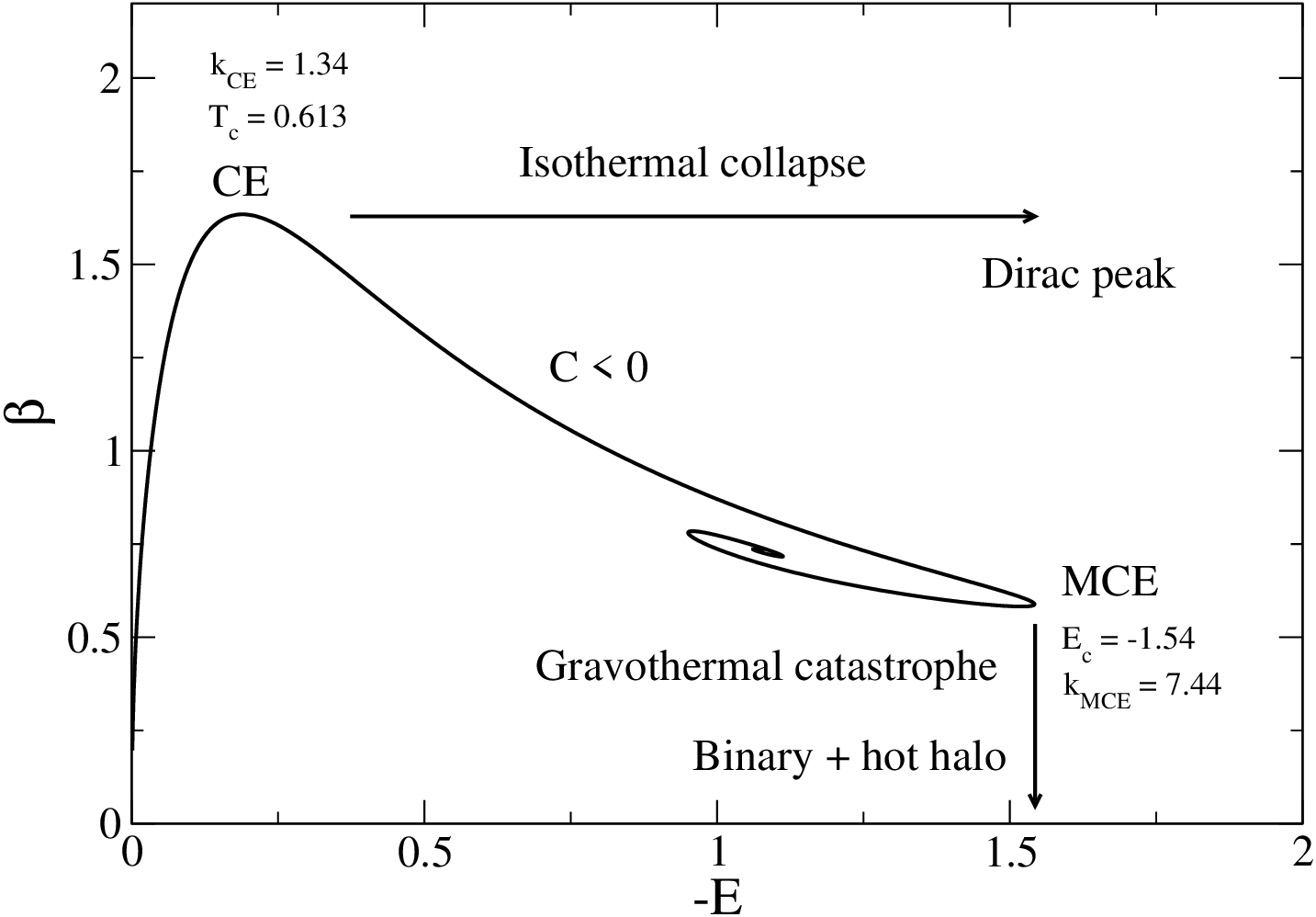}
\caption{Series of equilibria of the classical King model. It has a snail-like
(spiral) structure but only the part of the curve up to CE
in the canonical ensemble and up to  MCE in the microcanonical ensemble is
stable (the region between CE and MCE where the specific heat is negative
corresponds to a region of ensemble inequivalence).
}
\label{ckm}
\end{center}
\end{figure}

Even when self-gravitating systems are confined in boxes, or when evaporation is
properly taken into account by using the King model, a second
difficulty arises which is now related to the fact that self-gravitating systems
have the tendency to collapse \cite{bt}.  In the box model, it is found that
statistical equilibrium states exist only above a critical energy
$E_c=-0.335\, GM^2/R$ in MCE  and above a critical temperature $T_c=GMm/(2.52Rk_B)$ in CE, discovered by Emden \cite{emden}. The series of equilibria of classical isothermal spheres has the form of a spiral and these critical point correspond to turning points of energy and temperature. Stable configurations  have a density contrast $\rho(0)/\rho(R)<709$ in MCE and $\rho(0)/\rho(R)<32.1$ in CE \cite{antonov,lbw,katz,paddy,aa}. There are more stable states in MCE than in CE due to ensemble inequivalence. These configurations are metastable (local entropy maxima in MCE and local free energy maxima in CE)\footnote{There is no global maximum of entropy or free energy.  In MCE, one can always increase the entropy at fixed mass and energy by
forming a binary star surrounded by a hot halo. The entropy diverges when the binary is made tighter and tighter, and the halo hotter and hotter (see Appendix A of \cite{sc}). In CE, one can always increase the free energy by approaching the particles at the same point. The free
energy diverges when a Dirac peak containing all the particles is formed
(see Appendix B of \cite{sc}). The absence of a strict statistical equilibrium state in  a box can also be directly inferred from the divergence of the density of states  in MCE and from the divergence of the partition function in CE \cite{paddy,kiessling}.} but their lifetime is considerable since it scales as
$e^N$ (except close to the critical point) \cite{metastable}. For globular clusters with $N\sim 10^6$ this lifetime is so great that metastable states can be considered as stable states.\footnote{When the system is  in a metastable state, the spontaneous formation of  a binary star surrounded by a hot halo  in MCE ($S\rightarrow +\infty$)  or the spontaneous formation of a Dirac peak in CE ($J\rightarrow +\infty$) is a very rare event since its probability scales as $e^{-N}$. Indeed, in order to leave a metastable state, the system has to overcome a huge barrier of entropy or free energy whose hight scales as $N$. This requires very particular correlations and takes too much time to be physically relevant.}
Similar results are obtained with the classical King model (see Fig. \ref{ckm}) as shown by Katz \cite{katzking} and further analyzed in \cite{paper1} (Paper I). Therefore, a self-gravitating
system can reach a statistical equilibrium state described by the King model (truncated Boltzmann distribution) at sufficiently high energies and at
sufficiently high temperatures, even if there is no statistical equilibrium
state in a strict sense \cite{paddy,katzrevue,ijmpb}. However, for $E<E_c$ and $T<T_c$, there is no statistical equilibrium state anymore and the system undergoes gravitational collapse.\footnote{In practice, the energy and the temperature  slowly decrease with time due to collisions and evaporation until a point at which there is no equilibrium state anymore (see Appendix \ref{sec_dts}). In that case, the system collapses. This corresponds to a saddle-node bifurcation.}  This is called gravothermal catastrophe
\cite{lbw} in MCE and isothermal collapse
\cite{aa} in CE.  In MCE, the gravothermal catastrophe leads to a binary star surrounded by a hot halo \cite{cohn}. In CE, the isothermal collapse  leads to a Dirac peak containing all the particles \cite{post}. Therefore, the result of the
gravitational collapse is to form a singularity: a ``binary star $+$ hot halo''
in MCE and a ``Dirac peak'' in CE
\cite{ijmpb}.

The previous results are valid for classical particles such as stars in
globular clusters. If we now consider a gas of self-gravitating fermions,
gravitational collapse stops when the system becomes degenerate as a consequence
of the Pauli exclusion principle.
In that case, the singularity (tight binary or Dirac peak) is smoothed-out and replaced by
a compact object which is a completely degenerate  ``fermion ball'' similar to a white dwarf star.
At finite temperature, this compact object is surrounded by a
dilute atmosphere (vapor) so that the whole
configuration has a ``core-halo'' structure. Therefore, when quantum mechanics is properly accounted for, the system
is stabilized against gravitational collapse. In that case, there exist an equilibrium state  for any accessible value of energy and
temperature.\footnote{For self-gravitating fermions, there exists a strict statistical equilibrium state (global maximum of entropy or global maximum of free energy) for all accessible values of energy and temperature. There may also exist metastable states (local maxima of entropy or local maxima of free energy) that are as much, or even more, relevant than fully stable states. Indeed, the choice of the equilibrium state depends on a notion of ``basin of attraction'' and the metastable states may be reached more easily from generic initial conditions than the fully stable states that require very particular correlations. For example, in order to pass from the gaseous phase to the condensed phase, the system must cross a huge barrier of entropy or free energy and evolve through an intermediate phase in which some particles must approach very close to each other. Inversely, to pass from the condensed phase to the  gaseous phase, the system must cross a huge barrier of entropy or free energy and evolve through an intermediate phase in which some particles must escape from the condensate. The probability of such events is extremely low so that, in practice, the system remains in the metastable phase \cite{ijmpb}.} We can therefore study phase transitions between a ``gaseous phase''
unaffected by quantum mechanics and a ``condensed phase'' dominated by quantum mechanics. The nature of these phase transitions has been discussed in detail by Chavanis  \cite{pt,dark,ispolatov,rieutord,ptd,epjb14} (see a review in \cite{ijmpb}) in the case where the fermions are confined within a box.\footnote{Similar phase transitions are obtained if, instead of quantum particles, we consider classical particles and regularize the gravitational potential at short distances
\cite{follana,ispolatov,destri,nardini} or take into account the finite size of
the particles by considering a hard spheres gas
\cite{aronson,stahl,paddy,pt,champion,epjb14}. Even if the details of the phase
transitions depend on the specific form of the small-scale regularization, the
phenomenology of these phase transitions 
is relatively universal as described in \cite{ijmpb}.}  In the present paper, we
extend this study to the fermionic King model. This extension is interesting
because the fermionic King model is more realistic than box models. Furthermore,
the fermionic King model may have applications in astrophysics and cosmology.
Indeed, it may provide a realistic model of dark matter halos made of massive
neutrinos.

The observation of the rotation curves of galaxies has revealed that the galaxies are surrounded by a halo of dark
matter \cite{persic}. The nature of dark matter remains unknown and constitutes
one of the greatest challenge of modern cosmology.  The cold dark matter (CDM)
model is successful to describe the large scale structures of the universe but
it encounters many problems at the scale of galactic or sub-galactic structures.
In particular, CDM simulations \cite{nfw} lead to $r^{-1}$ cuspy density
profiles at galactic centers (in the scales of the order of $1$ kpc and smaller)
while most rotation curves indicate a smooth core density  \cite{observations}.
On the other hand, the predicted number of satellite galaxies around each
galactic halo is far beyond what we see around the Milky Way  \cite{satellites}.
It  is therefore  necessary to develop new models of dark matter in order to
solve these problems (cusp problem and missing satellite problem).

Several authors
\cite{gr,stella,mr,gao,mr2,merafina,kull,bilic1,csmnras,mnras,tv,bilic2,bv,
bilic3,viollier,bilic4,pt,dark,ijmpb,vega,vega2,vega3,vega4,vega5} have proposed
to describe dark matter as a gas of fermions.\footnote{There are also several
studies describing dark matter as a gas of bosons (see Paper I for a detailed
list of references).} They argue that the Pauli exclusion principle  avoids
density cusps at the halo center and solves the problems of the CDM model.
Quantum mechanics may be particularly relevant for compact dwarf halos. By
assuming that the smallest known dark matter halos are completely degenerate, de
Vega and Sanchez \cite{vega,vega2} have obtained a maximum  bound of about $2\,
{\rm keV}$ on the mass of the fermions. The dark matter particle may be a
sterile neutrino. In these studies,  the usual Fermi-Dirac distribution is used.
However, this distribution leads to models of dark matter halos with an infinite
mass. This prevents one from determining the caloric curves and studying the
thermodynamical stability of the configurations. In order to improve the
picture, we propose to describe dark matter halos by the fermionic King model.
This model takes into account the evaporation of high energy particles and has a
finite mass. This model was introduced by  Ruffini and Stella \cite{stella} as
an heuristic extension of the classical King model to the case of fermions. It
was introduced independently by Chavanis \cite{mnras,dubrovnik} who derived it
from a kinetic theory based on the fermionic Landau equation. As explained in
Paper I, the fermionic King model can describe either a gas of fermions (e.g.
massive neutrinos) at statistical equilibrium or a collisionless system of
particles (classical or quantum) experiencing a process of violent relaxation of
Lynden-Bell's type \cite{lb,csr,csmnras,mnras,dubrovnik}. In Paper I, we have
considered large dark matter halos for which quantum mechanics is negligible. 
We have shown that such halos are relatively well described by the
classical  King model at, or close to, the limit of microcanonical stability. At
that point, the King profile can be approximated by the modified Hubble profile
\cite{bt}. It has an isothermal core, an isothermal halo, and a polytropic
envelope of index $n=5/2$. The density profile is flat in the core and decreases
as $r^{-3}$ in the halo. As a result, the modified Hubble profile is
relatively similar  to the Burkert profile \cite{observations} that gives a good
fit of many dark matter halos.\footnote{The modified Hubble profile is a
particular case of the family of density profiles (I-107) introduced empirically
by de Vega and Sanchez \cite{vega4,vega5} (it corresponds to $\alpha=3$ in our
notations). They mention that the profiles with $\alpha\sim 3$ give a good fit
of the observations of dark matter halos. Our approach, that is based on the
physically motivated King model, provides a justification of their empirical
results.} For large dark matter halos, the problems of the CDM model  (density
cusps and missing satellites) are solved by thermal effects, not by quantum
mechanics. This corresponds to warm dark matter (WDM). However, quantum
mechanics becomes important for dwarf and intermediate size halos. In this
paper, in order to describe all types of dark matter halos, we consider the
fermionic King model with an arbitrary level of quantum degeneracy.

The paper is organized as follows. In Sec. \ref{sec_fkm}, we introduce
the fermionic King model. This distribution function is
appropriate to describe dark matter halos if they are made of massive neutrinos
at statistical equilibrium or if they have experienced a violent relaxation of
Lynden-Bell's type. In Sec. \ref{sec_kingfermi}, we discuss general properties
of the fermionic King model. We show that it generically has a polytropic core
of index $n=3/2$, a classical isothermal halo, and a polytropic envelope of
index $n=5/2$. In Sec. \ref{sec_ptfk}, we study the nature of phase transitions
in the fermionic King model depending on the value of the degeneracy parameter
(i.e. the size of the system). We emphasize the importance of metastable states
in systems with long-range interactions. In Sec. \ref{sec_dprc}, we plot the
density profiles and the rotation curves of the fermionic King model. We discuss
their ability at describing dark matter halos. We follow the series of
equilibria for increasing concentration parameter. We argue that large dark
matter halos are non degenerate, intermediate size halos are partially
degenerate, and dwarf halos are completely degenerate. We show that
gravitational  collapse in MCE results in the formation a small degenerate
object  containing a small mass accompanied by the expulsion of a hot envelope
containing a large mass. By contrast, gravitational collapse in CE leads to a
small degenerate object containing  almost all the mass surrounded by a tenuous
envelope. In Secs. \ref{sec_harbor}-\ref{sec_diffsl}, we argue that large dark
matter halos cannot
contain a fermion ball because this nucleus-halo structure is thermodynamically
unstable. This may explain why black holes are observationally favored over
fermion balls at the center of galaxies.  In Appendix \ref{sec_dts}, we discuss
subtle issues concerning the dynamical and thermodynamical stability of the
fermionic King model. In Appendix \ref{sec_vs}, we give arguments according to
which MCE is more appropriate than CE to describe dark matter halos. In Appendix
\ref{sec_fgh}, we introduce dimensionless quantities that can be compared to
observations. In Appendix \ref{sec_fbdm}, we compare fermionic and bosonic
models of dark matter and discuss the value of the mass of the dark
matter particle in each case. In Appendix \ref{sec_degclass}, we determine
whether dark matter halos are classical or quantum objects depending on their
mass. In Appendix \ref{sec_temp}, we determine the temperature of dark matter
halos. In Appendix \ref{sec_relat}, we determine the maximum mass of
relativistic compact objects. In Appendix \ref{sec_bh}, we argue that large
halos may contain a
central black hole and we determine the minimum halo mass above which black
holes can
form. In Appendix \ref{sec_scen}, we discuss general scenarios of formation of
dark matter halos depending on the nature of the dark matter particle.

\section{Models of dark matter halos based on statistical mechanics}
\label{sec_fkm}

We  consider the possibility that dark matter halos can be described by the fermionic King model defined by \cite{stella,mnras,dubrovnik}:
\begin{equation}
f=\eta_0\frac{1-e^{\beta(\epsilon-\epsilon_m)}}{1+e^{\beta\epsilon+\alpha}} \quad {\rm if} \quad \epsilon\le \epsilon_m,
\label{fkm1}
\end{equation}
and $f=0$ if $\epsilon\ge \epsilon_m$. Here, $f({\bf r},{\bf v})$ gives the mass
density of particles with position ${\bf r}$ and velocity ${\bf v}$,  $\rho({\bf
r})=\int f({\bf r},{\bf v})\, d{\bf v}$ gives the mass density of particles with
position ${\bf r}$, $\eta_0=gm^4/h^3$ is the maximum accessible value of the
distribution function fixed by the Pauli exclusion principle ($m$ is the mass of
the particles, $h$ is the Planck constant, and $g=2s+1$ is the spin multiplicity
of the quantum states\footnote{In the numerical applications, we shall take
$s=1/2$ and $g=2$.}), $\epsilon=v^2/2+\Phi({\bf r})$ is the individual energy of
the particles by unit of mass, $\Phi({\bf r})$ is the gravitational potential
determined by the Poisson equation $\Delta\Phi=4\pi G\rho$, $\epsilon_m$ is the
escape energy, $\beta=m/k_B T$ is the inverse temperature, and
$\epsilon_F=-\alpha/\beta$ is the chemical potential (Fermi
energy).\footnote{The preceding relations are written in the case where dark
matter is a quantum gas made of fermions (e.g. massive neutrinos) at statistical
equilibrium. They remain valid if dark matter is a collisionless gas undergoing
a process of violent relaxation of Lynden-Bell's type
\cite{lb,csr,csmnras,mnras,dubrovnik}. In that case, the physical meaning of
$\eta_0$ and $\beta$ is different as explained in Paper I.}

For $\epsilon_m\rightarrow +\infty$, we recover the Fermi-Dirac distribution $f=\eta_0/(1+e^{\beta\epsilon+\alpha})$.

In the non-degenerate limit $\alpha\rightarrow +\infty$, we recover the classical King model $f=\eta_0 e^{-\beta\epsilon_m-\alpha}\left\lbrack e^{-\beta(\epsilon-\epsilon_m)}-1\right\rbrack$ which reduces to the  Boltzmann distribution $f=\eta_0 e^{-(\beta\epsilon+\alpha)}$ for $\epsilon_m\rightarrow +\infty$.

The fermionic King model (\ref{fkm1}) can be derived from a kinetic theory based on the fermionic Landau equation by looking for a quasi stationary state of this equation such that $f=0$ at $\epsilon=\epsilon_m$ \cite{mnras,dubrovnik}.

The fermionic King model (\ref{fkm1}) will be of the form of Eq.
(I-17)\footnote{Here and in the following (I-x) refers to Eq. (x) of Paper I.}
provided that $\beta\epsilon_m+\alpha$ can be treated as a constant along the
series of equilibria. We write this constant as
\begin{equation}
A\equiv \eta_0 e^{-\beta\epsilon_m-\alpha}.
\label{fkm2}
\end{equation}
Using Eq. (\ref{fkm2}), the fermionic King model can be rewritten as
\begin{equation}
f=A\frac{\frac{\eta_0}{A}-e^{\beta\epsilon+\alpha}}{1+e^{\beta\epsilon+\alpha}}.
\label{fkm3}
\end{equation}
It is of the form of Eq. (I-17) with ${\cal F}(x)=(\mu-e^x)/(1+e^x)$ where we
have introduced 
the degeneracy parameter $\mu=\eta_0/A$. The function ${\cal F}(x)$ vanishes at
$x_0=\ln\mu$ and we check that Eq. (\ref{fkm2}) satisfies the general relation
$\beta\epsilon_m+\alpha=x_0$ of Paper I. Using Eq. (\ref{fkm2}), the fermionic
King model can also be rewritten as
\begin{equation}
f=A\frac{e^{-\beta(\epsilon-\epsilon_m)}-1}{1+\frac{A}{\eta_0}e^{-\beta(\epsilon-\epsilon_m)}}.
\label{fkm4}
\end{equation}
It is of the form of Eq. (I-19) with ${\cal F}_s(x)=(e^{-x}-1)/(1+e^{-x}/\mu)$.
By construction ${\cal F}_s(0)=0$.

In the non degenerate limit $\alpha\rightarrow +\infty$, we recover the
classical King model. Using Eq. (\ref{fkm3}) it can be written as
\begin{equation}
f=A\left\lbrack \frac{\eta_0}{A}e^{-(\beta\epsilon+\alpha)}-1\right\rbrack,
\label{fkm5}
\end{equation}
corresponding to ${\cal F}(x)=\mu e^{-x}-1$. Alternatively, using Eq.
(\ref{fkm4}), it can be written as
\begin{equation}
f=A\left \lbrack e^{-\beta(\epsilon-\epsilon_m)}-1\right\rbrack
\label{fkm6}
\end{equation}
corresponding to ${\cal F}_s(x)=e^{-x}-1$.

Finally, following the method of Paper I, we can determine the generalized
entropy associated with the King model. For the fermionic King model we get
\begin{eqnarray}
C(f)=A \left \lbrack \left (1+\frac{f}{A}\right ) \ln \left (1+\frac{f}{A}\right )-\frac{f}{A}\right\rbrack\nonumber\\
+\eta_0 \left \lbrack \left (1-\frac{f}{\eta_0}\right ) \ln \left (1-\frac{f}{\eta_0}\right )+\frac{f}{\eta_0}\right\rbrack
-\ln\left (\frac{\eta_0}{A}\right )f
\label{fkm7}
\end{eqnarray}
and for the classical King model we obtain
\begin{equation}
C(f)=A \left \lbrack \left (1+\frac{f}{A}\right ) \ln \left (1+\frac{f}{A}\right )-\frac{f}{A}\right\rbrack-\ln\left (\frac{\eta_0}{A}\right )f.
\label{fkm8}
\end{equation}

We note that our description of the self-gravitating Fermi gas 
is based on a mean field approximation and on the Thomas-Fermi (TF)
approximation where the quantum potential (Heisenberg uncertainly principle) is
neglected. For box-confined configurations, Hertel and Thirring \cite{htmath,ht}
have established that these approximations are rigorously valid in a proper
thermodynamic limit where $N\rightarrow +\infty$.

\section{The fermionic King model}
\label{sec_kingfermi}

In this section, we apply the general formalism developed in Paper I to the case of the fermionic King model.

\subsection{The distribution function}
\label{sec_df}

\begin{figure}
\begin{center}
\includegraphics[clip,scale=0.3]{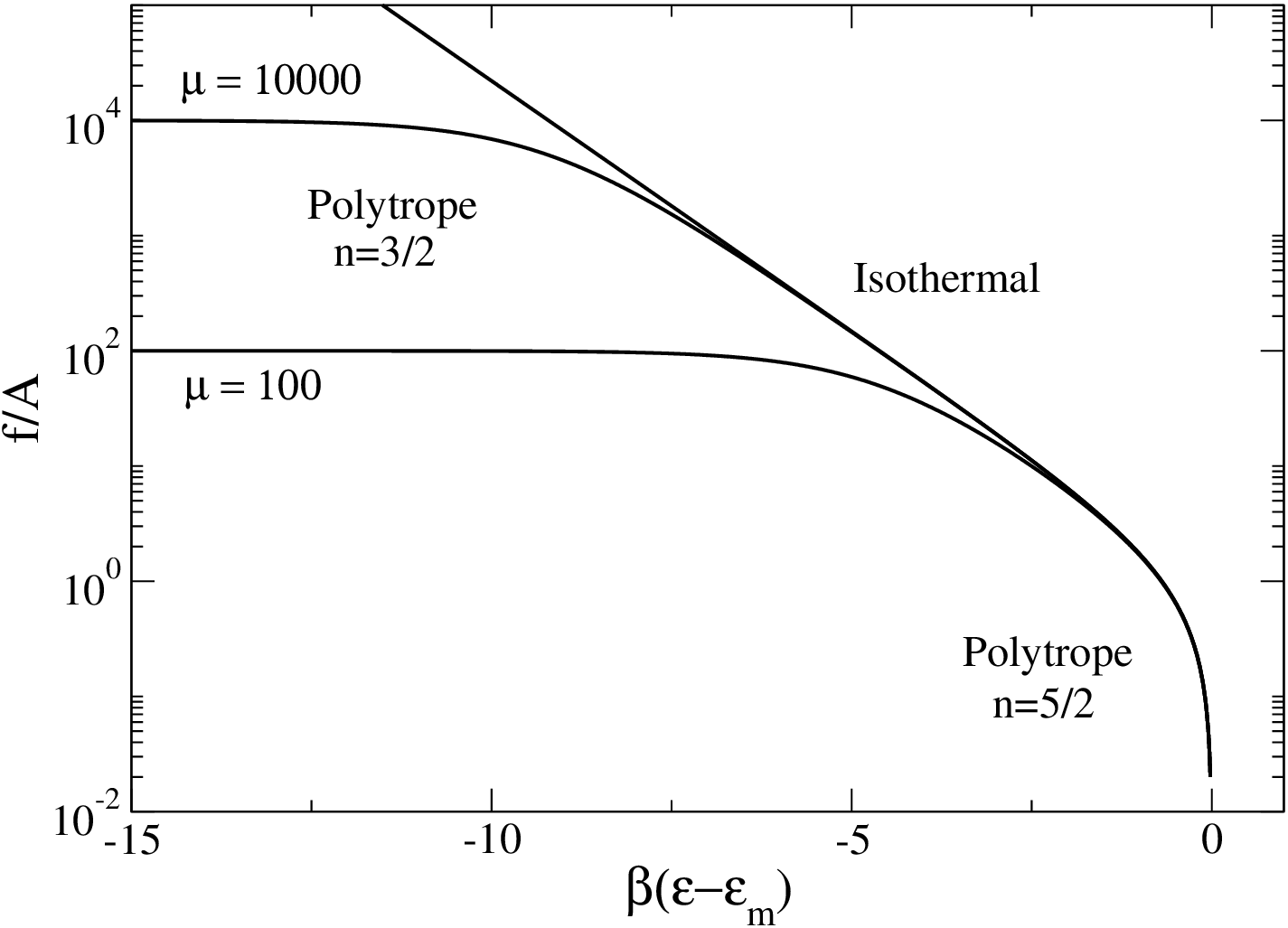}
\caption{The distribution function $f(\epsilon)$ in scaled variables showing
the polytropic core, the isothermal halo, and the polytropic envelope.}
\label{fepsmu}
\end{center}
\end{figure}

The fermionic King model is defined by
\begin{equation}
f=A\frac{e^{-\beta(\epsilon-\epsilon_m)}-1}{1+\frac{1}{\mu}e^{-\beta(\epsilon-\epsilon_m)}}\quad {\rm if} \quad \epsilon\le \epsilon_m,
\label{df1}
\end{equation}
\begin{equation}
f=0 \quad {\rm if} \quad \epsilon\ge\epsilon_m,
\label{df2}
\end{equation}
where $\epsilon_m$ is the escape energy at which the particles leave the system
($f=0$) and $\mu=\eta_0/A$ is the degeneracy parameter. For $\epsilon\rightarrow
-\infty$, the fermionic King distribution tends to a constant value
$f\rightarrow \mu A=\eta_0 $ so it is equivalent to a polytropic distribution of
index $n=3/2$ \cite{bt}. For intermediate energies, it can be approximated by
the Boltzmann distribution $f\sim A e^{-\beta(\epsilon-\epsilon_m)}$. For
$\epsilon\rightarrow \epsilon_m^{-}$, it reduces to $f\sim
A\beta(\epsilon_m-\epsilon)/(1+1/\mu)$ so it is equivalent to a polytropic
distribution of index $n=5/2$ \cite{bt}. Therefore, the fermionic King model
generically
describes a cluster with a polytropic core of index $n=3/2$, a classical
isothermal halo, and a polytropic envelope of index $n=5/2$. The proportion of
these different regions depends on the concentration parameter
$k=\beta(\epsilon_m-\Phi_0)$ as shown in the sequel. The distribution function
$f(\epsilon)$ is represented in Fig. \ref{fepsmu}.

The fermionic King distribution is of the form of Eq. (I-19) with
\begin{equation}
{\cal F}_s(x)=\frac{e^{-x}-1}{1+\frac{1}{\mu}e^{-x}}.
\label{df3}
\end{equation}
The asymptotic behaviors of the functions  $I_n(z)$ defined in Paper I with Eq. (\ref{df3}) for small and large values of $z$ are easily obtained. For $z\rightarrow 0$, using the fact that ${\cal F}_s(x)\sim -x/(1+1/\mu)$ for $x\rightarrow 0$, we get
\begin{equation}
I_n(z)\sim \frac{1}{1+\frac{1}{\mu}}\frac{8\pi}{(2n+1)(2n+3)}z^{(2n+3)/2}.
\label{df6}
\end{equation}
For $z\rightarrow +\infty$, using the fact that ${\cal F}_s(x)\rightarrow \mu$ for $x\rightarrow -\infty$, we get
\begin{equation}
I_n(z)\sim \frac{4\pi \mu}{2n+1} z^{(2n+1)/2}.
\label{df8}
\end{equation}

\subsection{The equation of state}
\label{sec_eos}

For $\chi\rightarrow +\infty$, using Eqs. (I-27), (I-37) and (\ref{df8}), we find that
\begin{equation}
\rho\sim A\left (\frac{2}{\beta}\right )^{3/2} \frac{4\pi}{3}\mu \chi^{3/2},\quad p\sim \frac{1}{3} A\left (\frac{2}{\beta}\right )^{5/2} \frac{4\pi}{5}\mu \chi^{5/2},
\label{eos1}
\end{equation}
leading to the polytropic equation of state
\begin{equation}
p\sim \frac{1}{5}\left (\frac{3}{4\pi\eta_0}\right )^{2/3}\rho^{5/3}.
\label{eos2}
\end{equation}
This equation of state is valid at high densities.

For $\chi\rightarrow 0$, using Eqs. (I-27), (I-37) and (\ref{df6}), we find that
\begin{eqnarray}
\rho\sim A\left (\frac{2}{\beta}\right )^{3/2} \frac{8\pi}{15}\frac{1}{1+\frac{1}{\mu}}\chi^{5/2},\nonumber\\ p\sim \frac{1}{3} A\left (\frac{2}{\beta}\right )^{5/2} \frac{8\pi}{35}\frac{1}{1+\frac{1}{\mu}}\chi^{7/2},
\label{eos3}
\end{eqnarray}
leading to the polytropic equation of state
\begin{equation}
p\sim \frac{1}{7}\left (\frac{15}{4\pi A\beta}\right )^{2/5}\left (1+\frac{1}{\mu}\right )^{2/5}\rho^{7/5}.
\label{eos4}
\end{equation}
This equation of state is valid at low densities.

For intermediate densities, we obtain an isothermal equation of state $p\sim \rho/\beta$.

\begin{figure}
\begin{center}
\includegraphics[clip,scale=0.3]{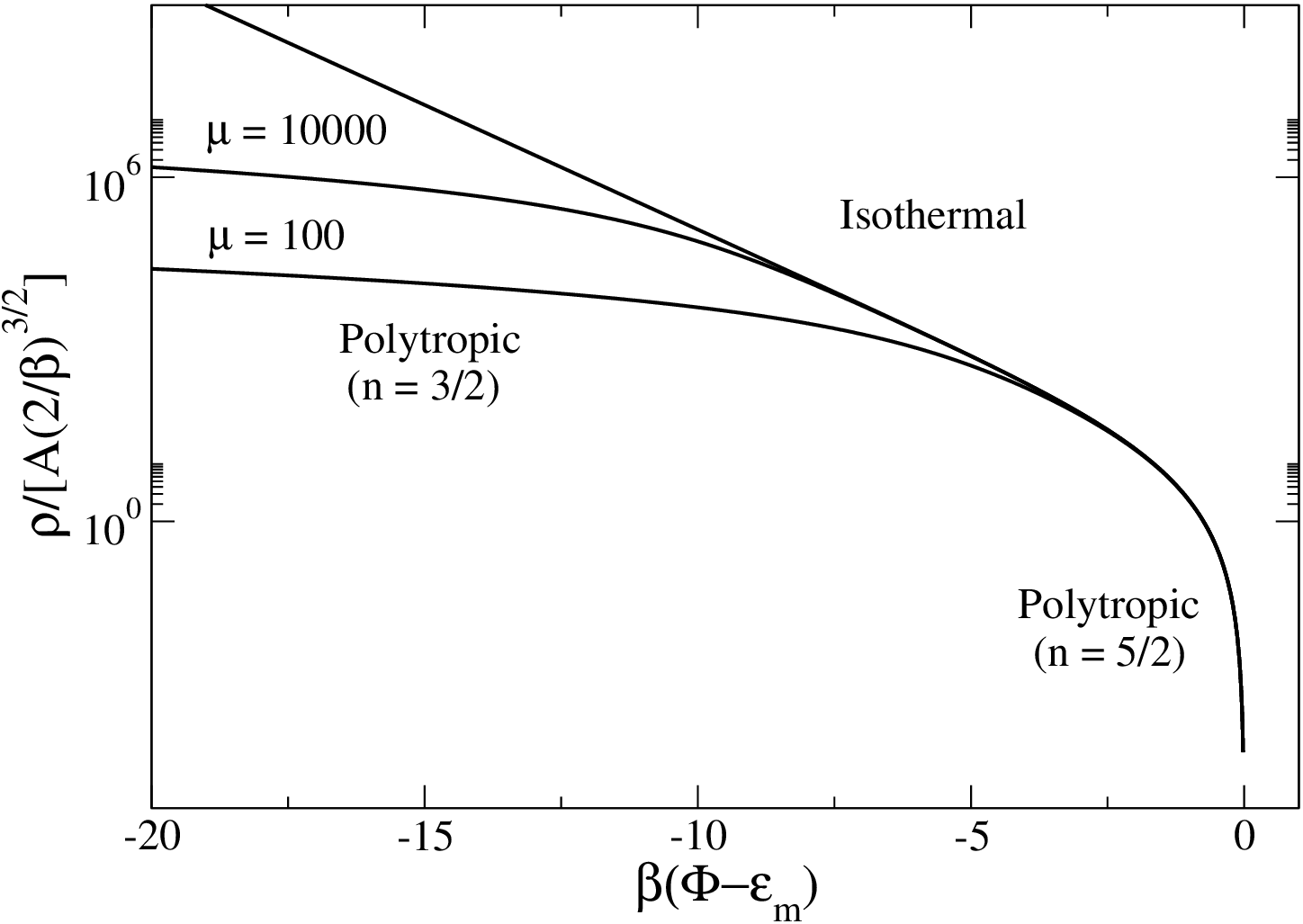}
\caption{The density $\rho(\Phi)$ in scaled variables showing
the polytropic core, the isothermal halo, and the polytropic envelope.}
\label{rhophimu}
\end{center}
\end{figure}

For $\Phi\rightarrow -\infty$, the density is related to the
gravitational potential by  $\rho(\Phi)\propto
(-\Phi)^{3/2}$ which corresponds to a polytropic distribution of index $n=3/2$. For intermediate values of $\Phi$, we get the
Boltzmann distribution $\rho(\Phi)\propto
e^{-\beta\Phi}$. For $\Phi\rightarrow \epsilon_m^{-}$, the density is related to
the gravitational potential by $\rho(\Phi)\propto
(\epsilon_m-\Phi)^{5/2}$ which corresponds to a polytropic distribution of index
$n=5/2$. The relation $\rho(\Phi)$ is
represented in Fig. \ref{rhophimu}.

\subsection{The polytropic limit $k\rightarrow 0$}
\label{sec_poly}

In the limit $k\rightarrow 0$, the function $\chi$ is always small, so we can
use the approximation (\ref{df6}) everywhere. As a result, the King model is
equivalent to a pure polytrope ($p=K\rho^{1+1/n}$) of index $n=5/2$ and
polytropic constant $K=(1/7)\left ({15}/{4\pi A\beta}\right
)^{2/5}(1+{1}/{\mu})^{2/5}$. The degeneracy parameter $\mu$ affects the value of
the polytropic constant but its effect is weak when $\mu\gg 1$. The differential
equation (I-33) reduces to 
the Lane-Emden equation (I-77). The results of Paper I can be easily generalized
to account for the $\mu$-dependence of the different quantities when
$k\rightarrow 0$. We get $\tilde{R}\sim {20.0}(1+1/\mu)^2/{\tilde\beta^2}$,
$\tilde\beta\sim 5.77 \, (-\tilde E)^{1/2}(1+1/\mu)$, $\tilde{R}\sim
-3/(5\tilde{E})$, $\tilde\beta=2.02\,  k^{1/3}(1+1/\mu)^{2/3}$, $\tilde{R}\sim
4.90\, k^{-2/3}(1+1/\mu)^{2/3}$, and $\tilde{E}\sim -0.123\,
k^{2/3}(1+1/\mu)^{-2/3}$. We also note that $\epsilon\rightarrow -3/5$,
$\eta\sim -\xi_1 \theta'_1 k\sim 0.409\, k$, ${\cal K}\rightarrow 4.93$,
$\tilde\rho_0\sim 0.584\, k^2 (1+1/\mu)^{-2}$,  ${\tilde \sigma}_0^2\sim 0.141\,
k^{2/3}(1+1/\mu)^{-2/3}$, and $\beta\sigma_0^2\sim (2/7)k$.

\subsection{The completely degenerate limit $k\rightarrow +\infty$}
\label{sec_compdeg}

In the limit $k\rightarrow +\infty$,  we can use the approximation (\ref{df8})
everywhere. As a result, the fermionic King model is equivalent to a pure
polytrope ($p=K\rho^{1+1/n}$) of index $n=3/2$ and polytropic constant
$K=(1/5)\left ({3}/{4\pi \eta_0}\right )^{2/3}$. This corresponds to the
completely degenerate limit, valid at $T=0$, in which the distribution function
is $f=\eta_0 H(\epsilon-\epsilon_F)$ where $H(x)$ is the Heaviside function and
$\epsilon_F=-\alpha/\beta$ is the Fermi energy. We note that, in this limit, the
Fermi energy coincides with the escape energy since Eq. (\ref{fkm2}), which can
be rewritten as $\epsilon_m-\epsilon_F=(1/\beta)\ln\mu$,  reduces to
$\epsilon_F=\epsilon_m$ for $\beta\rightarrow +\infty$.  Defining
$\theta=\chi/k$ and $\xi=\zeta/\sqrt{k}$, we find that the differential equation
(I-33) reduces to the Lane-Emden equation
\begin{equation}
\frac{1}{\xi^2}\frac{d}{d\xi}\left (\xi^2\frac{d\theta}{d\xi}\right )=-\theta^{3/2}
\label{poly1}
\end{equation}
\begin{equation}
\theta(0)=1,\qquad \theta'(0)=0,
\label{poly2}
\end{equation}
corresponding to a polytrope $n=3/2$ \cite{chandra}. Solving this equation numerically, we obtain $\xi_1=3.65$ and $\theta'_1=-0.203$.  Using the theory of polytropes, we can analytically obtain the radius and the energy of the completely degenerate cluster at $T=0$. This corresponds to the ground state of the self-gravitating Fermi gas. Proceeding as in Paper I, we get
\begin{equation}
MR^3=\frac{\chi}{\eta_0^2 G^3}
\label{poly7}
\end{equation}
and
\begin{equation}
E=-\frac{3G^2M^{7/3}\eta_0^{2/3}}{7\chi^{1/3}},
\label{poly9}
\end{equation}
where $\chi=9\,\omega_{3/2}/(2048\pi^4)=5.97\, 10^{-3}$. Introducing the dimensionless variables defined in Paper I, we obtain
\begin{equation}
\tilde{E}_{min}=-\frac{3}{7\chi^{1/3}}\frac{1}{(8\sqrt{2}\pi)^{2/3}}\mu^{2/3}=-0.219\, \mu^{2/3},
\label{poly10b}
\end{equation}
\begin{equation}
\tilde{R}=\chi^{1/3}(8\sqrt{2}\pi)^{2/3}\mu^{-2/3}=1.96\, \mu^{-2/3}.
\label{poly10bb}
\end{equation}
These two quantities are related to each other by $\tilde E=-3/(7\tilde R)$. From Eqs. (I-42), (\ref{poly1}) and (\ref{poly2}), we also find that
\begin{equation}
\tilde\beta\sim \left (\frac{3}{4\pi\mu}\right )^{2/3}(-\xi_1^2 \theta'_1)^{4/3}\, k\sim 1.45\,  \mu^{-2/3} k\rightarrow +\infty.
\label{poly11}
\end{equation}
Finally, we note that $\epsilon\rightarrow -3/7$, $\eta\sim -\xi_1 \theta'_1 k\sim 0.741\, k\rightarrow +\infty$, ${\cal K}\rightarrow (5/2)(1-\xi_1\theta'_1)=4.35$, $\tilde\rho_0\rightarrow (4\pi\mu/3)^2/(-\xi_1^2\theta'_1)^2=2.40\mu^2$, ${\tilde \sigma}_0^2\rightarrow 0.276 \mu^{2/3}$, and $\beta\sigma_0^2\sim (2/5)k$.

\subsection{The degeneracy parameter $\mu$}

An important quantity in the problem is the degeneracy parameter $\mu=\eta_0/A$. Since $A$ has the dimension of a typical distribution function $\langle f\rangle$, the degeneracy parameter can be rewritten as $\mu=\eta_0/\langle f\rangle$. It represents the ratio between the maximum distribution function $\eta_0$ fixed by the Pauli exclusion principle and the typical distribution function of the system $\langle f\rangle$. If we write $\langle f\rangle\sim M R^{-3} V^{-3}$ where $M$ is the typical mass of the system, $R$ its typical radius and $V$ its typical velocity, and use a virial type relation $V\sim (GM/R)^{1/2}$, we obtain $\langle f\rangle\sim G^{-3/2}M^{-1/2}R^{-3/2}$. As a result, the degeneracy parameter $\mu\sim \eta_0 G^{3/2}M^{1/2}R^{3/2}$ coincides, up to a multiplicative constant, with the degeneracy parameter $\mu_{box}$ introduced in the study of box-confined self-gravitating fermions (see Sec. 5.5 of \cite{ijmpb}). As discussed in more detail in \cite{ijmpb}, $\mu$ is a measure of the size of the system. Large values of $\mu$ correspond to large dark matter halos and small values of $\mu$ correspond to small dark matter halos. We shall keep this interpretation in mind in our analysis.

\section{Phase transitions in the framework of the fermionic King model}
\label{sec_ptfk}

A detailed study of phase transitions in the self-gravitating Fermi gas has been performed by Chavanis \cite{pt,dark,ispolatov,rieutord,ptd,epjb14} (see a review in \cite{ijmpb}) in the case where the fermions are confined within a box. In this section, we extend this study to the case of the fermionic King model.  This extension is interesting because this model has a finite mass so it does not require the introduction of an artificial box.

\subsection{Series of equilibria}
\label{sec_proper}

\begin{figure}
\begin{center}
\includegraphics[clip,scale=0.3]{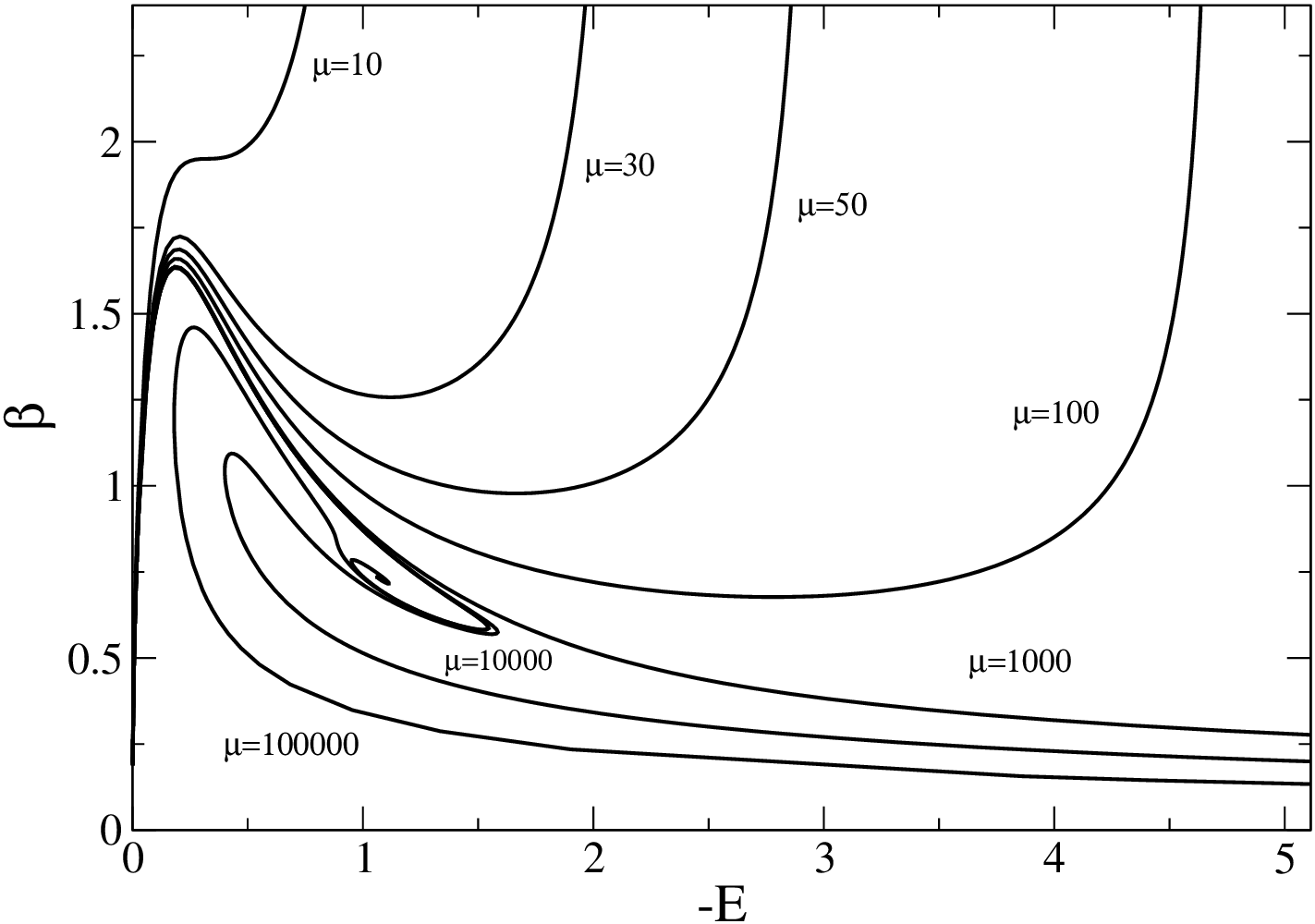}
\caption{Series of equilibria of the fermionic King model for different values of $\mu$. The thick line corresponds to the classical King model ($\mu\rightarrow +\infty$).  Note that for
large values of $\mu$, the minimum energy $E_{min}(\mu)$ corresponding to
$T=0$ is outside the frame of the Figure.}
\label{calomulti}
\end{center}
\end{figure}

\begin{figure}
\begin{center}
\includegraphics[clip,scale=0.3]{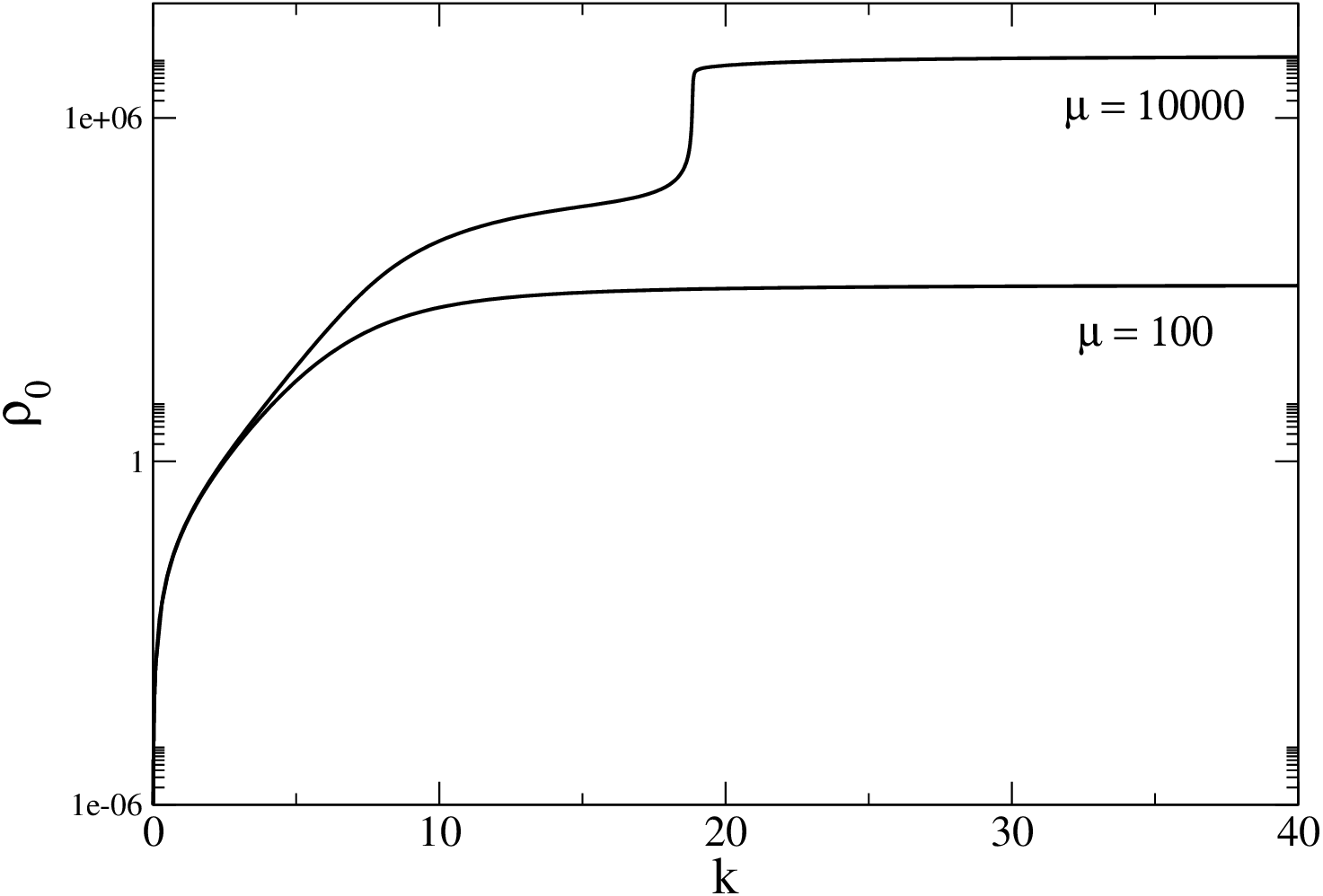}
\caption{Central density normalized by $(4\pi G)^3A^2M^2$ as a function of $k$ in
semi-logarithmic scales.}
\label{concentration}
\end{center}
\end{figure}

The series of equilibria $\beta(E)$ of the fermionic King model is represented in Fig. \ref{calomulti} for different values of $\mu$. The method of construction of the series of equilibria is described in Paper I for a general distribution function of the form $f=f(\epsilon)$ with $f'(\epsilon)<0$. We recall that $A$ is fixed along the series of equilibria and that the thermodynamical parameters $\beta$ and $E$ correspond to the dimensionless parameters $\tilde\beta$ and $\tilde E$ of Paper I. On the other hand, $S$ and $J$ refer to $S/M$ and $J/M$. We also recall that, for each value of $\mu$, the series of equilibria is parameterized by the concentration
parameter $k=\beta(\epsilon_m-\Phi_0)$ going from $k=0$ at $(E, \beta)=(0,0)$ to $k=+\infty$ at  $(E, \beta)=(E_{min},+\infty)$ where $E_{min}(\mu)$ is the minimum accessible energy (ground state) corresponding to $T=0$ [see Eq. (\ref{poly10b})]. The concentration parameter $k$ is a monotonically increasing function of the normalized central density as shown in Fig. \ref{concentration}.

The shape of the series of equilibria $\beta(E)$ of the fermionic King model crucially depends on the value of the degeneracy parameter $\mu$ as shown in Fig. \ref{calomulti}. In the non degenerate limit $\mu\rightarrow +\infty$, we recover the spiral corresponding to the classical King model (see  Fig. \ref{ckm}). However, for smaller values of $\mu$, we see
that the effect of quantum mechanics  (Pauli exclusion principle)  is to unwind the spiral.  Depending on
the value of $\mu$, the series of equilibria can
have different shapes. In the following, we consider two typical
series of equilibria, one corresponding to a relatively large value of $\mu$ equal to $10^4$
(Sec. \ref{sec_large}) and one corresponding to a relatively small
value of $\mu$ equal to $100$ (Secs. \ref{sec_small} and \ref{sec_small2}).

\subsection{Large halos in MCE: $\mu=10^4$}
\label{sec_large}

\begin{figure}
\begin{center}
\includegraphics[clip,scale=0.3]{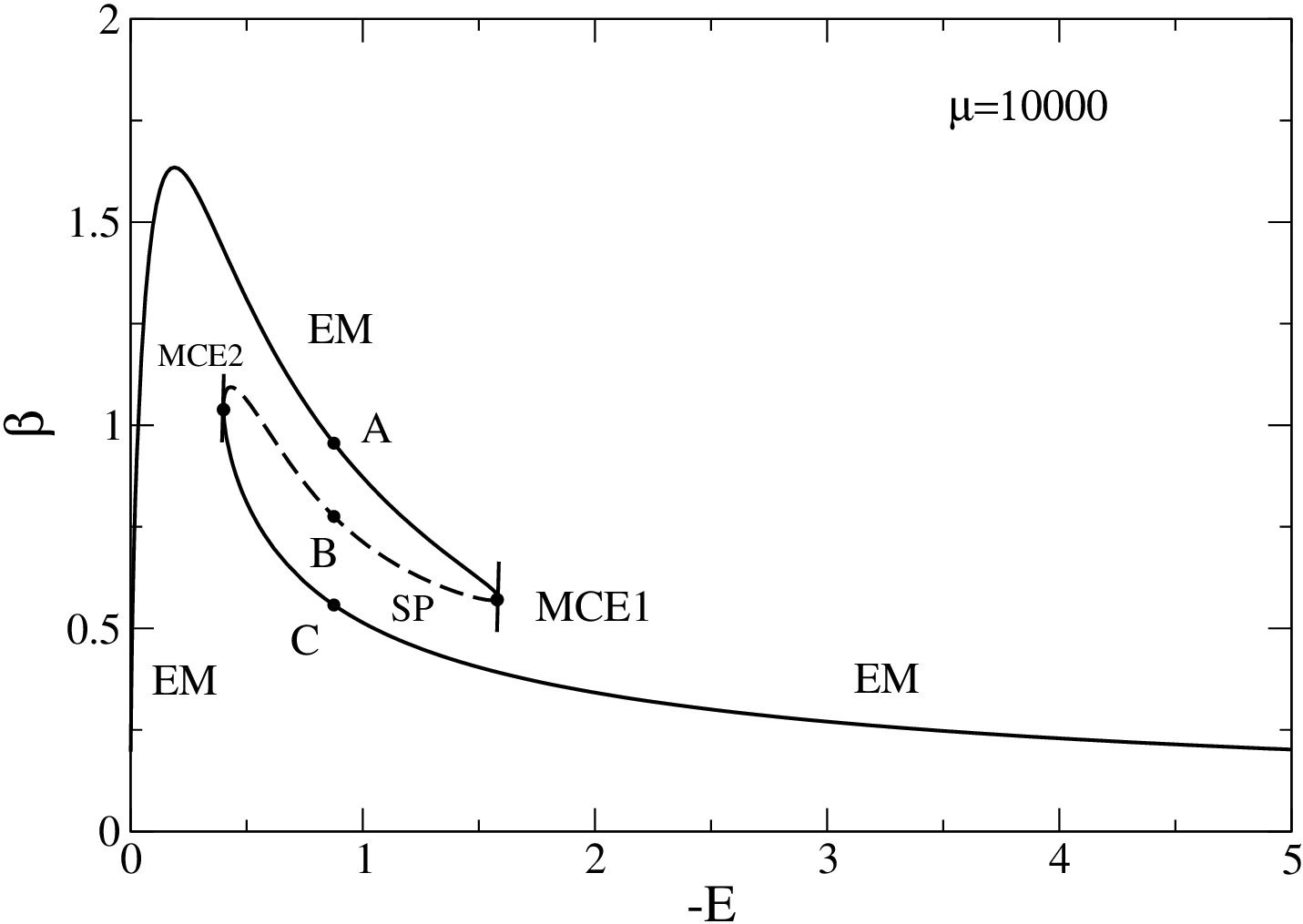}
\caption{Series of equilibria of the fermionic King model
 with $\mu=10^4$.}
\label{ETmicro5instable}
\end{center}
\end{figure}

For $\mu=10^4$ (large halos), the series of equilibria of the fermionic King model  is represented in Fig.
\ref{ETmicro5instable}. It has a $Z$-shape structure.

In this section and in the next one, we assume that the system is isolated. In that case, the control parameter is
the energy $E$ and the relevant statistical ensemble is MCE.  In MCE, we must determine maxima of entropy at
fixed mass and energy.  Since the  curve $\beta(E)$ is multi-valued, phase transitions occur in MCE. Using the Poincar\'e theorem (see Paper I), we deduce that all the states on the upper branch of the series of equilibria are  entropy maxima (EM) until the first turning point of energy MCE1. For large values of $\mu$, this critical energy is close to the energy $E_{c}=-1.54$ corresponding to the classical King model ($\mu\rightarrow +\infty$). At that
point, the curve turns clockwise so that a mode of stability is
lost. This mode of stability is regained at the second turning point
of energy MCE2 at which the curve turns anti-clockwise. The
corresponding energy $E_{*}(\mu)$ depends on the value of
$\mu$  and tends to $E_{*}(\mu)\rightarrow 0$ for $\mu\rightarrow +\infty$. The configurations on the  branch
between MCE1 and MCE2 are saddle points (SP) of entropy
while the configurations on the lower branch after MCE2 are entropy
maxima (EM).

\begin{figure}
\begin{center}
\includegraphics[clip,scale=0.3]{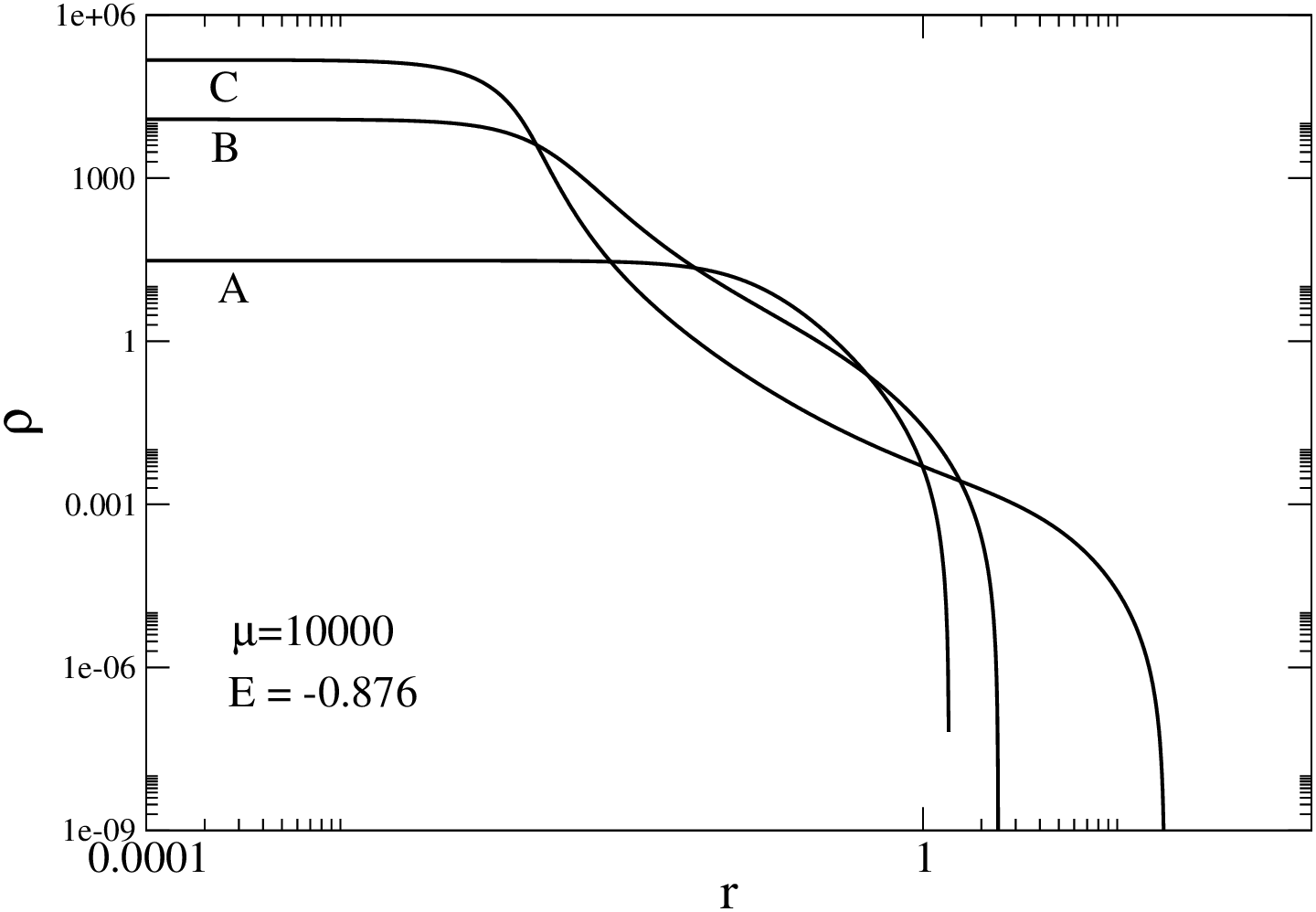}
\caption{Normalized density profiles corresponding to the different phases of the fermionic
King model with $\mu=10^4$ and $E=-0.876$. Here and in the following figures, the radial distance is scaled by $1/(4\pi G M^{1/3}A^{2/3})$
and the density by $(4\pi G)^3A^2M^2$. The ``core-halo'' structure of solutions B and C comprising a dense
degenerate nucleus (fermion ball) surrounded by an atmosphere is clearly visible.}
\label{profilesREAL}
\end{center}
\end{figure}

\begin{figure}
\begin{center}
\includegraphics[clip,scale=0.3]{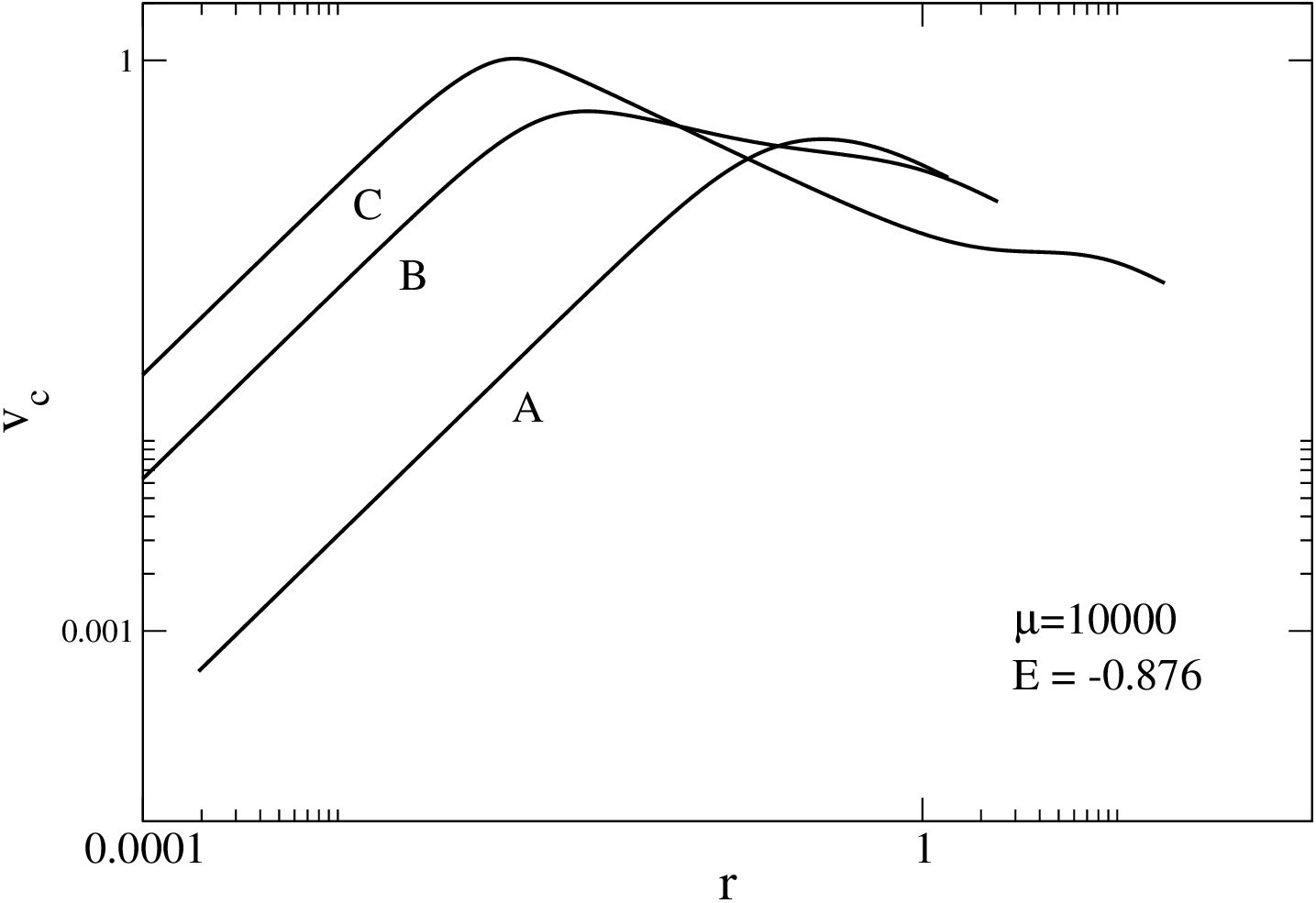}
\caption{Normalized rotation curves corresponding to the different phases of the fermionic
King model with $\mu=10^4$ and $E=-0.876$. Here and in the following figures, the radial distance is scaled by $1/(4\pi G M^{1/3}A^{2/3})$
and the circular velocity by $4\pi GA^{1/3}M^{2/3}$.}
\label{profilesVREAL}
\end{center}
\end{figure}

The solutions on the upper branch are stable (EM). They are non degenerate
and have a smooth density profile. They form the ``gaseous phase'' unaffected by quantum mechanics (see solution $A$
in Figs. \ref{profilesREAL} and \ref{profilesVREAL}).
The solutions on the lower branch are also stable  (EM). They  have a core-halo structure
consisting of  a degenerate nucleus (fermion ball) surrounded by a dilute  atmosphere (vapor).  They
form the ``condensed phase'' dominated by quantum mechanics (see solution $C$ in Figs. \ref{profilesREAL} and \ref{profilesVREAL}).
The nucleus (condensate) is equivalent to a completely  degenerate self-gravitating fermion ball
 at $T=0$ with the maximum phase space density $\eta_0$. It is stabilized against gravitational collapse by the Pauli exclusion principle. The solutions on the intermediate  branch are unstable  (SP). They are similar to the solutions of the gaseous phase but they contain a small embryonic
degenerate nucleus playing the role of a ``germ'' in the language of phase transitions (see solution $B$ in
Figs. \ref{profilesREAL} and \ref{profilesVREAL}). These solutions form a
barrier of entropy that the
system has to cross in order to pass from the gaseous phase to the condensed phase,
or inversely (see Ref. \cite{ijmpb} for more details).

\begin{figure}
\begin{center}
\includegraphics[clip,scale=0.3]{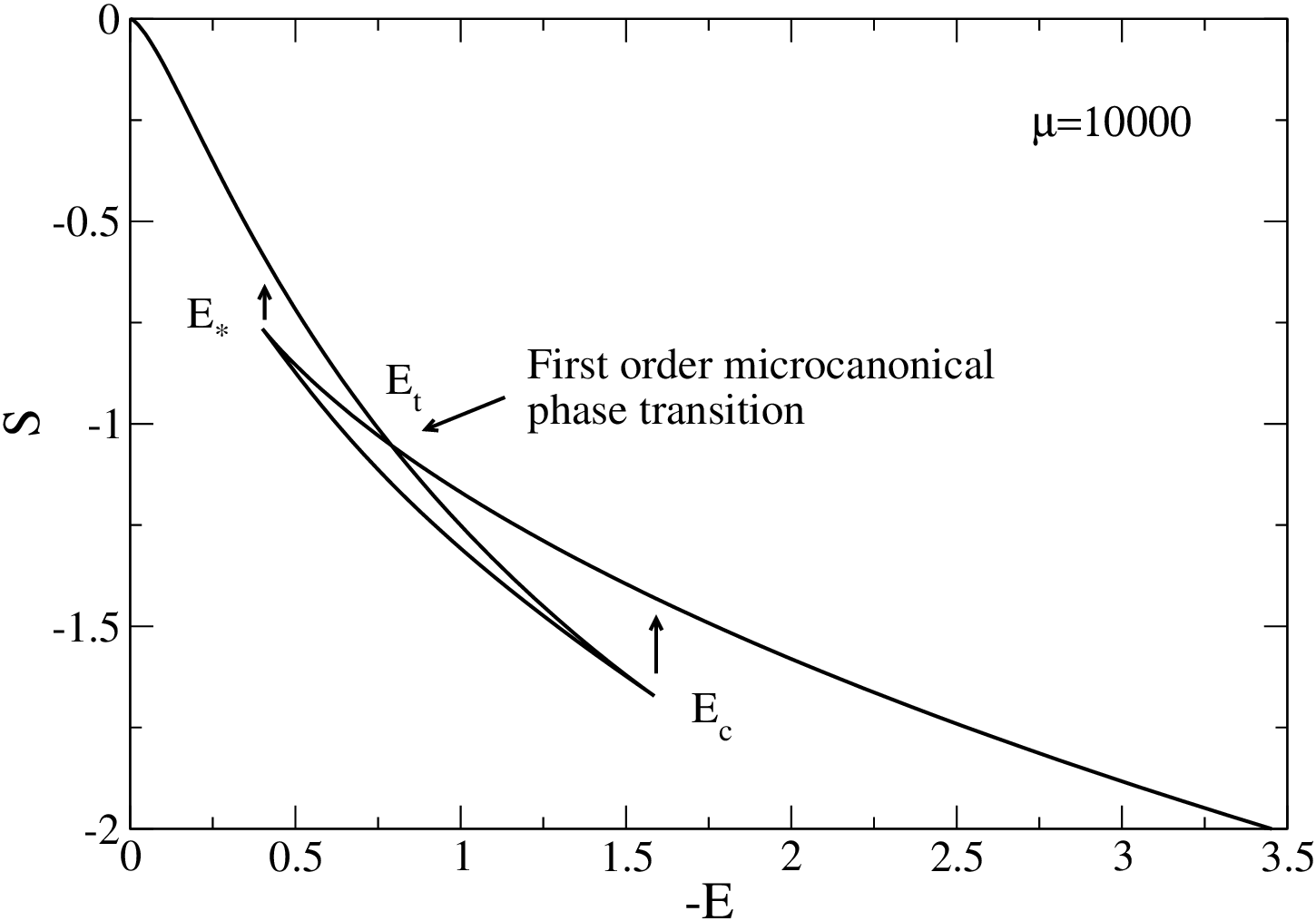}
\caption{Entropy of each phase versus
energy for $\mu=10^4$.}
\label{Smicro5}
\end{center}
\end{figure}

\begin{figure}
\begin{center}
\includegraphics[clip,scale=0.3]{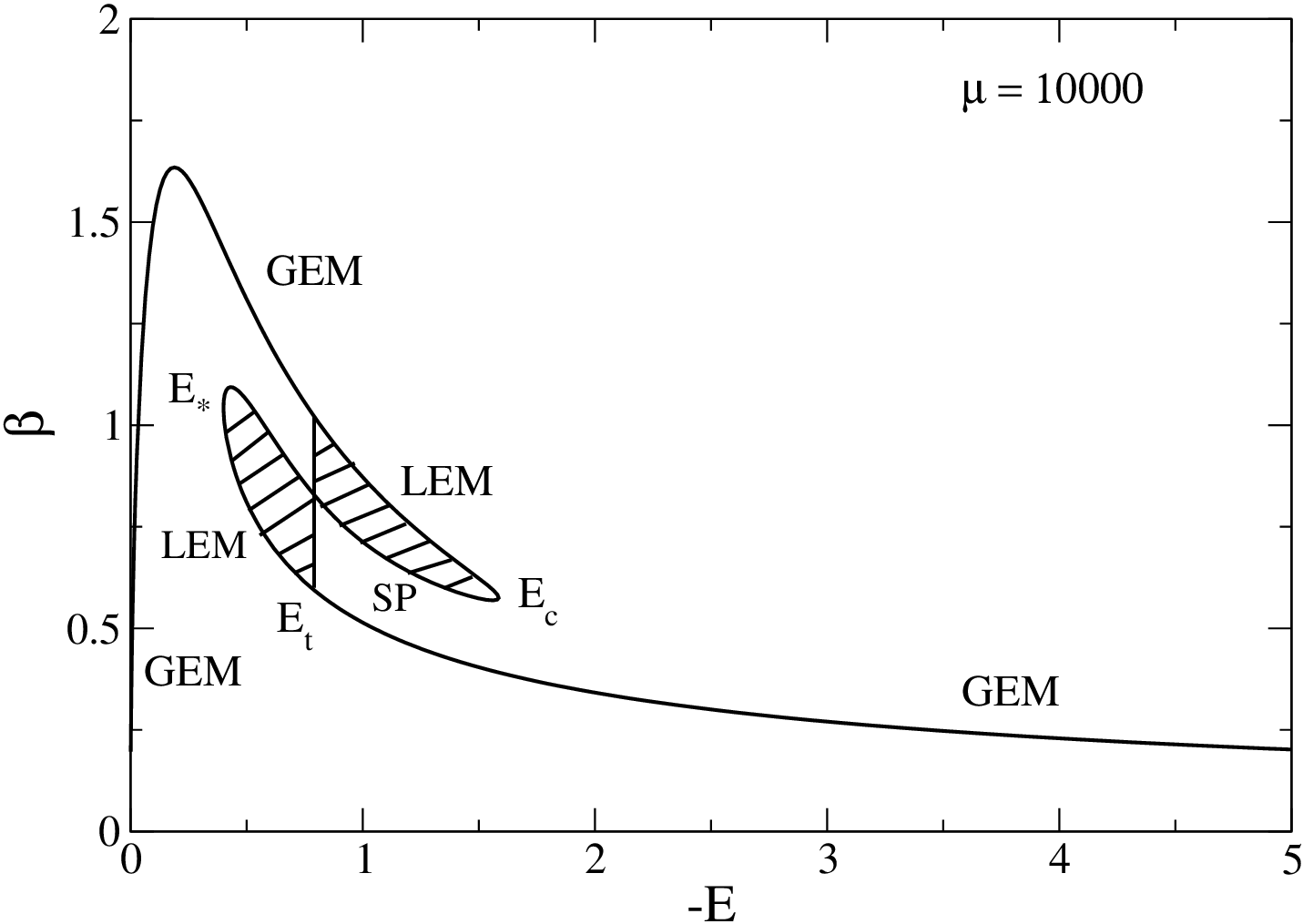}
\caption{Maxwell construction for $\mu=10^4$.}
\label{ETmicro5}
\end{center}
\end{figure}

\begin{figure}
\begin{center}
\includegraphics[clip,scale=0.3]{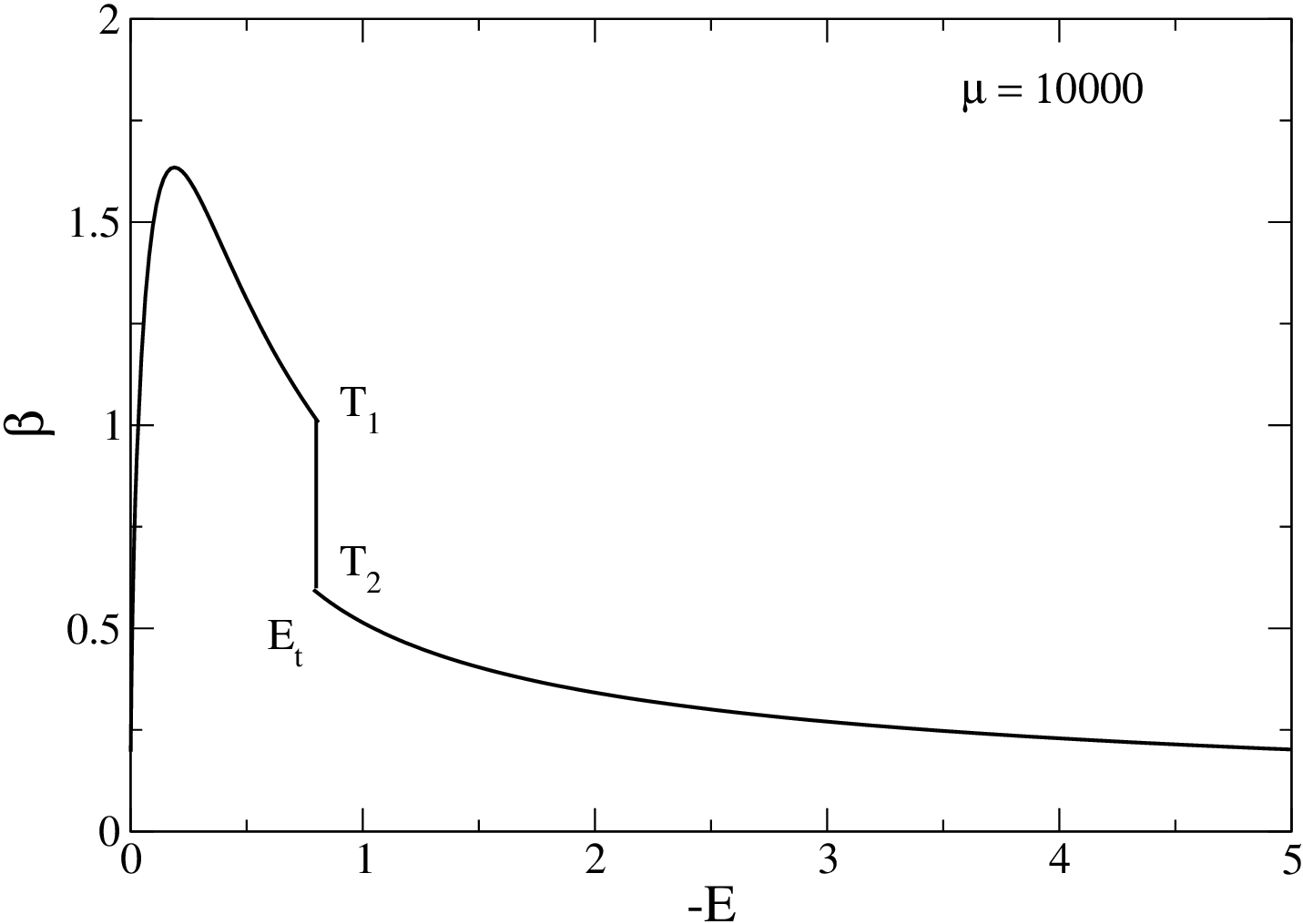}
\caption{Strict microcanonical caloric curve for $\mu=10^{4}$.}
\label{ETmicro5strict}
\end{center}
\end{figure}

If we compare the entropy of the solutions (see Fig. \ref{Smicro5}), we expect a first
order microcanonical phase transition to take place at a transition energy
$E_t(\mu)$ where the entropy of the gaseous phase and the entropy of the
condensed phase become equal.
The transition energy $E_t(\mu)$ may also be obtained by performing a Maxwell  construction
(see the vertical plateau in Fig. \ref{ETmicro5}) \cite{ijmpb}. For $E>E_t$ the gaseous phase is fully stable (global entropy maximum GEM at fixed
mass and energy) while the condensed phase is metastable (local entropy maximum LEM
at fixed mass and energy). For $E<E_t$ the gaseous phase is metastable (LEM) while
the condensed phase is fully stable (GEM). The strict caloric curve is obtained by keeping only the fully stable states (see Fig. \ref{ETmicro5strict}). It is marked by a discontinuity of the inverse
temperature $\beta=\partial S/\partial E$ at $E=E_t(\mu)$. Equivalently, the first derivative of the entropy is discontinuous
at the transition  (see Fig. \ref{Smicro5}). This characterizes a microcanonical first order phase transition. The specific heat $C=dE/dT$ is also discontinuous at the transition. If $E_{t}>E_{gas}$
(where $E_{gas}$ is the energy corresponding to the first turning
point of temperature) the specific heat passes from a positive to
a negative value. If $E_{t}<E_{gas}$, the specific heat is always
negative at the transition (the crossover occurs for
$\mu\simeq 8.02\, 10^5$ and $E_t=E_{gas}=-0.189$; see
the intersection between $E_t$ and $E_{gas}$ in Fig. \ref{phasemicro}).

\begin{figure}
\begin{center}
\includegraphics[clip,scale=0.3]{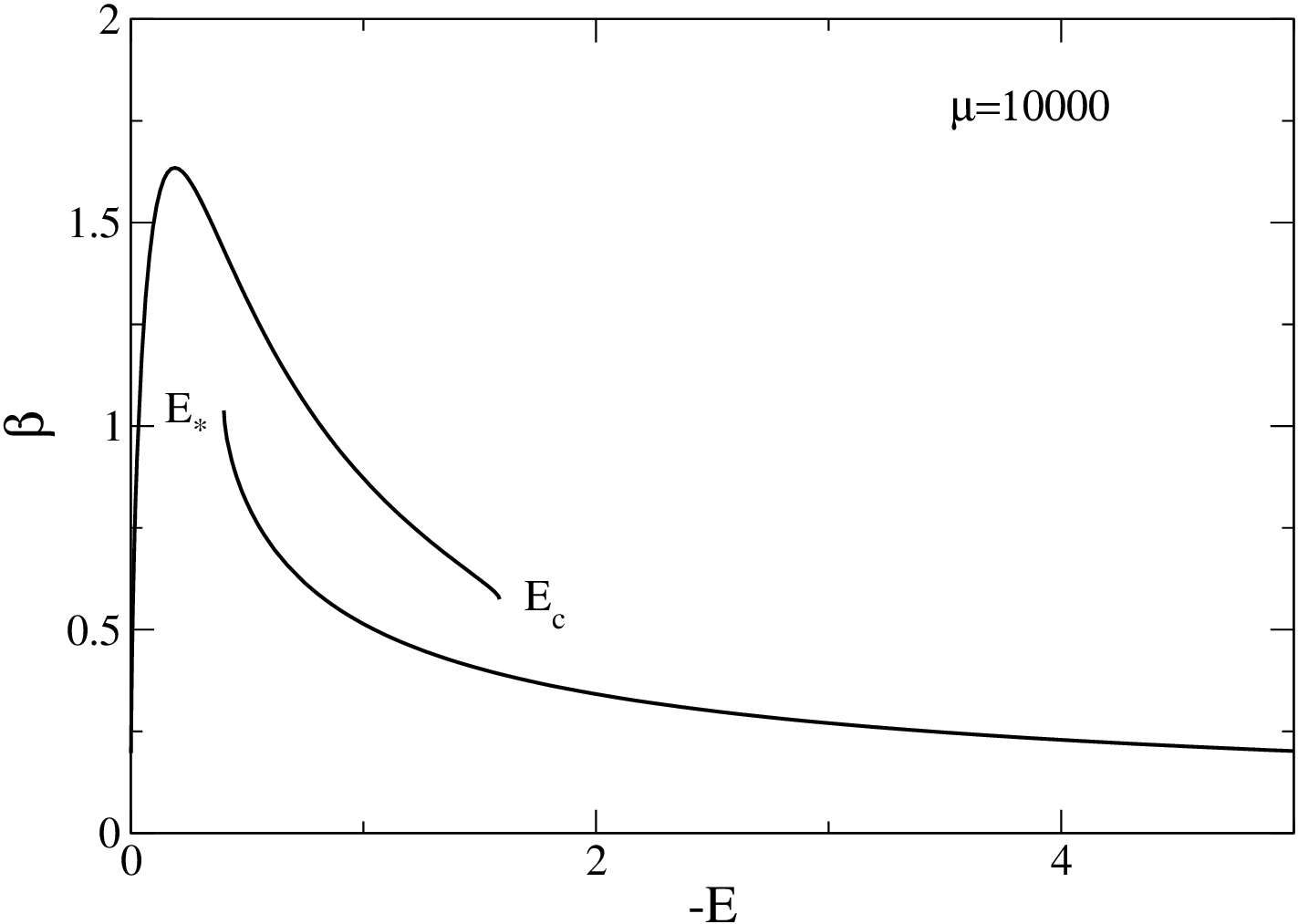}
\caption{Physical microcanonical caloric curve for $\mu=10^{4}$. }
\label{ETmicro5meta}
\end{center}
\end{figure}

However, for systems with long-range interactions, the metastable states are long-lived  because the probability that a fluctuation triggers a phase transition and drives the system towards the fully stable state is extremely weak.  Indeed, the
system has to cross the entropic barrier played by the solution on
the intermediate branch.\footnote{To pass from the gaseous phase to the condensed phase, the system  must spontaneously form a small nucleus  where the particles are closely packed together. To pass from the condensed phase to the gaseous phase, the system  must spontaneously form a massive atmosphere.}
 For systems with long-range interactions, the hight of the entropic barrier scales linearly with the number $N$ of particles  and, consequently, the probability of transition scales like $e^{-N}$.  For $N\gg 1$, this is a very rare event. Therefore, the metastable states are
extremely robust.  They have considerably large lifetimes scaling as $e^N$.  The microcanonical first order phase
transition at $E_t$ does {\it not} take place in practice and, for
sufficiently large values of $N$, the system remains frozen in
the metastable phase past the transition energy $E_t$. Accordingly, the strict caloric curve of Fig.
\ref{ETmicro5strict} is not physical. The physical microcanonical caloric
curve is the one shown in Fig. \ref{ETmicro5meta} which takes the metastable states
into account. It is obtained from the series of
equilibria of Fig. \ref{ETmicro5instable} by discarding only the
unstable saddle points of entropy that form the intermediate branch.

\begin{figure}
\begin{center}
\includegraphics[clip,scale=0.3]{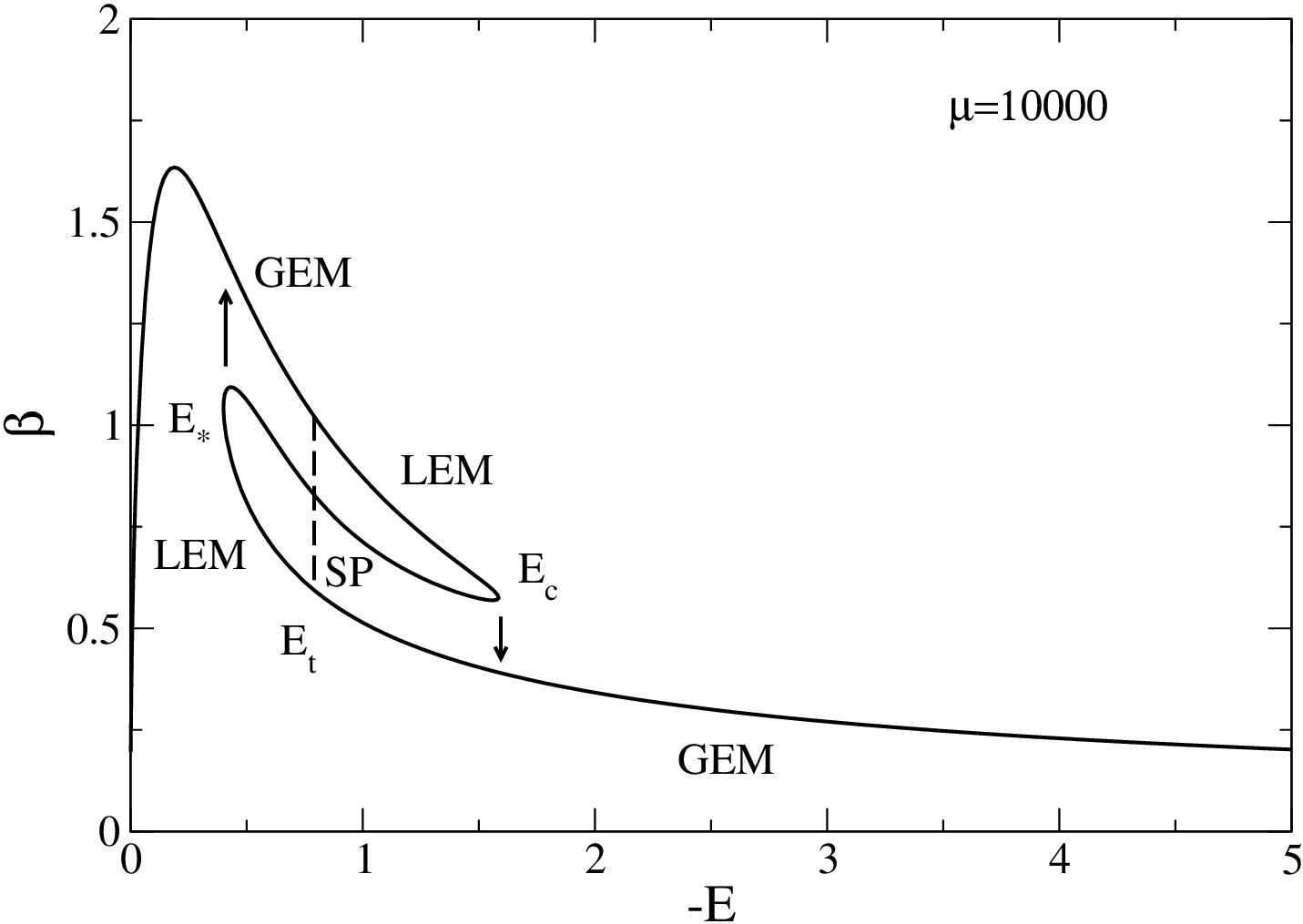}
\caption{Summary of the phase transitions of the fermionic King model
with $\mu=10^{4}$ in MCE. }
\label{ETmicro5summary}
\end{center}
\end{figure}

The phase transitions of the fermionic King model are summarized in Fig. \ref{ETmicro5summary}.
At $E=0^{-}$, the system is in the gaseous phase where quantum mechanics is completely negligible. At some transition energy $E_{t}$, a first order phase transition is expected to occur and drive the system towards the condensed phase dominated by quantum mechanics.  However,
gaseous states are still metastable, and long-lived,  beyond this
point so the first order phase transition does  not take place in practice. If we decrease the
energy, the system remains in the gaseous phase until the critical energy $E_c(\mu)$ at
which the gaseous phase disappears. This is similar to a  spinodal point in the language of phase transitions.  For $E<E_c(\mu)$,
the system undergoes a gravitational collapse (gravothermal catastrophe). This corresponds to a
saddle-node bifurcation. However, the
collapse stops when the core of the system becomes degenerate.
In that case, it ends up in the condensed phase. The system has a ``core-halo'' structure with a
degenerate nucleus surrounded by a non-degenerate atmosphere. The condensate results from the balance between the gravitational attraction  and the pressure due to the Pauli exclusion principle. This is a very compact object equivalent to a completely degenerate fermion ball at $T=0$. Since the collapse is
accompanied by a discontinuous jump of entropy (see Fig. \ref{Smicro5}),
this is sometimes called a microcanonical zeroth order phase
transition. If we now increase the energy, the system remains in the condensed
phase until the critical energy $E_*(\mu)$ at which the condensed phase
disappears. Indeed, the first order phase transition expected at $E_t(\mu)$ does not take place in practice due to the long lifetime
of the metastable states.  For $E>E_*(\mu)$, the system undergoes an ``explosion'' reversed to
the collapse
and returns to the gaseous phase. In this sense, we can describe an hysteretic
cycle in MCE (see the arrows in Figs. \ref{Smicro5} and \ref{ETmicro5summary}).
These microcanonical phase transitions exist only above a microcanonical
 critical point $\mu_{MCP}=1980$ (see Sec. \ref{sec_maxwell}).

\subsection{Small halos in MCE: $\mu=100$}
\label{sec_small}

\begin{figure}
\begin{center}
\includegraphics[clip,scale=0.3]{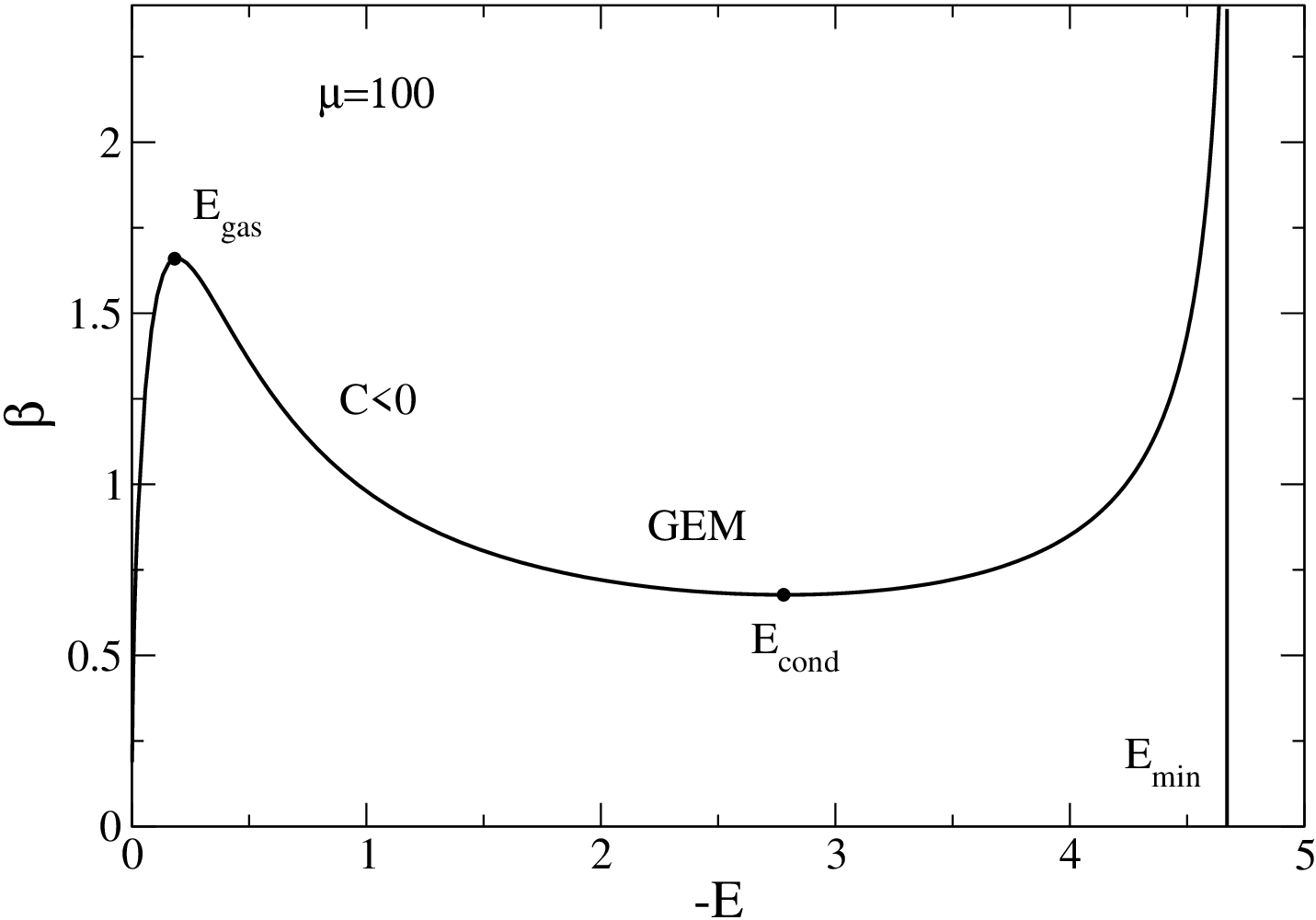}
\caption{Series of equilibria of the fermionic King model with $\mu=100$.}
\label{ETmicro3}
\end{center}
\end{figure}

\begin{figure}
\begin{center}
\includegraphics[clip,scale=0.3]{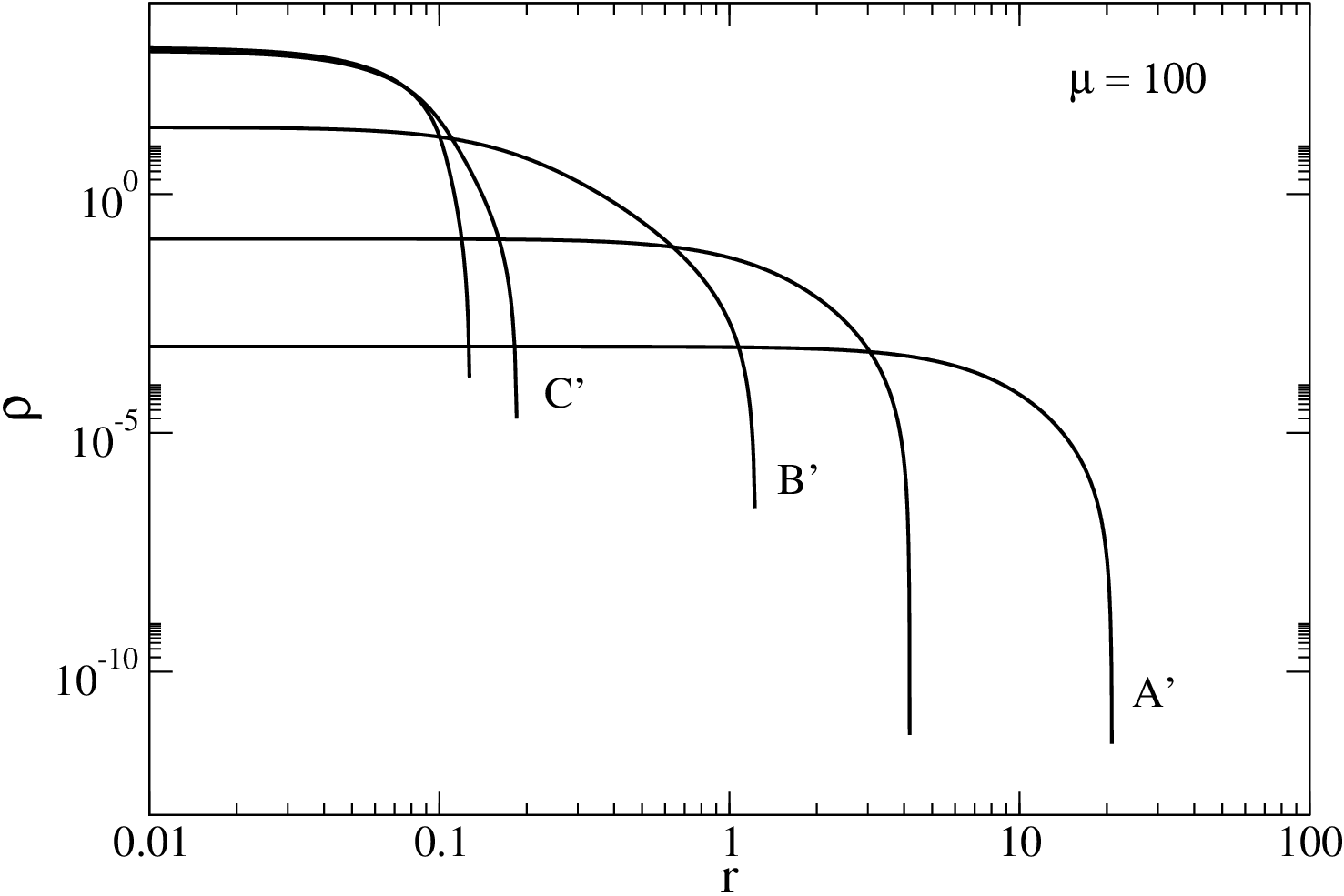}
\caption{Normalized density profiles along the series of equilibria with $\mu=100$ in
logarithmic scales. We have selected $k=0.142$ ($E=-0.0329$, $\beta=1.03$; solution A'),
$k=1.31$ ($E=-0.181$, $\beta=1.66$), $k=4.99$ ($E=-0.914$, $\beta=1.03$; solution B'),
$k=18.0$ ($E=-4.27$, $\beta=1.03$; solution C'), and $k=41$ ($E=-4.65$,
$\beta=2.51$). The central density increases as the energy decreases.}
\label{densityMu100}
\end{center}
\end{figure}

\begin{figure}
\begin{center}
\includegraphics[clip,scale=0.3]{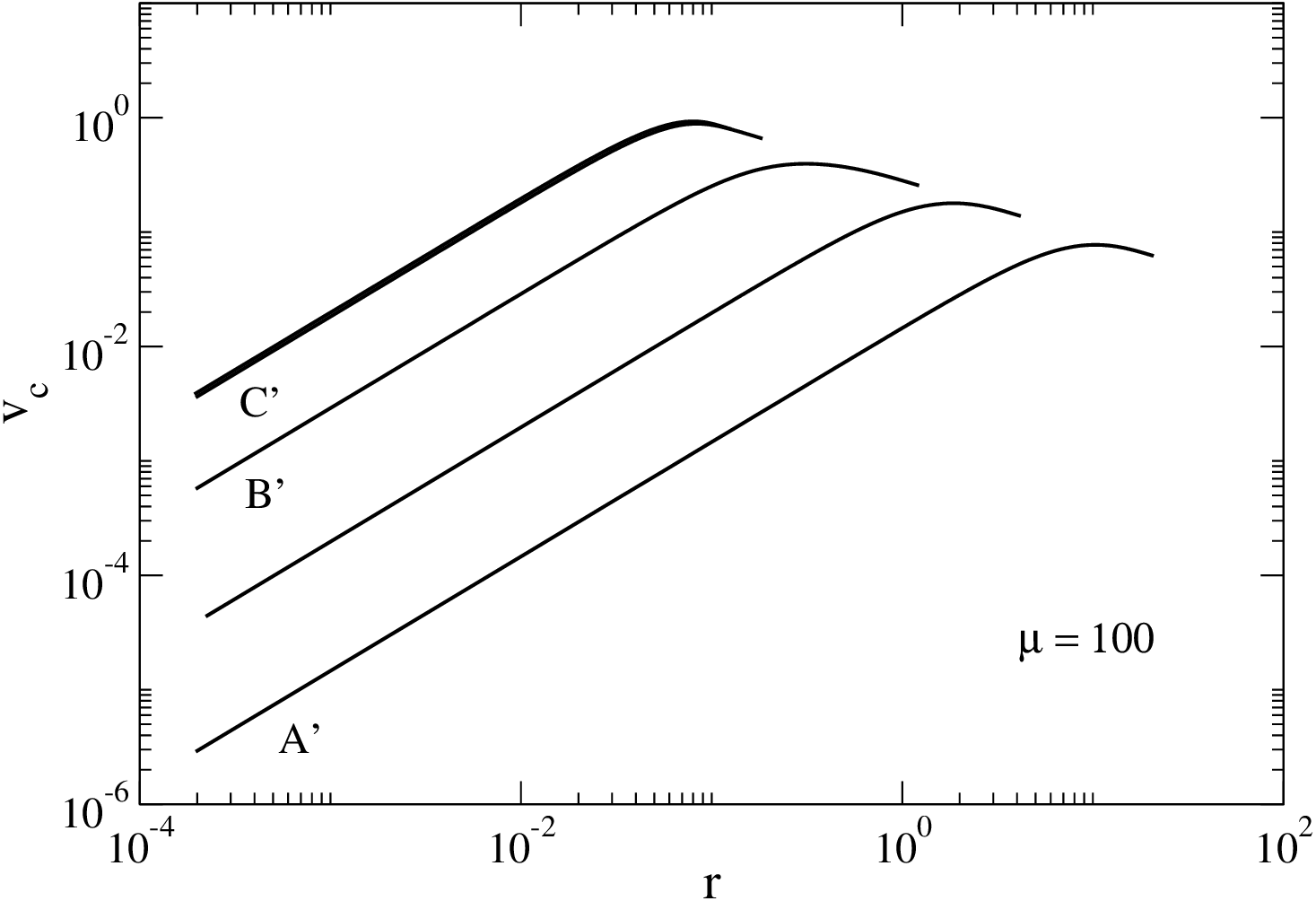}
\caption{Normalized rotation curves along the series of equilibria with $\mu=100$ in
logarithmic scales.}
\label{velocityMu100}
\end{center}
\end{figure}

For $\mu=100$ (small halos), the series of equilibria is represented in Fig.
\ref{ETmicro3}. It has an $N$-shape structure. Since the  curve $\beta(E)$ is
univalued there is no phase transition in MCE. All the configurations are
fully stable (GEM). However, there is a sort of condensation
(clustering) as the energy is progressively decreased (see Figs. \ref{densityMu100} and
\ref{velocityMu100}). At high
energies, the equilibrium states are non degenerate. At intermediate
energies, between the energies $E_{gas}$ and $E_{cond}$ corresponding to the
extrema of temperature, the caloric curve displays a region of negative specific heats
($C=dE/dT<0$). In this region, the equilibrium states
have a ``core-halo'' structure with a partially degenerate nucleus
and a non degenerate enveloppe (atmosphere). As energy is further decreased, the nucleus becomes more and more
degenerate and contains more and more mass. At the minimum energy
$E_{min}$, corresponding to $T=0$, all the mass is in the completely
degenerate nucleus.  In that case, the atmosphere has been
swallowed and the system reduces to a pure fermion ball with a maximum phase space density
$\eta_0$ fixed by the Pauli exclusion principle.

\begin{figure}
\begin{center}
\includegraphics[clip,scale=0.3]{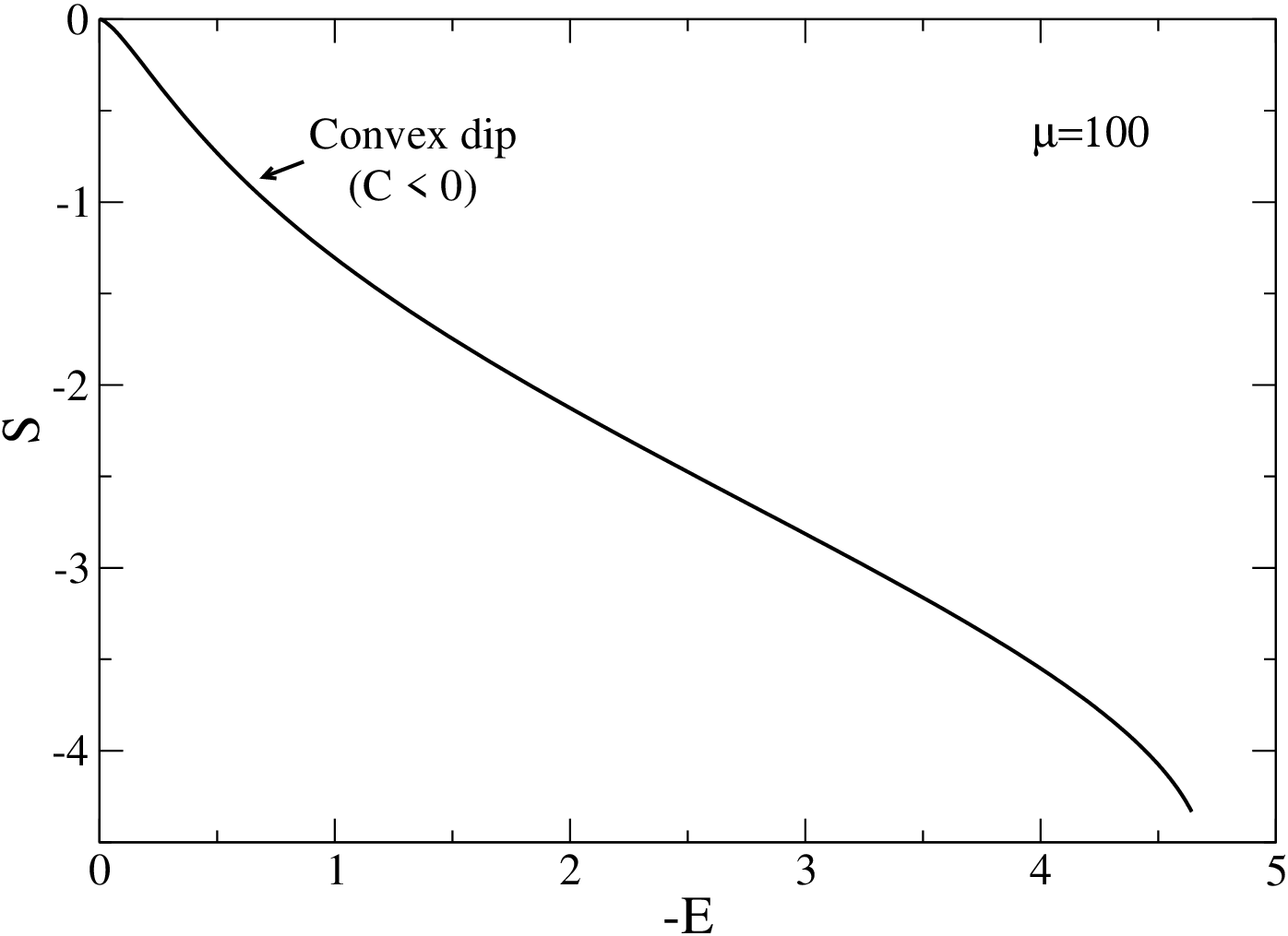}
\caption{Entropy versus
energy for $\mu=100$.}
\label{Smicro3}
\end{center}
\end{figure}

The entropy versus energy curve $S(E)$ is represented in Fig. \ref{Smicro3}. Since
$S''(E)=-1/(CT^{2})$, the entropy displays a convex intruder
($S''>0$) in the region of negative specific heat ($C<0$).  For systems with long-range interactions,
for which the energy is non-additive, the region of negative specific heat in the caloric curve (see Fig. \ref{ETmicro3}) and the
convex intruder   in the entropy versus energy curve (see Fig. \ref{Smicro3}) are allowed in  MCE.

\subsection{Small halos in CE: $\mu=100$}
\label{sec_small2}

\begin{figure}
\begin{center}
\includegraphics[clip,scale=0.3]{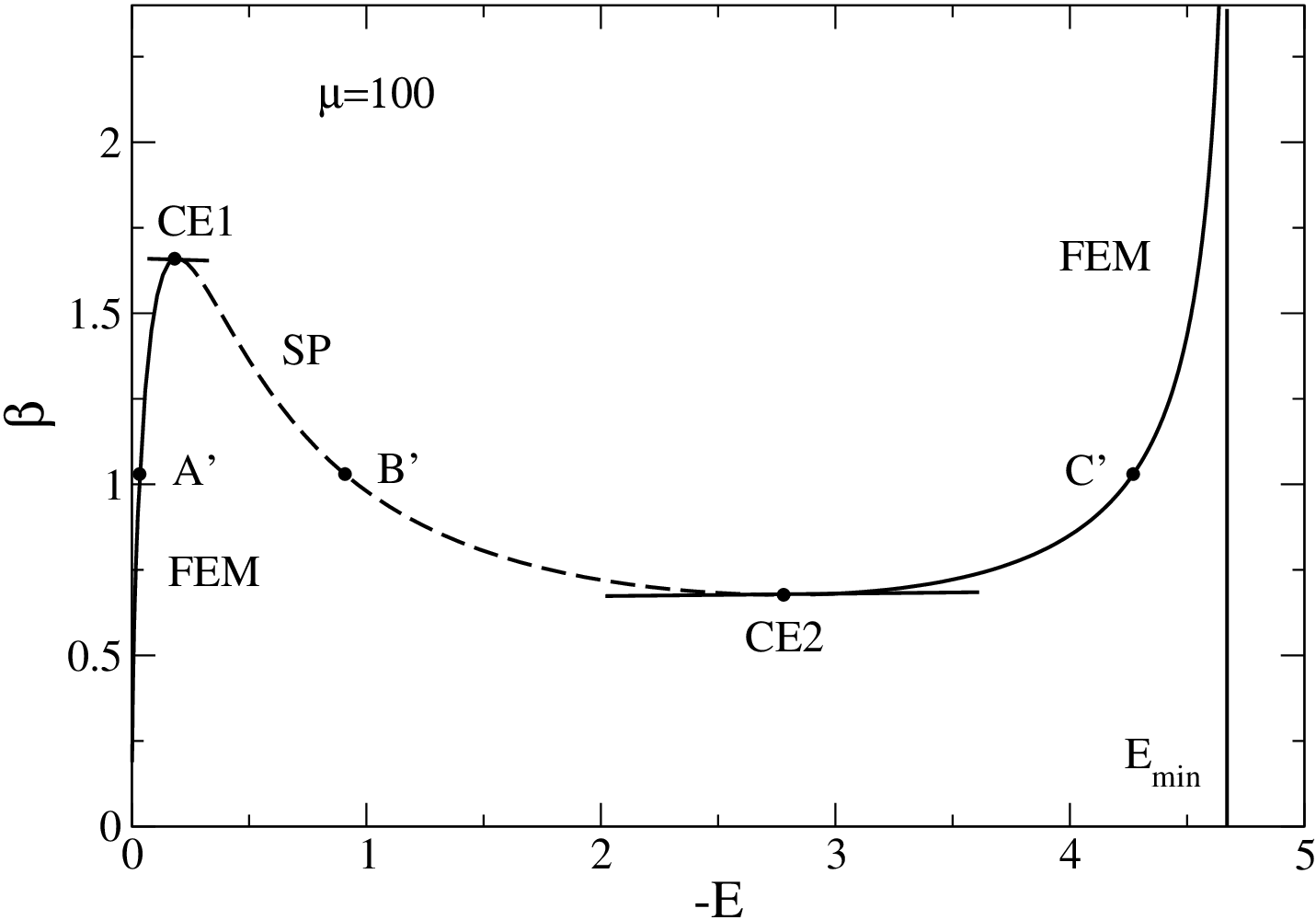}
\caption{Series of equilibria of the fermionic King model with $\mu=100$. }
\label{ETcano3instable}
\end{center}
\end{figure}

In this section, we consider a dissipative system in contact with a thermal bath imposing its temperature $T$.
In that case, the control parameter is
the temperature $T$ and the relevant statistical ensemble is CE.  In CE, we must determine maxima of free energy at
fixed mass. Considering again the case $\mu=100$ (small halos), we note that the series of
equilibria $E(\beta)$ represented in Fig. \ref{ETcano3instable} is multi-valued. This gives rise to
canonical phase transitions. Using the Poincar\'e theorem, we deduce that all the states on the left branch of the series of equilibria are free energy maxima (FEM) until the first turning point of temperature CE1. For large values of $\mu$, this critical temperature is close to the temperature $T_{c}=0.613$ corresponding to the classical King model ($\mu\rightarrow +\infty$). At that
point, the curve turns clockwise so that a mode of stability is
lost. This mode of stability is regained at the second turning point
of temperature CE2 at which the curve turns anti-clockwise. The
corresponding temperature $T_{*}(\mu)$ depends on the value of
$\mu$  and tends to $T_{*}(\mu)\rightarrow +\infty$ for $\mu\rightarrow +\infty$. The configurations on the  branch
between CE1 and CE2 are saddle points (SP) of free energy while the configurations on the right branch after CE2 are free energy
maxima (FEM).

The configurations on the left branch are stable (FEM). They form the ``gaseous phase'' (see solution A' in Figs. \ref{densityMu100} and
\ref{velocityMu100}).  The solutions on the right branch are also stable  (FEM).  They form the ``condensed phase'' (see solution C' in Figs. \ref{densityMu100} and \ref{velocityMu100}).  The solutions on the intermediate  branch are unstable  (SP).  These solutions (see solution B' in Figs. \ref{densityMu100} and \ref{velocityMu100}) form a  barrier of free energy that the system has to cross in order to pass from the gaseous phase to the condensed phase, or inversely \cite{ijmpb}.

\begin{figure}
\begin{center}
\includegraphics[clip,scale=0.3]{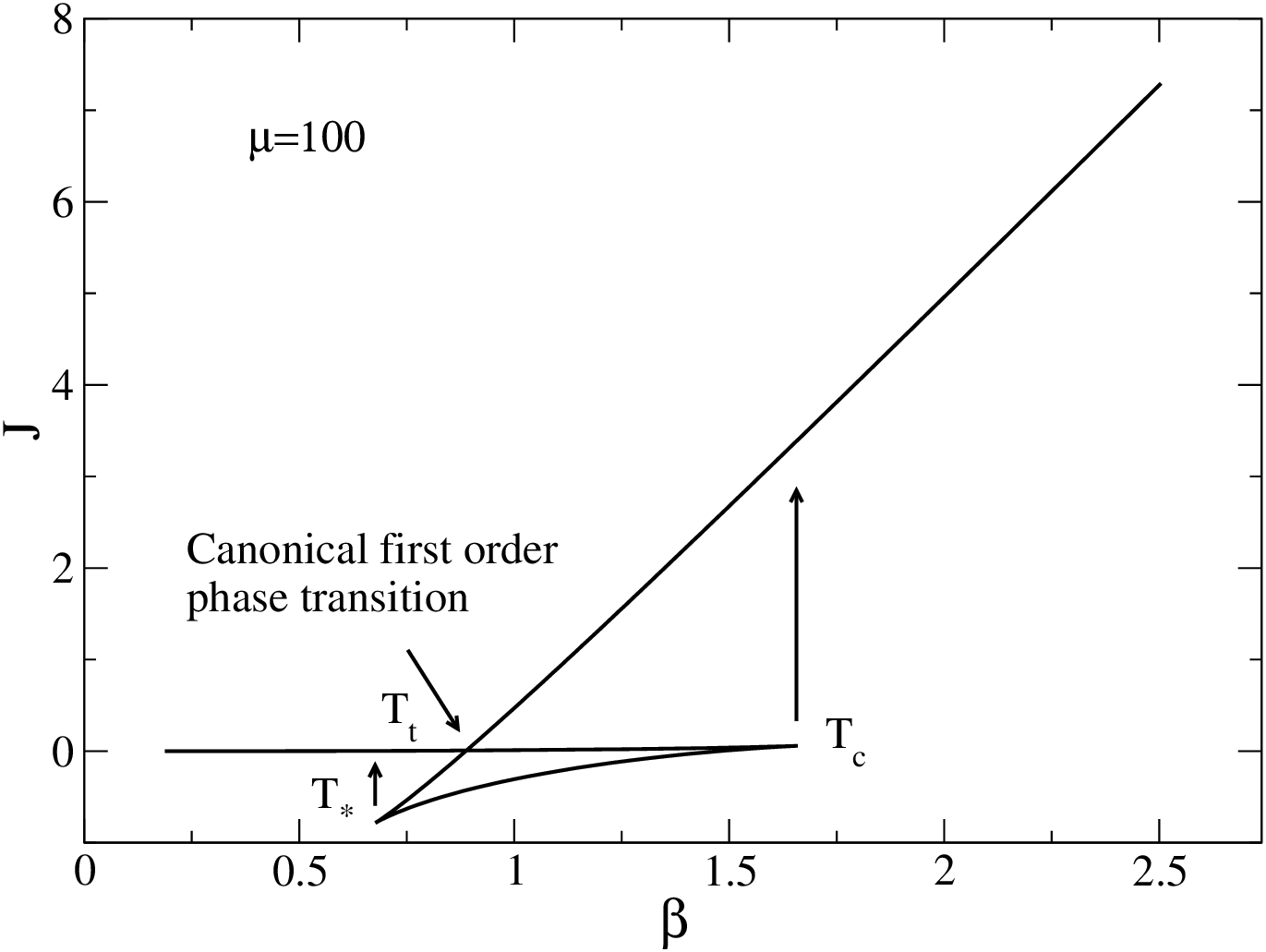}
\caption{Free energy of each phase versus
temperature for $\mu=100$.}
\label{Fcano3}
\end{center}
\end{figure}

\begin{figure}
\begin{center}
\includegraphics[clip,scale=0.3]{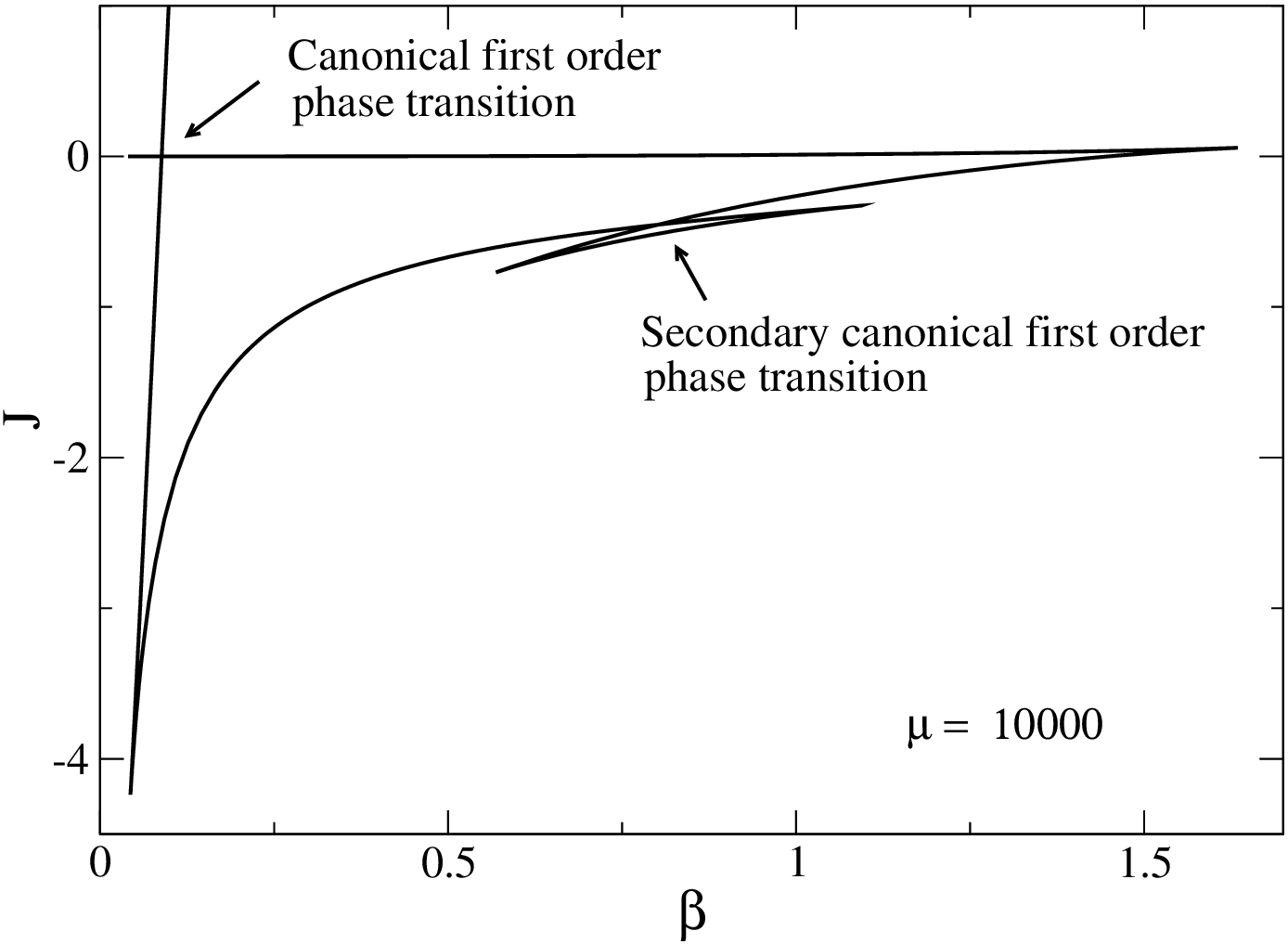}
\caption{Free energy versus
temperature for $\mu=10^4$. }
\label{secondaryCPT}
\end{center}
\end{figure}

\begin{figure}
\begin{center}
\includegraphics[clip,scale=0.3]{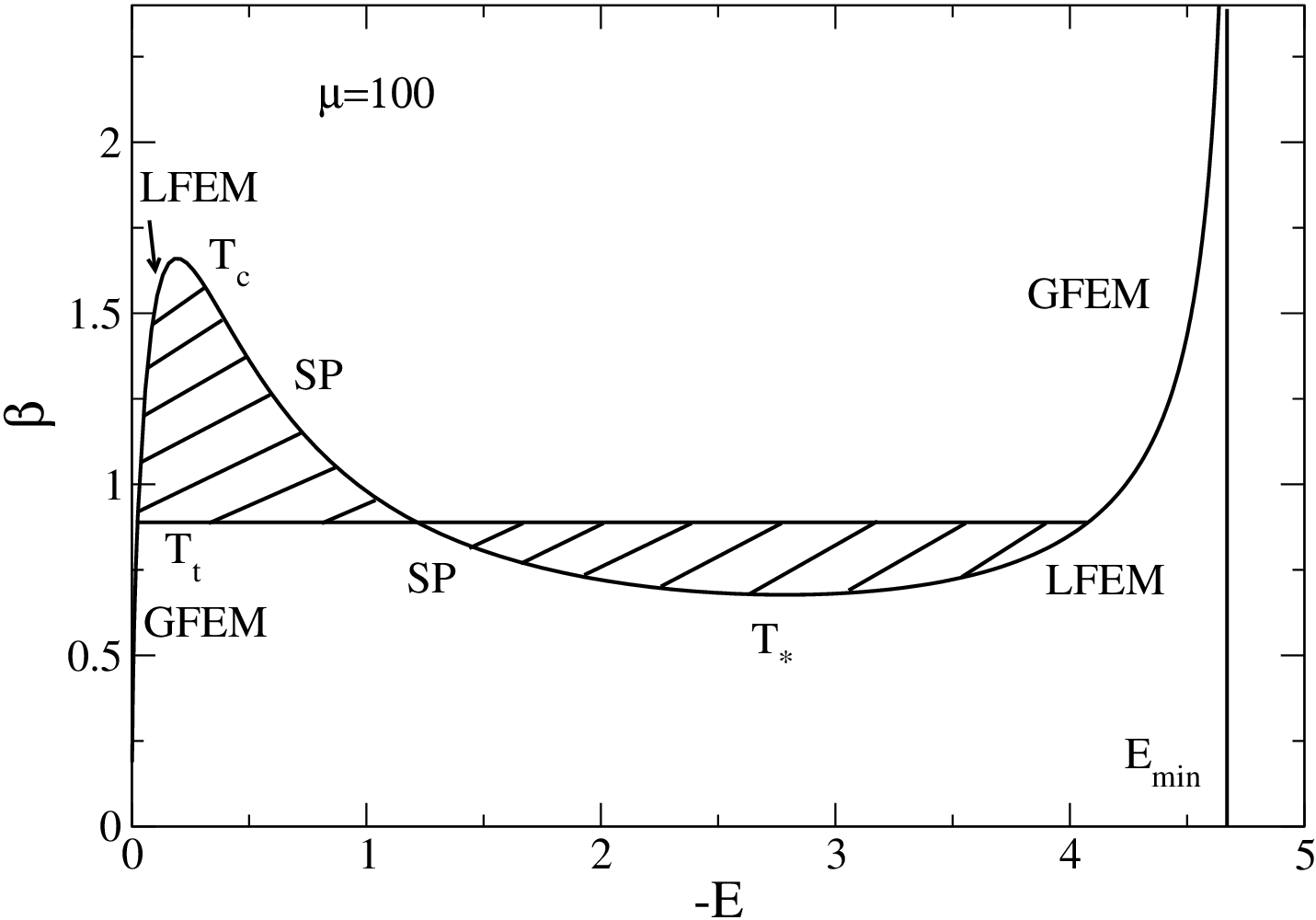}
\caption{Maxwell construction for $\mu=100$.}
\label{ETcano3}
\end{center}
\end{figure}

\begin{figure}
\begin{center}
\includegraphics[clip,scale=0.3]{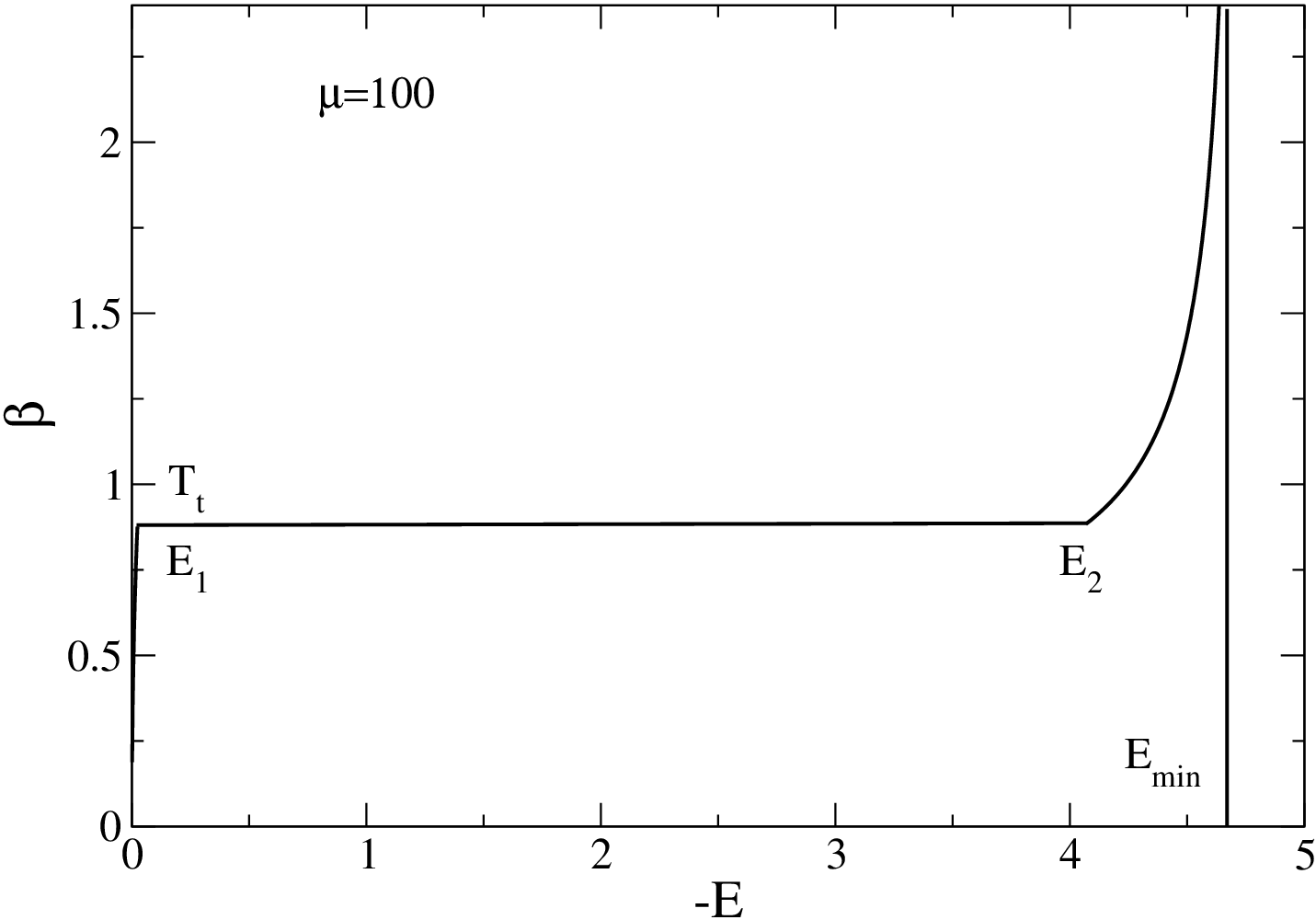}
\caption{Strict canonical caloric curve for $\mu=100$.}
\label{ETcano3strict}
\end{center}
\end{figure}

If we compare the free energy of the configurations (see Fig. \ref{Fcano3}), we expect a canonical first
order phase transition to take place at a transition temperature
$T_t(\mu)$ where the free energy of the gaseous phase and the free energy of the
condensed phase become equal.\footnote{We note that for large values of $\mu$, secondary canonical first order
phase transitions appear, as shown in Fig. \ref{secondaryCPT}, due to the winding of the
series of equilibria (see Fig. \ref{calomulti}). However, this is essentially a mathematical curiosity because these phase transitions
take place between unstable saddle points of free energy. As a result they may not be physical.}
The transition temperature $T_t(\mu)$ may also be obtained by performing a Maxwell  construction
(see the horizontal plateau in Fig. \ref{ETcano3}) \cite{ijmpb}. For $T>T_t$ the gaseous phase is fully stable (global free energy maximum GFEM at fixed
mass) while the condensed phase is metastable (local free energy maximum LFEM
at fixed mass). For $T<T_t$ the gaseous phase is metastable (LFEM) while
the condensed phase is fully stable (GFEM). The strict caloric curve in CE is obtained by keeping only the fully stable states (see Fig. \ref{ETcano3strict}). It is marked by a discontinuity of the energy $E=-\partial J/\partial \beta$ at $T=T_t(\mu)$. Equivalently, the first derivative of the free energy is discontinuous  at the transition  (see Fig. \ref{Fcano3}). This characterizes a canonical first order phase transition. The specific heat $C=dE/dT$ is also discontinuous at the transition.

It is instructive to compare the strict canonical caloric curve of
Fig. \ref{ETcano3strict} with the microcanonical caloric curve of Fig.
\ref{ETmicro3} for the same value of the degeneracy parameter
$\mu=100$. We see that the region of negative specific heats in MCE
is replaced by an isothermal phase transition (plateau) in CE that connects the gaseous phase (left branch) to the condensed phase (right branch). This corresponds to a situation of strict ensemble inequivalence: the energies between $E_1$ and $E_2$ are accessible in MCE but not in CE.

\begin{figure}
\begin{center}
\includegraphics[clip,scale=0.3]{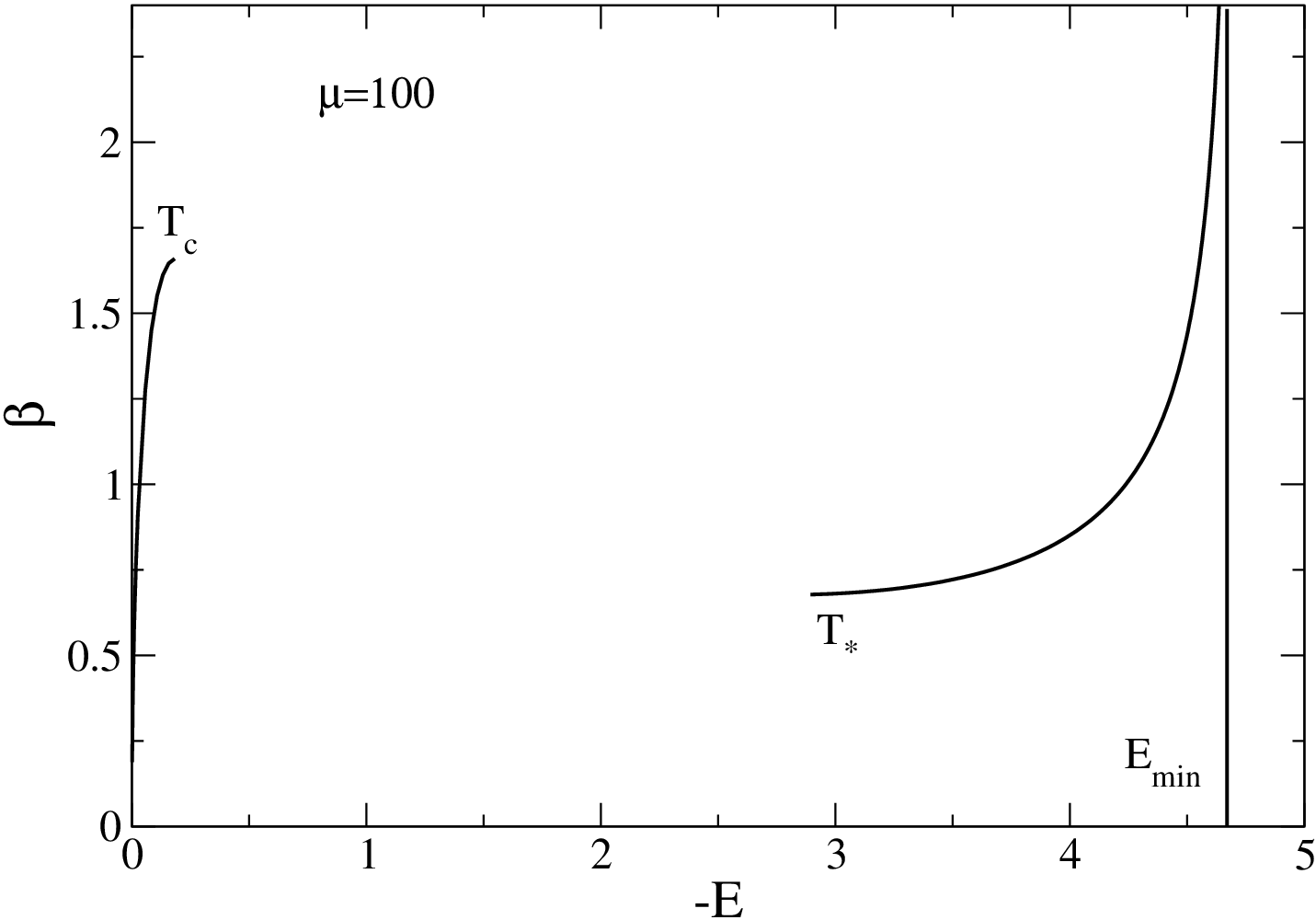}
\caption{Physical canonical caloric curve for $\mu=100$.}
\label{ETcano3meta}
\end{center}
\end{figure}

However, for systems with long-range interactions, the metastable states must be considered as stable states as explained previously.  The canonical first order phase transition at $T_t$ does {\it not} take place in practice because, for
sufficiently large values of $N$, the system remains frozen in
the metastable phase past the
transition temperature $T_t$. Therefore, the strict caloric curve of Fig.
\ref{ETcano3strict} is not physical. The physical canonical caloric
curve is the one shown in Fig. \ref{ETcano3meta}  which takes the metastable states
into account. It is obtained from the series of
equilibria of Fig. \ref{ETcano3instable} by discarding only the
unstable saddle points of free energy that form the intermediate branch.
These configurations  lie in the region of negative specific heats that is forbidden
in CE.  As a result, the region of physical ensemble inequivalence corresponds to the energies between $E_{gas}$ and $E_{cond}$. These energies are accessible in the microcanonical ensemble but not in the canonical ensemble (compare Figs. \ref{ETmicro3} and \ref{ETcano3meta}).

\begin{figure}
\begin{center}
\includegraphics[clip,scale=0.3]{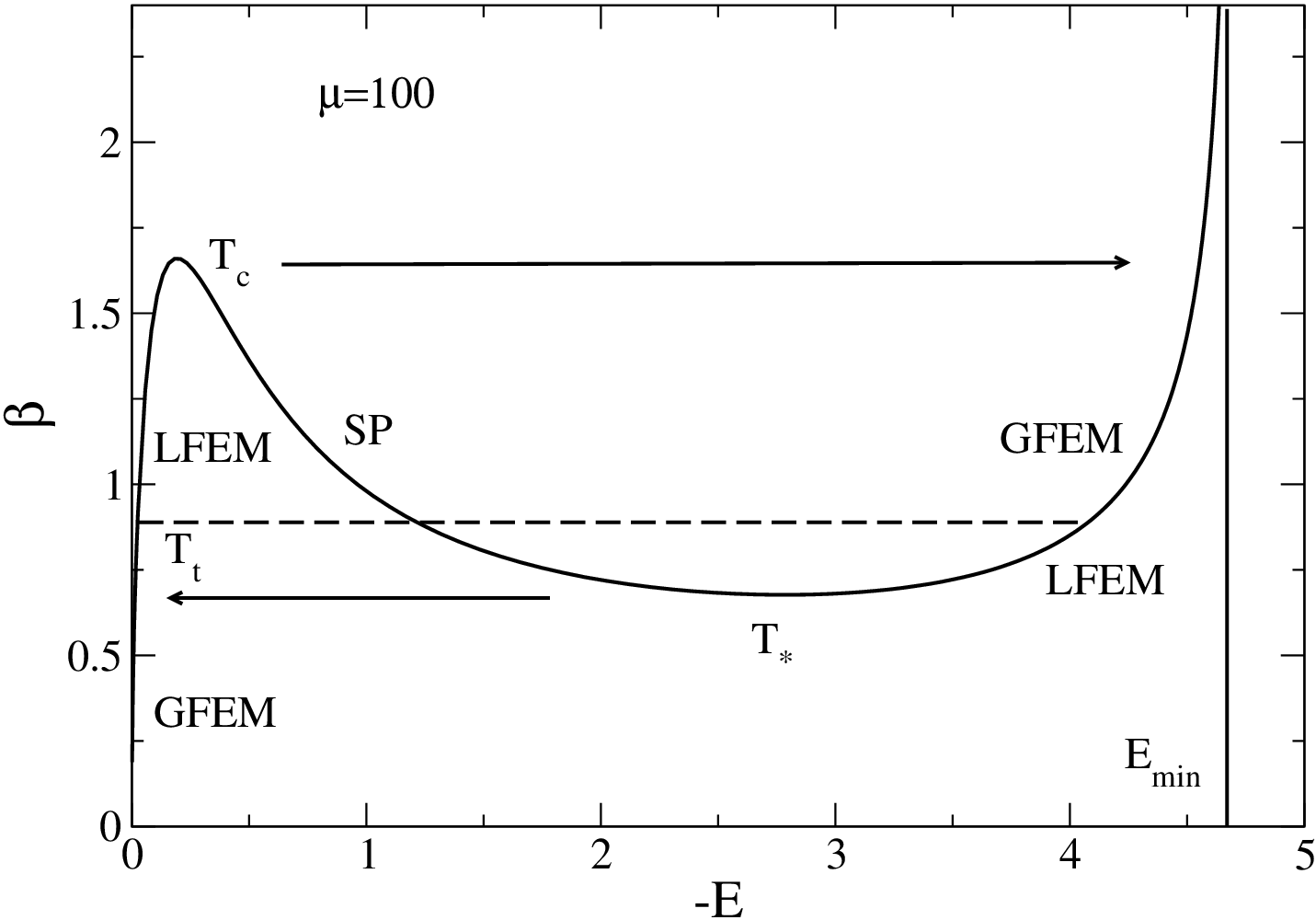}
\caption{Summary of the phase transitions of the fermionic King model with $\mu=100$ in CE. }
\label{ETcano3summary}
\end{center}
\end{figure}

The phase transitions of the fermionic King model in CE are summarized in Fig. \ref{ETcano3summary}.
For $T\rightarrow +\infty$, the system is in the gaseous phase where quantum mechanics is
completely negligible. At some transition temperature $T_{t}$, a first order phase transition is expected to occur and drive the system towards the condensed phase.  However,
gaseous states are still metastable, and long-lived,  beyond this
point so the first order phase transition does  not take place in practice. Therefore, if we decrease the
temperature,  the system remains in the gaseous phase until the critical temperature $T_c(\mu)$ at
which the gaseous phase disappears. This is similar to a  spinodal point in the language of phase transitions.   For $T<T_c(\mu)$,
the system undergoes a gravitational collapse (isothermal collapse). This corresponds to a saddle-node bifurcation. However, the
collapse stops when the core of the system becomes degenerate.
In that case, it
ends up in the condensed phase. The system has a ``core-halo'' structure with a
degenerate nucleus surrounded by a non-degenerate atmosphere.  The condensate results from the balance between the gravitational attraction  and the pressure due to the Pauli exclusion principle. This is a very compact object equivalent to a completely degenerate fermion ball at $T=0$. Since the
collapse is
accompanied by a discontinuous jump of free energy (see Fig. \ref{Fcano3}),
this is sometimes called a canonical zeroth order phase
transition. If we now increase the temperature, the system remains in the condensed
phase until the critical temperature $T_*(\mu)$ at which the condensed phase
disappears.  Indeed, the first order phase transition expected at $T_t(\mu)$ does not take place in practice due to the long lifetime
of the metastable states. For $T>T_*(\mu)$, the system undergoes an ``explosion'' reversed to
the collapse
and returns to the gaseous phase. In this sense, we can describe an hysteretic
cycle in the canonical ensemble (see the arrows in Figs. \ref{Fcano3} and \ref{ETcano3summary}).
Preliminary numerical simulations illustrating this hysteretic cycle for
self-gravitating fermions 
have been performed in \cite{ribot}. These canonical phase transitions exist
only above a canonical critical point $\mu_{CCP}=10.1$ (see Sec.
\ref{sec_maxwell}).

\subsection{Microcanonical and canonical critical points}
\label{sec_maxwell}

The deformation of the series of equilibria of the fermionic King model as a
function of the degeneracy parameter $\mu$ ($\sim$ system's size) is represented
in Fig. \ref{calomulti}. There exist two critical points in the problem,
one in each ensemble.

\begin{figure}
\begin{center}
\includegraphics[clip,scale=0.3]{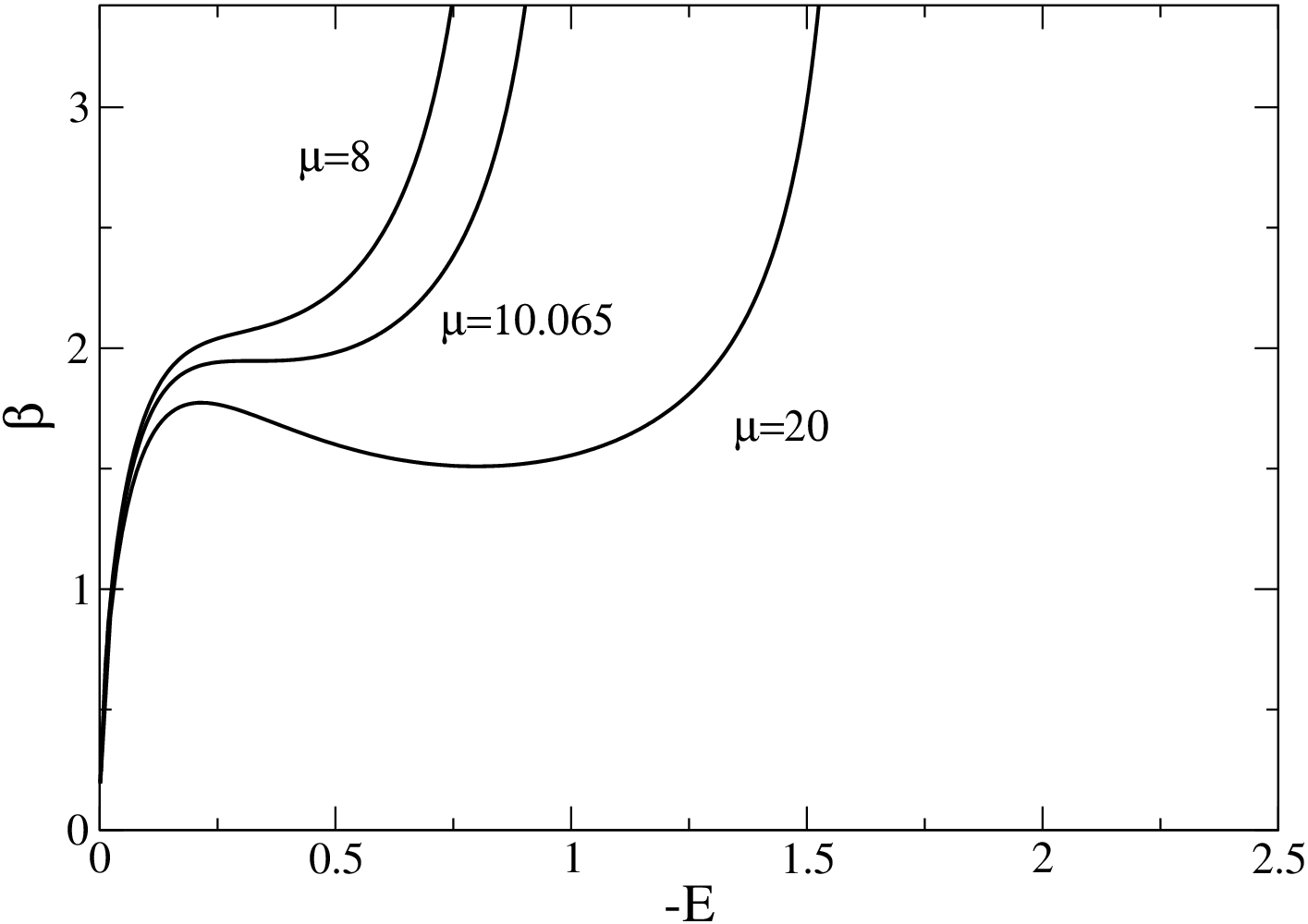}
\caption{Enlargement of the caloric curve near the
canonical critical
point ($\mu_{CCP}=10.1$,
$E_{CCP}= -0.325$, $\beta_{CCP}= 1.95$).}
\label{ccp}
\end{center}
\end{figure}

\begin{figure}
\begin{center}
\includegraphics[clip,scale=0.3]{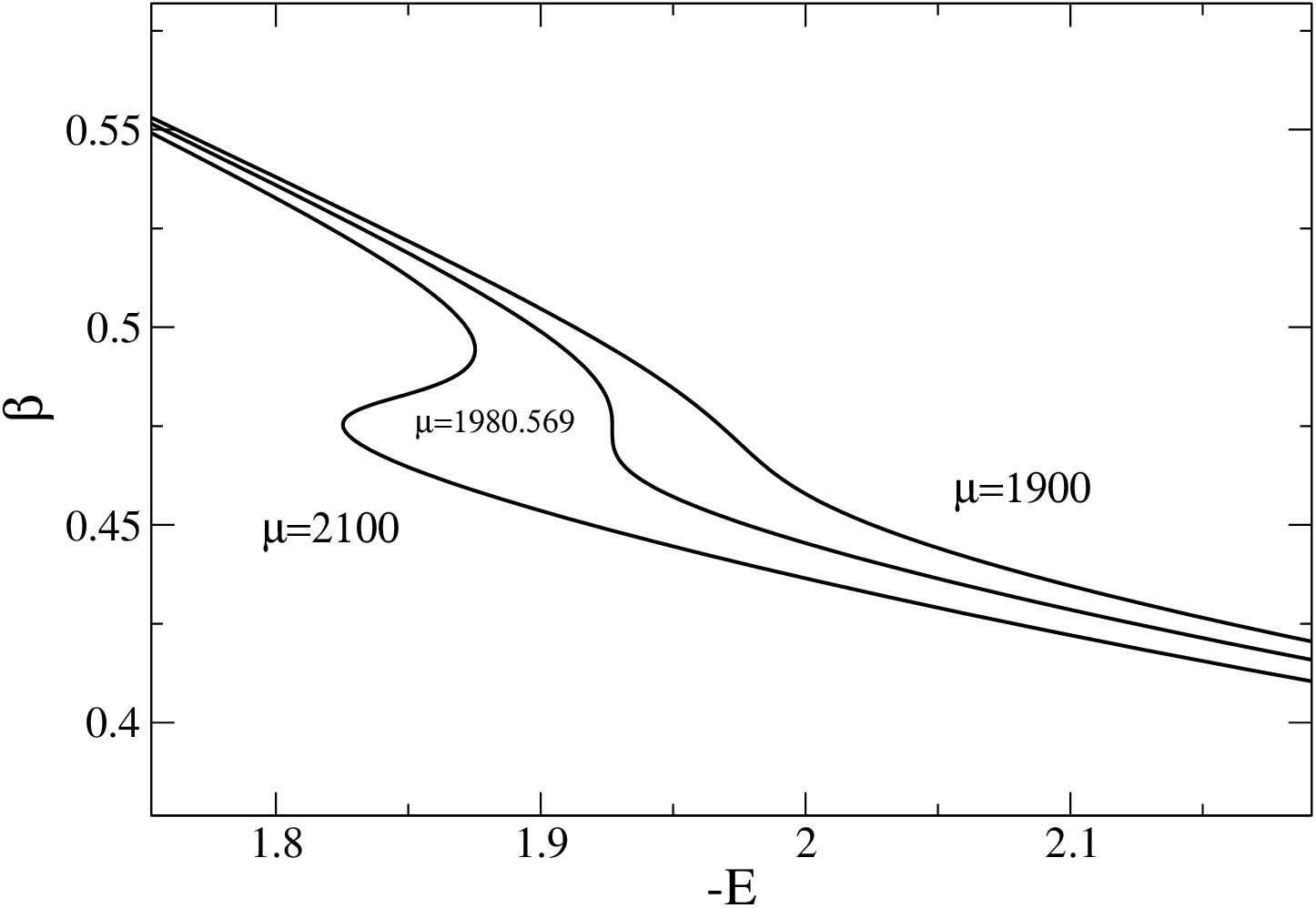}
\caption{Enlargement of the caloric curve near the
microcanonical critical point  ($\mu_{MCP}=1980$,
$E_{MCP}= -1.93$, $\beta_{MCP}= 0.474$).}
\label{mcp}
\end{center}
\end{figure}

For $\mu<\mu_{CCP}\simeq 10.1$, the curve
$\beta(E)$ is monotonic, so there is no phase transition. For
$\mu>\mu_{CCP}\simeq 10.1$, the curve $E(\beta)$ is multi-valued so that a
 canonical  phase transition takes place. At the canonical
critical point $\mu_{CCP}$, the caloric  curve $E(\beta)$
presents an inflection point and the canonical phase transition disappears (see
Fig. \ref{ccp}). At that point the specific heat is infinite. For
$\mu>\mu_{MCP}\simeq 1980$, the curve $\beta(E)$ is multivalued so
that a microcanonical phase transition takes place
(in addition to the canonical phase transition that exists
for any $\mu>\mu_{CCP}$).  At the
microcanonical critical point $\mu=\mu_{MCP}$,  the caloric curve $\beta(E)$ presents an
 inflection point and the microcanonical phase transition disappears (see Fig. \ref{mcp}). At that point
the specific heat vanishes.

Therefore, for $\mu>\mu_{MCP}$, the system exhibits a
microcanonical and a canonical phase transition, for
$\mu_{CCP}<\mu<\mu_{MCP}$ the system exhibits  only a canonical
phase transition, and for $\mu<\mu_{CCP}$ the system does not exhibit
any phase transition. We recall, however, that due to the presence of long-lived
metastable states, the first order phase transitions are not physically
relevant. Only the zeroth order phase transitions that occur at $E_{c}$ in MCE
and at $T_{c}$ in CE
(spinodal points) are relevant. We also recall that the secondary turning
points of energy and temperature that appear for large values of $\mu$ are not
physically relevant because they concern unstable saddle points. Only the first
and the last turning points of energy and temperature are physically relevant.
Therefore, despite the mathematical complexity of the spiral that appears for
large values of $\mu$, the physical nature of the phase transitions remains
relatively simple.

\subsection{Phase diagrams}
\label{sec_pd}

Typical caloric curves illustrating microcanonical   and canonical
 phase transitions are shown in Figs. \ref{ETmicro5}
and \ref{ETcano3} respectively. The phase diagram of the fermionic King model
can be directly deduced from these curves by identifying
characteristic energies and characteristic temperatures.  

\begin{figure}
\begin{center}
\includegraphics[clip,scale=0.3]{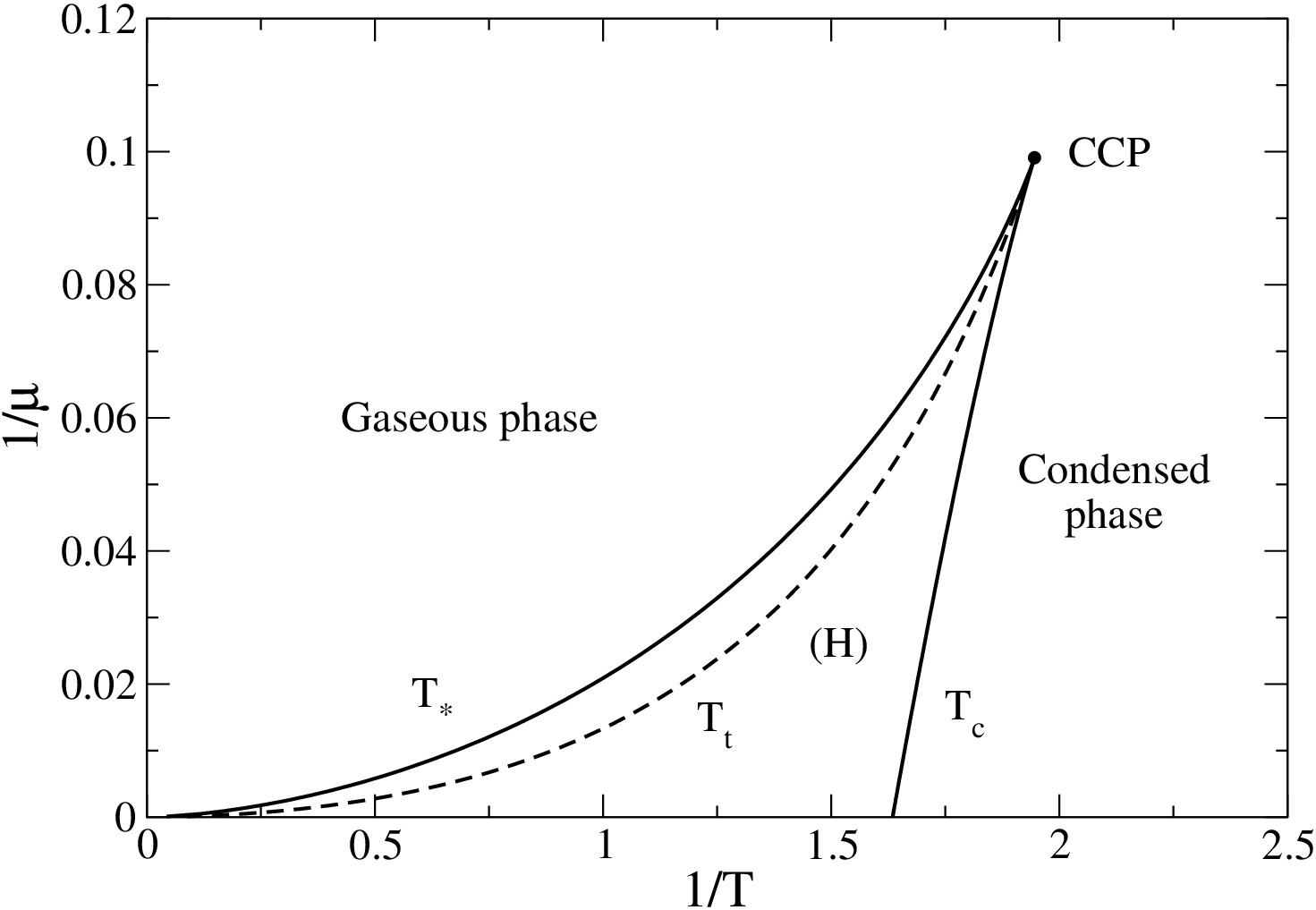}
\caption{Canonical phase diagram of the fermionic King model.  The $H$-zone
between $T_{c}$ and $T_{*}$ corresponds to an hysteretic zone where the actual phase depends on
the history of the system. If the system is initially prepared in a gaseous
state, it remain gaseous until the minimum temperature $T_{c}$ at which it
collapses and becomes condensed. Inversely, if the system is initially
prepared in a condensed state, it remains condensed until the maximum
temperature $T_{*}$ at which it explodes and becomes gaseous.}
\label{phasecano}
\end{center}
\end{figure}

In CE,
we note $T_{t}$ the temperature of transition (determined by the
equality of the free energies of the two phases), $T_{c}$ the end point of the metastable gaseous phase (first
turning point of temperature), and $T_{*}$ the end point of the
metastable condensed phase (last turning point of temperature). The
canonical phase diagram is represented in Fig. \ref{phasecano}. It shows in
particular the canonical critical point $\mu_{CCP}=10.1$ at which the
canonical phase transition disappears.

\begin{figure}
\begin{center}
\includegraphics[clip,scale=0.3]{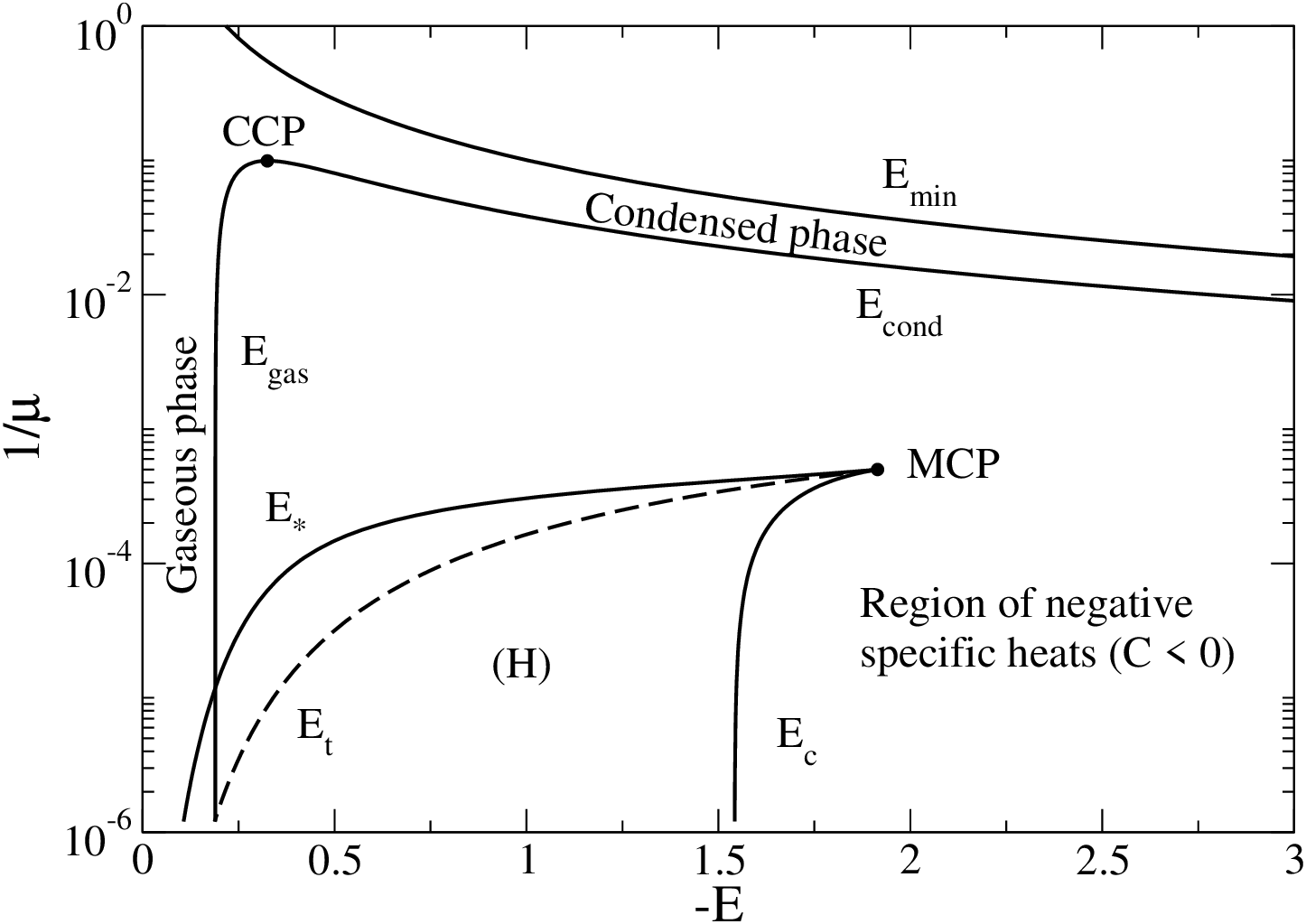}
\caption{Microcanonical phase diagram of the fermionic King model.  The
$H$-zone between $E_{c}$ and $E_{*}$  corresponds to an hysteretic zone where the actual phase
depends on the history of the system. The phase diagram in MCE is more
complex than in CE due to the existence of the negative specific heat
region that is forbidden in CE. This corresponds to a region of ensemble
inequivalence.
}
\label{phasemicro}
\end{center}
\end{figure}

In MCE, we note $E_{t}$ the energy of
transition (determined by the equality of the entropy of the two
phases), $E_{c}$  the end point of the metastable
gaseous phase (first turning point of energy), and $E_{*}$ the end
point of the metastable condensed phase (last turning point of
energy). We also denote by $E_{gas}$ the energy at which we enter in
the zone of negative specific heat (first turning point of
temperature) and $E_{cond}$ the energy at which we leave the zone of negative specific heat
(last turning point of
temperature). Finally, we introduce the minimum energy $E_{min}$ (ground
state). The microcanonical phase diagram is represented in Fig.
\ref{phasemicro}. It shows in particular the microcanonical critical
point $\mu_{MCP}=1980$ at which the microcanonical  phase
transition disappears.

\section{Density profiles and rotation curves of the fermionic King model}
\label{sec_dprc}

In this section, we study how the density profiles and the rotation curves of the fermionic King model depend on the values of $\mu$, $E$, and $T$. We also discuss their ability at describing dark matter halos. In Secs. \ref{sec_mumce}-\ref{sec_finmce} we consider MCE and in Secs. \ref{sec_muce}-\ref{sec_fince} we consider CE.

\subsection{The effect of increasing $\mu$ for  fixed $E>E_c$}
\label{sec_mumce}

We consider the series of equilibria of Fig. \ref{ETmicro5instable} corresponding to $\mu=10^4$. We take an energy $E=-0.876$ larger than the critical energy $E_c=-1.54$ of gravitational collapse (gravothermal catastrophe).  At that energy, the system can be found in three different states: a gaseous phase (solution A), an embryonic phase (solution B), and a condensed phase (solution C). We study the evolution of the  solutions A, B and C as $\mu$ increases.

\begin{figure}
\begin{center}
\includegraphics[clip,scale=0.3]{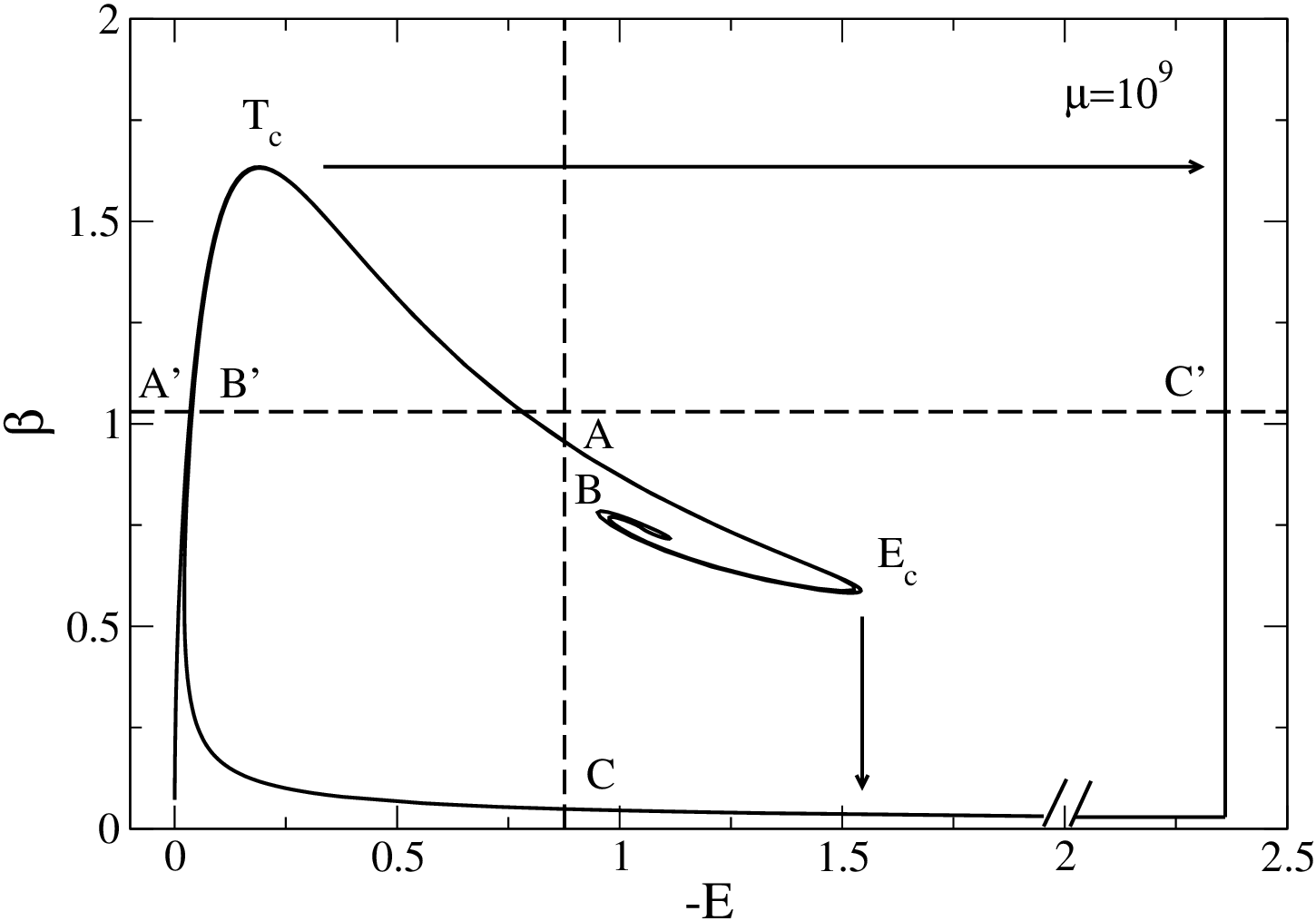}
\caption{Series of equilibria of the fermionic King model with $\mu=10^9$.
For large but finite values of $\mu$,
the series of equilibria winds up and makes several turns before
finally unwinding. A mode of stability is lost each time the curve winds up (rotates clockwise)
and a mode of stability is regained each time the curve unwinds (rotates
anti-clockwise). Therefore, only the part of the series of equilibria before
the first turning point and after the last turning point is stable.}
\label{mutresgrandprolongehenon}
\end{center}
\end{figure}

\begin{figure}
\begin{center}
\includegraphics[clip,scale=0.3]{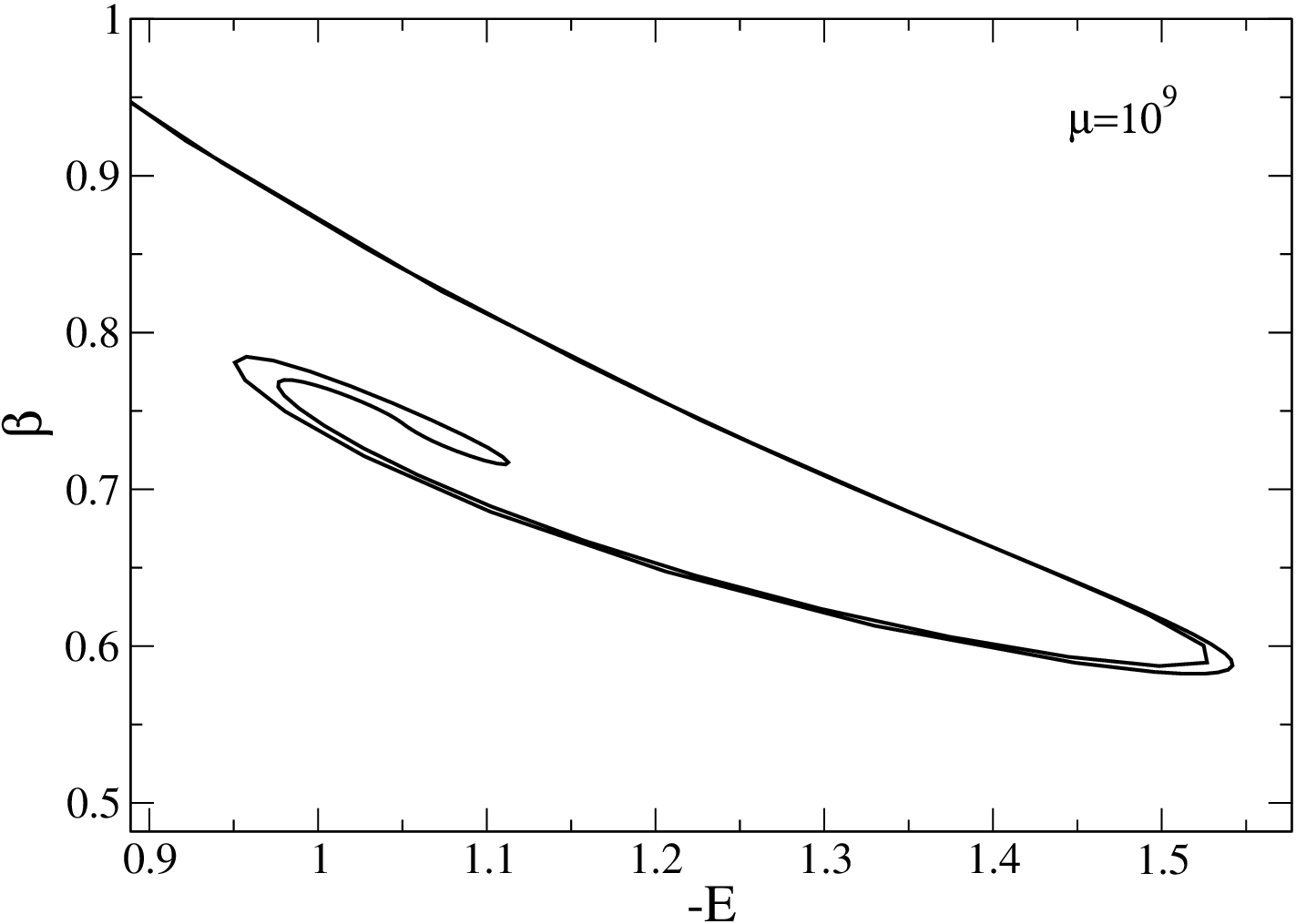}
\caption{Zoom on the spiral of Fig. \ref{mutresgrandprolongehenon}. The spiral rotates several times before unwinding. However, this is essentially a mathematical curiosity since the states on the spiral are unstable, hence unphysical.}
\label{doublespirale}
\end{center}
\end{figure}

For large values of $\mu$, the series of equilibria  rotates several times 
before unwinding (see Figs. \ref{mutresgrandprolongehenon} and 
\ref{doublespirale} obtained for $\mu=10^9\gg 1$). The branches A and B approach
each other while the branch C moves away. For $\mu\rightarrow +\infty$, the
branches A and B superimpose  while the branch C coincides with the $\beta=0$
axis (see Fig. \ref{mutresgrandprolongehenon}). In this limit, we recover the 
spiral corresponding to the classical King model (see Fig. \ref{ckm}).

\begin{figure}
\begin{center}
\includegraphics[clip,scale=0.3]{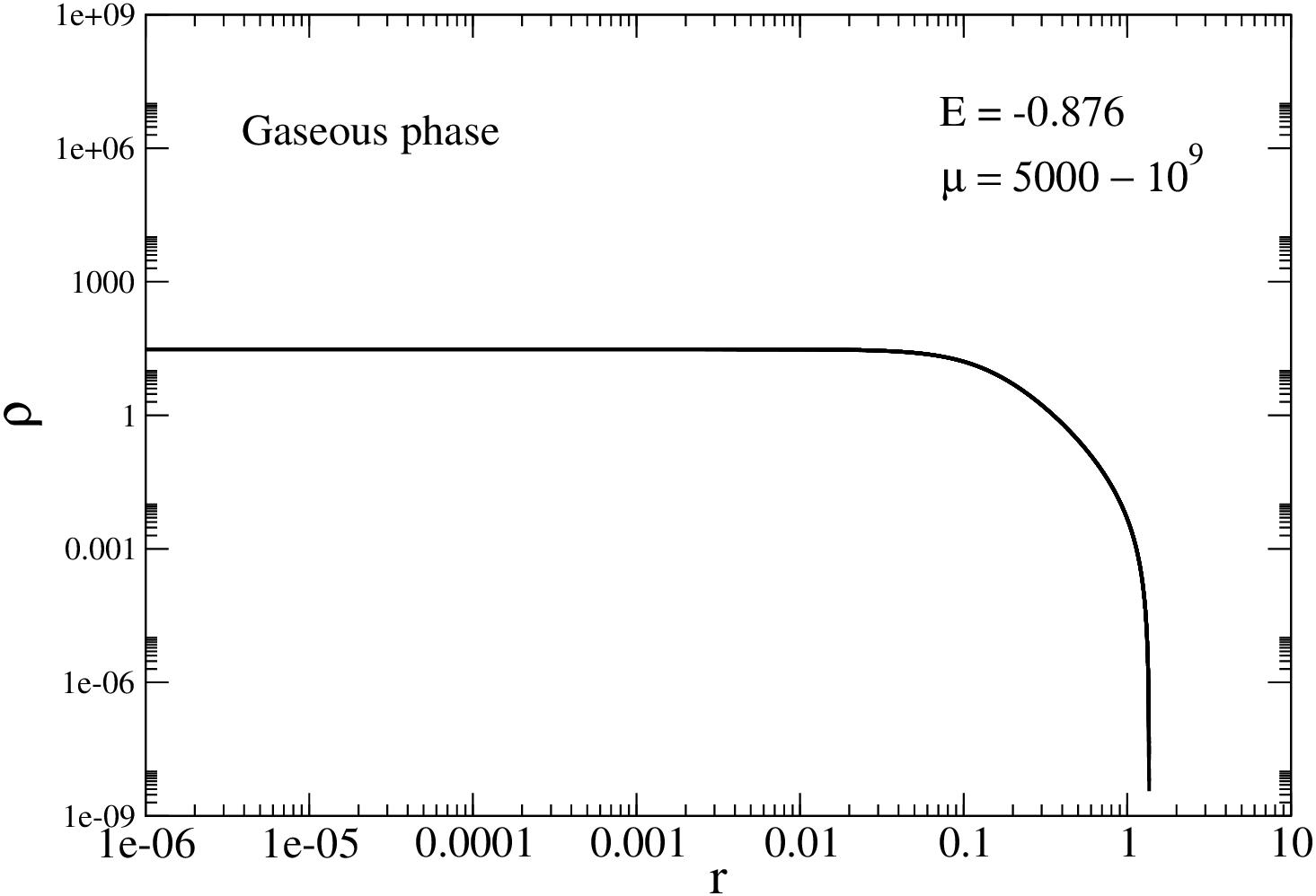}
\caption{Density profile of the gaseous phase (solution A)
for different values of $\mu$ in logarithmic scales (here and in the following figures we have selected $\mu=5000$, $10^4$,
$10^5$, $10^6$, $10^7$, $10^8$, and $10^9$). For sufficiently large values of $\mu$ the density profile of the gaseous phase
does not change. It reaches an asymptotic profile corresponding
to the classical King model.}
\label{gaseousRHO}
\end{center}
\end{figure}

\begin{figure}
\begin{center}
\includegraphics[clip,scale=0.3]{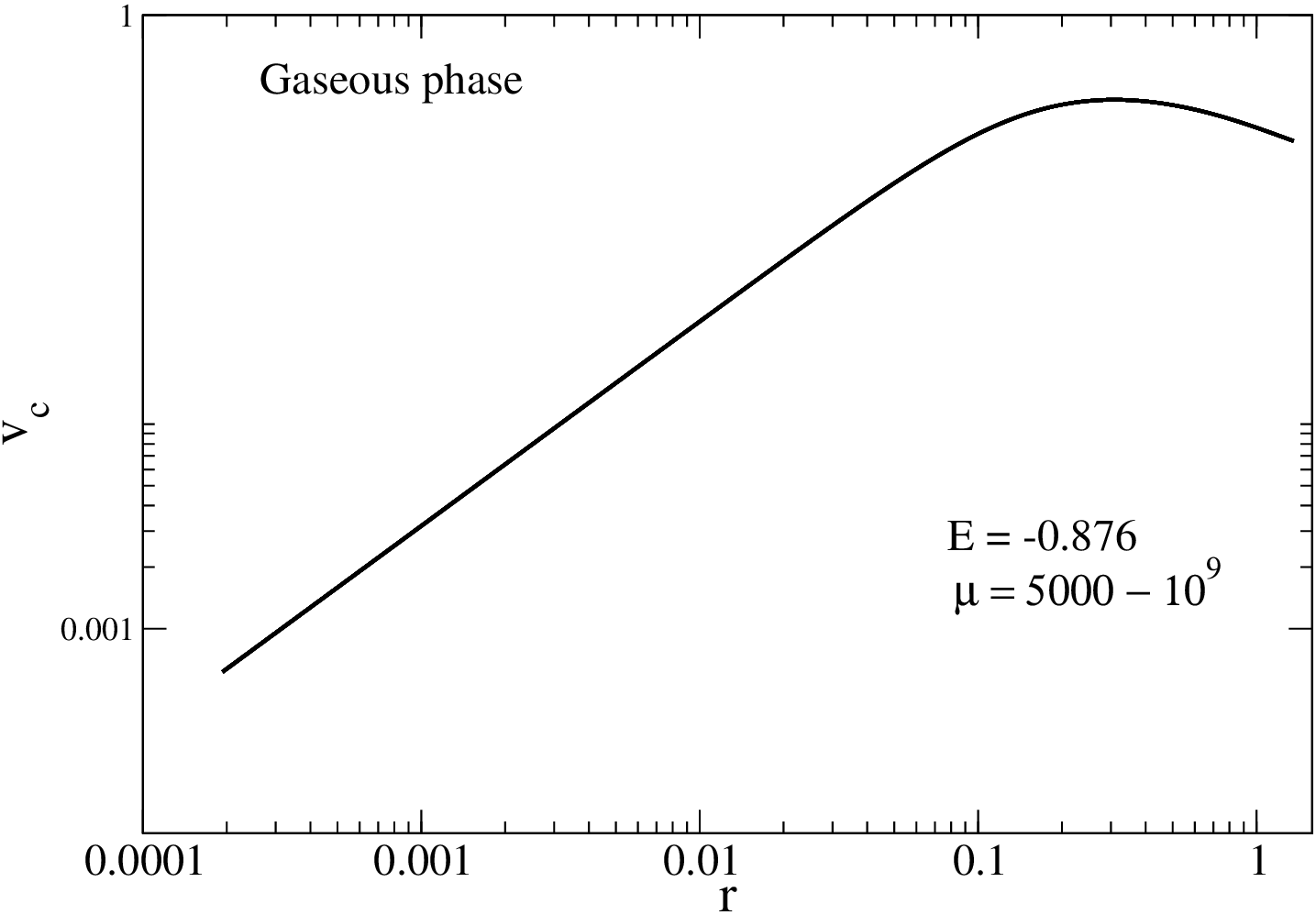}
\caption{Circular velocity profile of the gaseous phase
for different values of $\mu$ in logarithmic scale. }
\label{gaseousV}
\end{center}
\end{figure}

\begin{figure}
\begin{center}
\includegraphics[clip,scale=0.3]{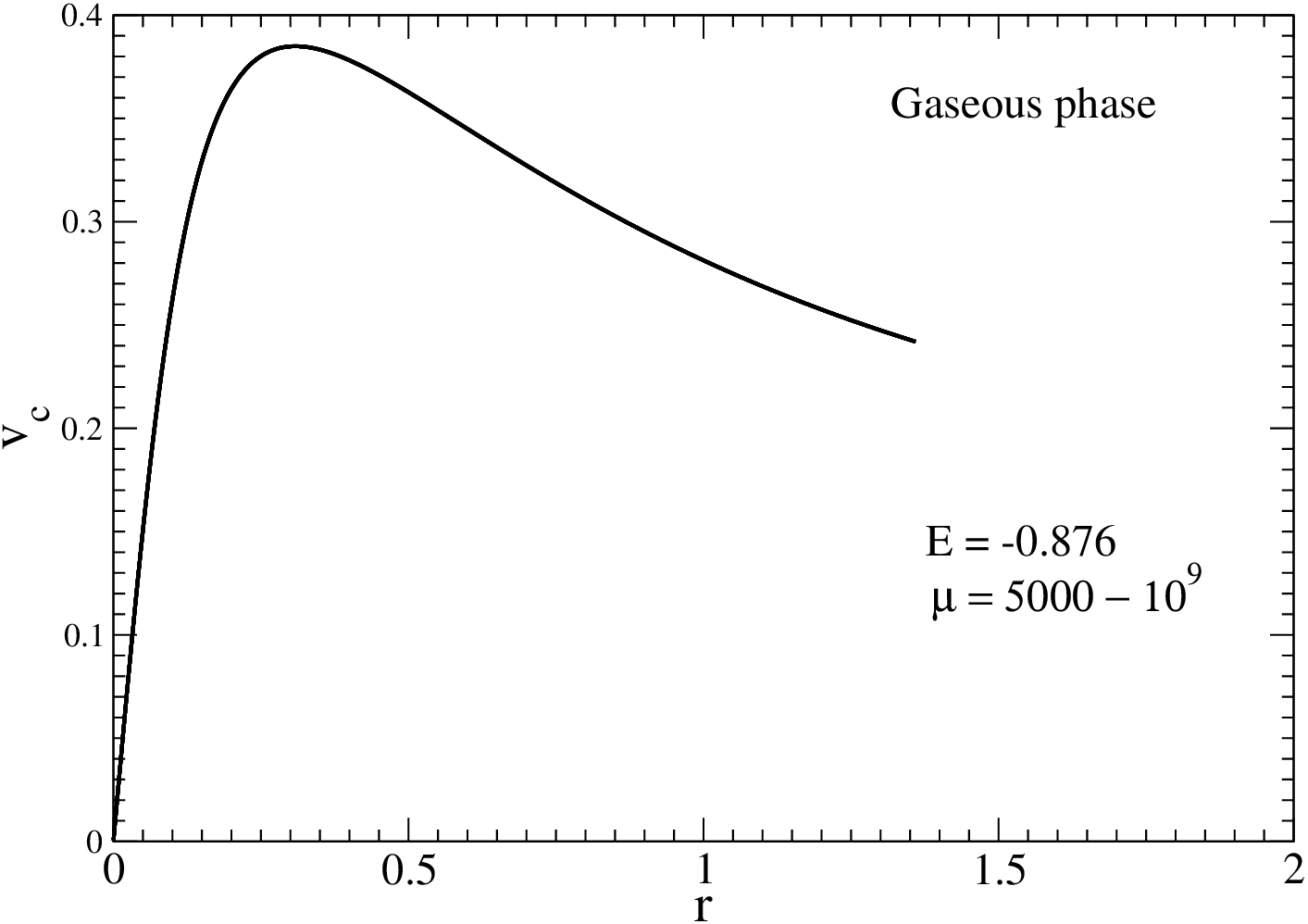}
\caption{Circular velocity profile of the gaseous phase
for different values of $\mu$ in linear scale. }
\label{gaseousVlin}
\end{center}
\end{figure}

\begin{figure}
\begin{center}
\includegraphics[clip,scale=0.3]{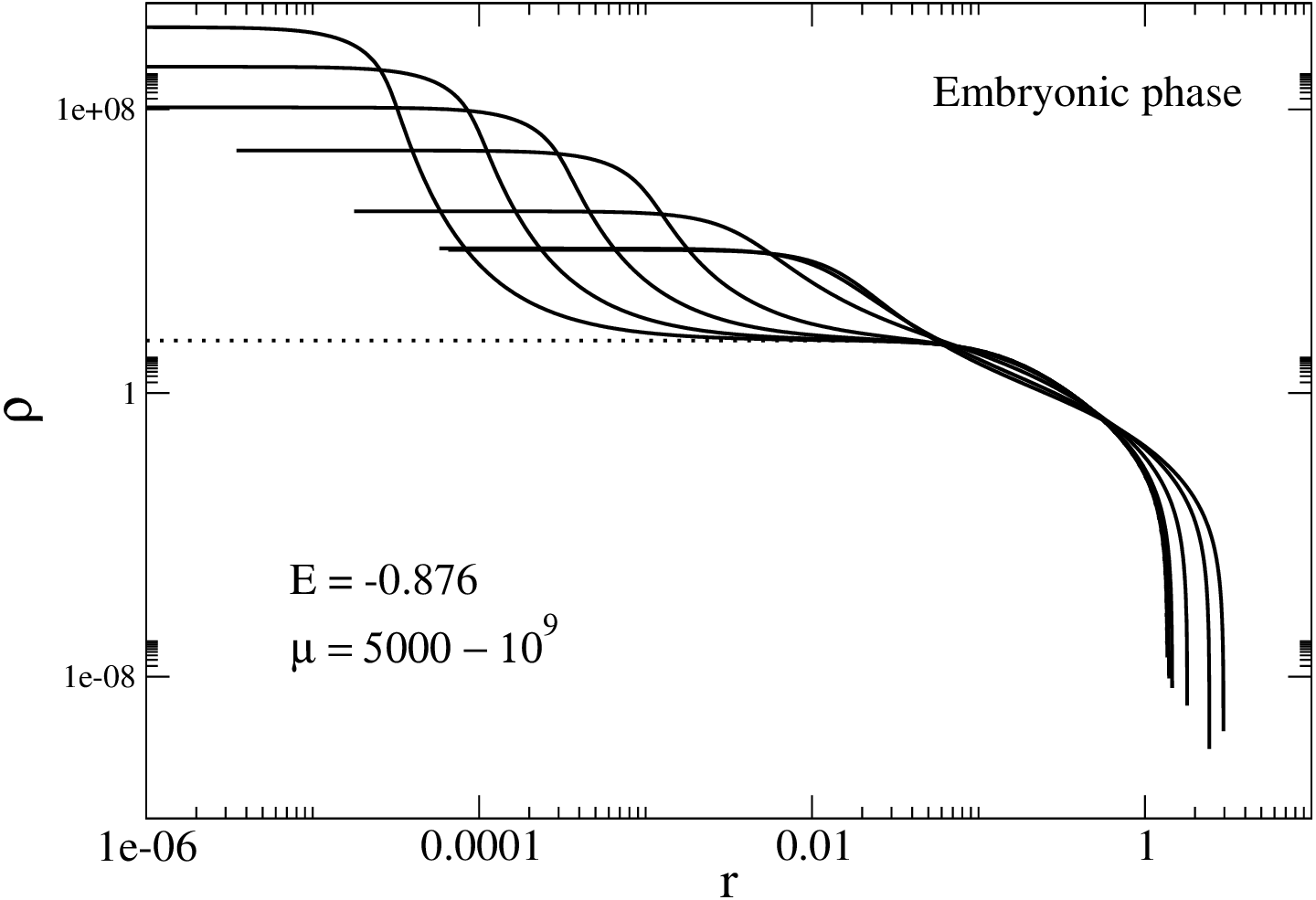}
\caption{Density profile of the embryonic phase (solution B) for different
values of $\mu$ in logarithmic scales (the central density increases with $\mu$). For increasing $\mu$, the solution B coincides with the solution A (gaseous phase; dotted line)
corresponding to the classical King model, except that it contains a small
embryonic degenerate nucleus (fermion ball) with
a small mass and a small absolute value of potential energy. This nucleus of
almost constant density is followed by a plateau as detailed in \cite{csmnras}.
}
\label{embryonRHOdot}
\end{center}
\end{figure}

\begin{figure}
\begin{center}
\includegraphics[clip,scale=0.3]{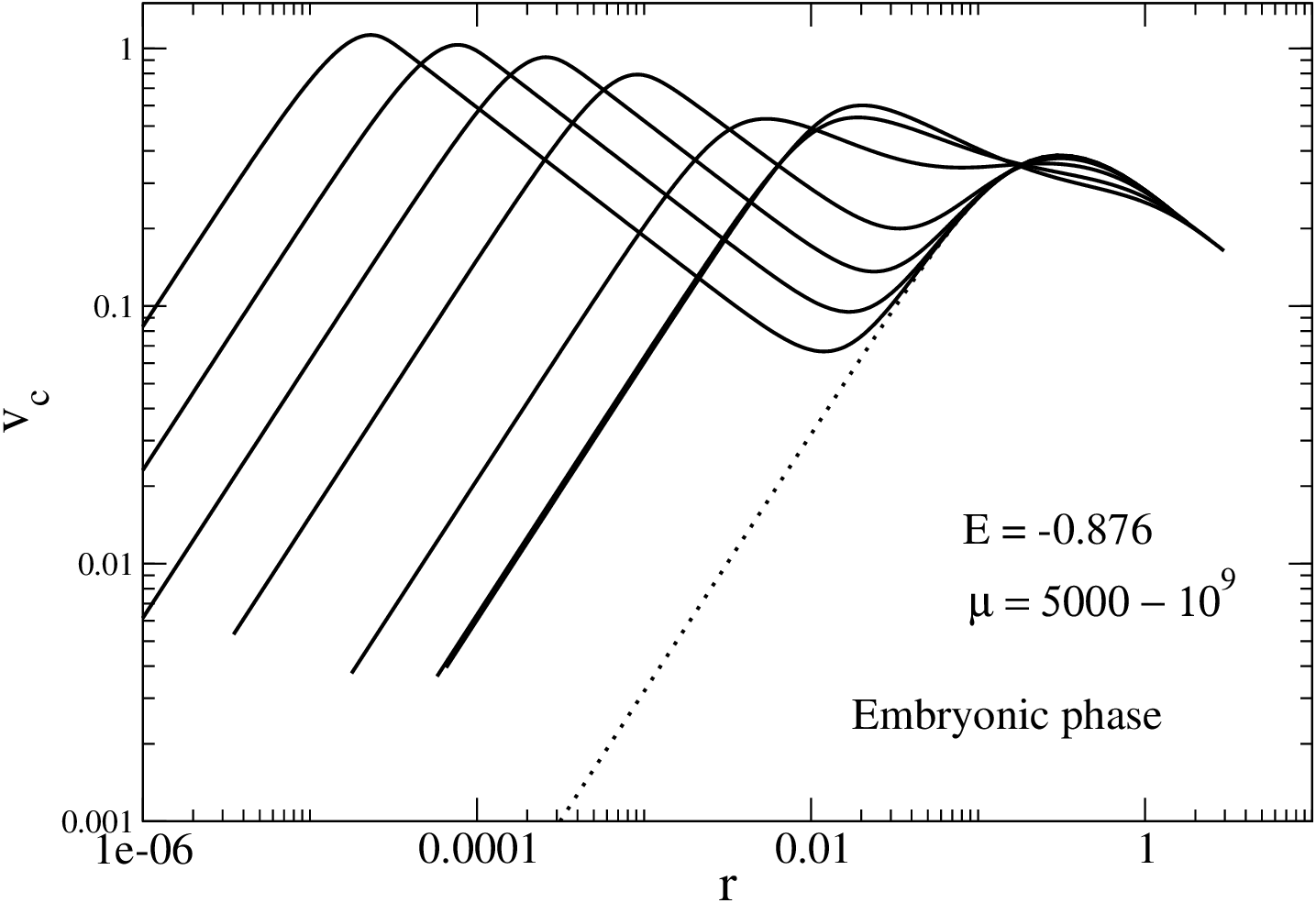}
\caption{Circular velocity profile of the embryonic phase for different
values of $\mu$ in logarithmic scales. For increasing $\mu$, the solution B
approaches the solution A (gaseous phase; dotted line) corresponding to the
classical King model, except at very small radii. The presence of a small
nucleus (fermion ball) where $v_c\propto r$, followed by a plateau where
$v_c\propto r^{-1/2}$, manifests itself by a secondary peak in the rotation
curve at the very
center
of the system (see Fig. \ref{embryonVlindot}). However,
these distances are probably not accessible to observations. Furthermore, these solutions are thermodynamically
unstable so this secondary peak may not be physical.}
\label{embryonVdot}
\end{center}
\end{figure}

\begin{figure}
\begin{center}
\includegraphics[clip,scale=0.3]{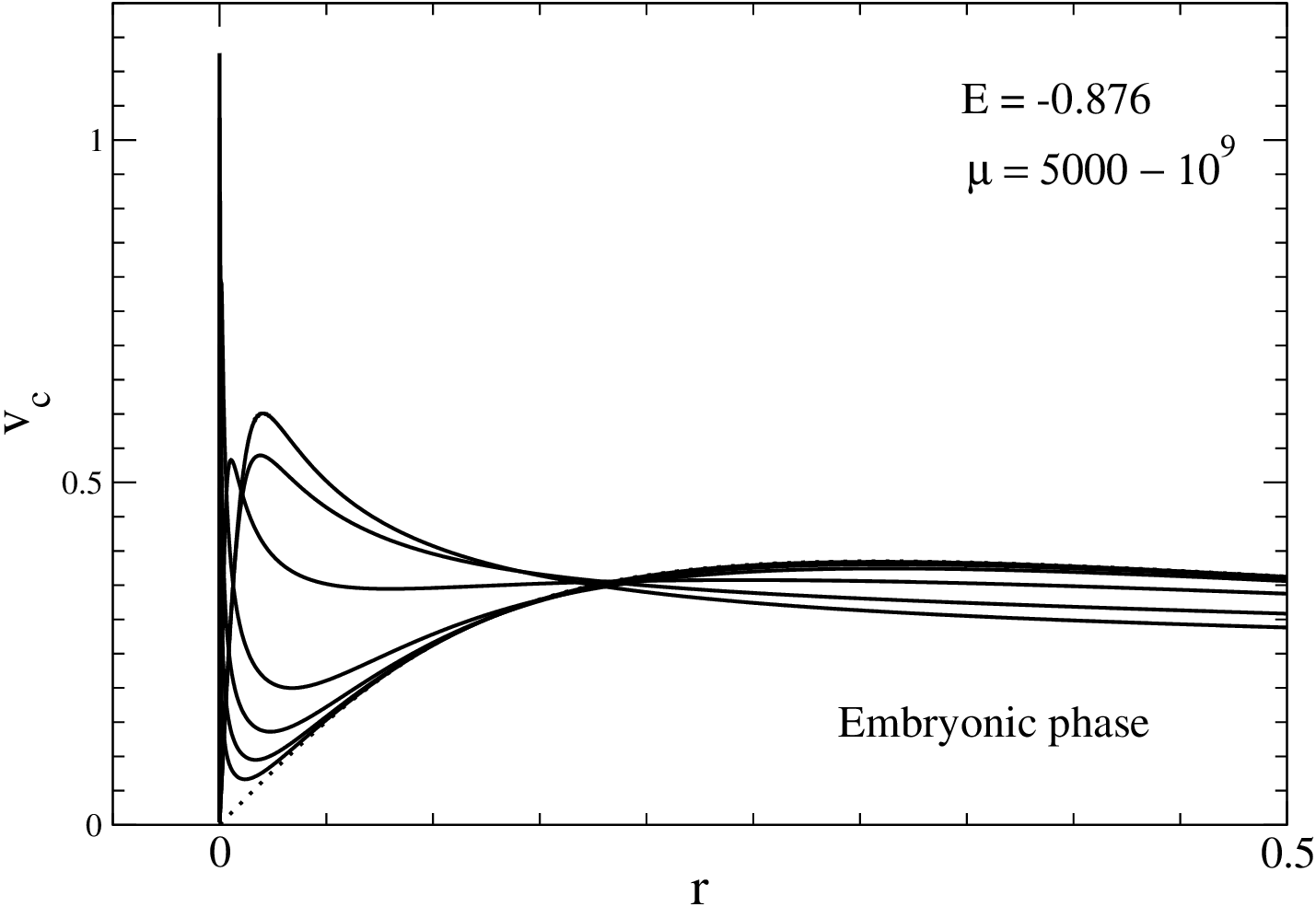}
\caption{Circular velocity profile of the embryonic phase for different
values of $\mu$ in linear scales. For large values of
$\mu$, we recover the classical King model (dotted line) except at the very center. The secondary peak due to the
degenerate nucleus (fermion ball) manifests itself by a spike near the origin. }
\label{embryonVlindot}
\end{center}
\end{figure}

\begin{figure}
\begin{center}
\includegraphics[clip,scale=0.3]{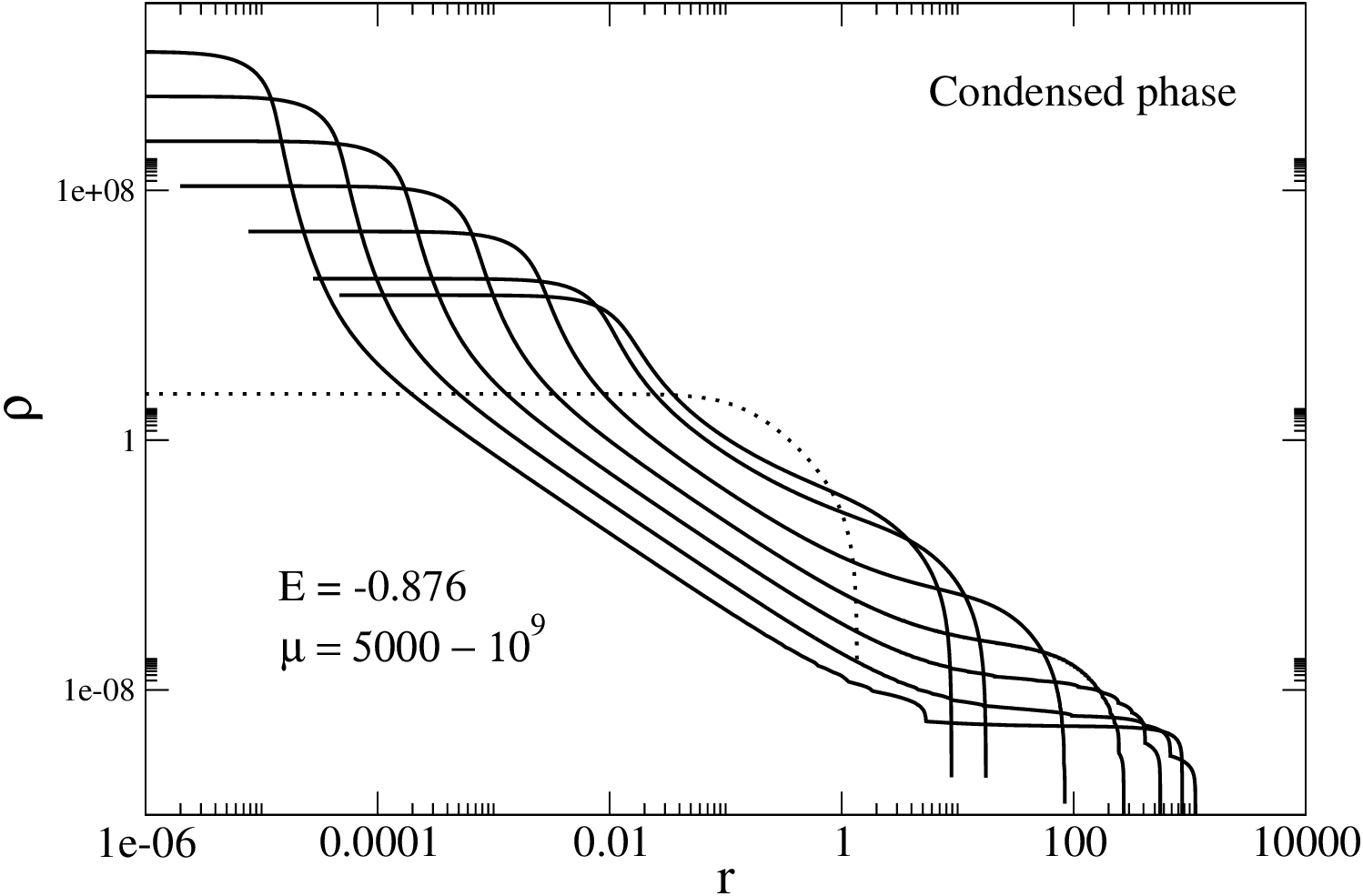}
\caption{Density profile of the condensed phase (solution C) for different
values of $\mu$ in logarithmic scales (the central density increases with $\mu$).
For increasing $\mu$,
the solution C contains a small degenerate nucleus with a relatively small mass
but a more and more negative
potential energy.  As a result, the halo becomes hotter and hotter in order to conserve the total energy.
This is why it forms a sort of plateau with constant density that extends at larger and larger distances. The resulting profile is very
different from solution A (gaseous phase;
dotted line) corresponding to the classical King model.}
\label{condensedRHOdot}
\end{center}
\end{figure}

\begin{figure}
\begin{center}
\includegraphics[clip,scale=0.3]{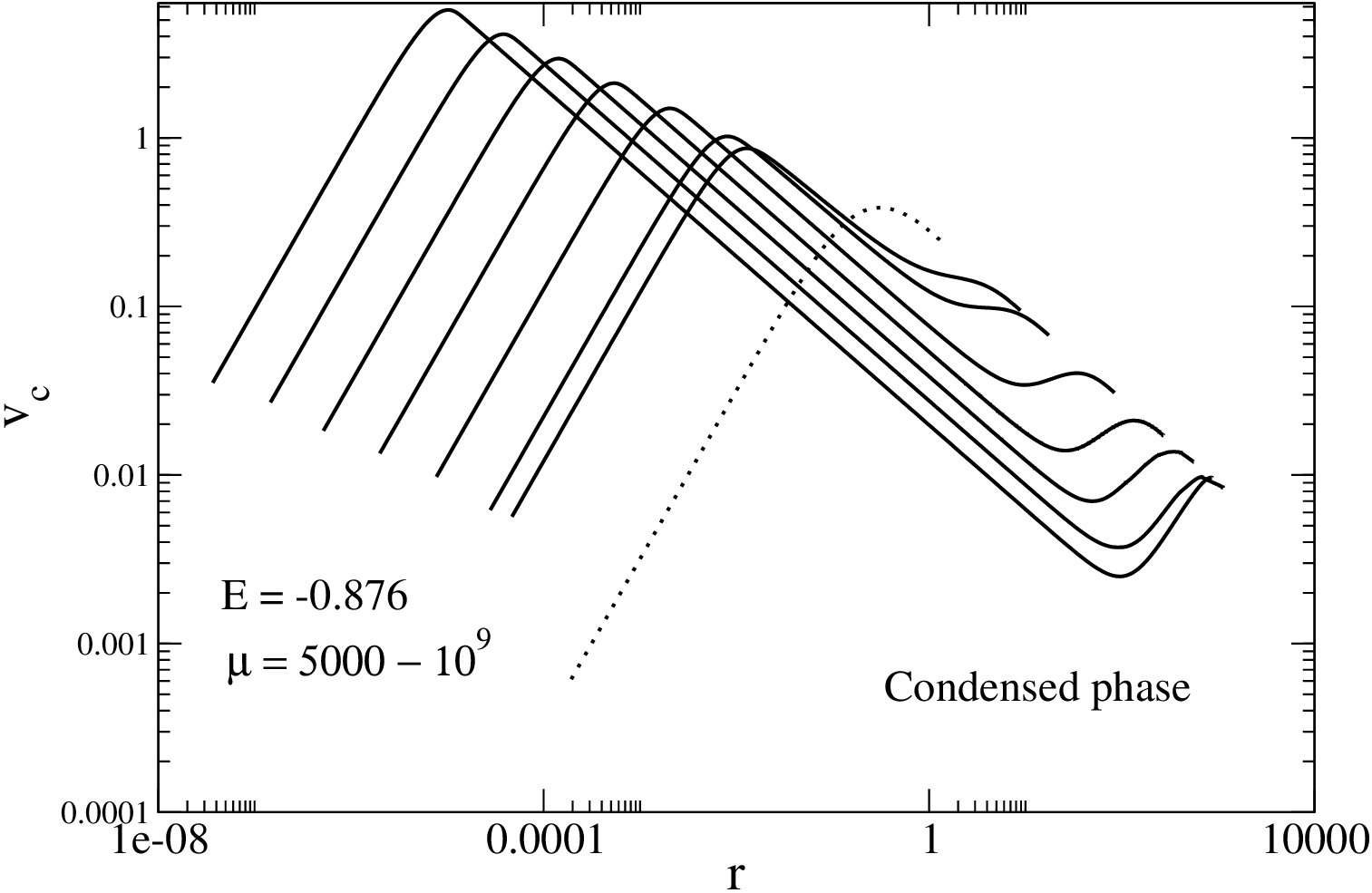}
\caption{Circular velocity profile of the condensed phase  for different values of $\mu$ in logarithmic
scales.  It is very
different from solution A (gaseous phase; dotted line)  corresponding to the classical King model. This is
because the halo is expelled at large distances as the nucleus becomes denser
and denser,  and more and more energetic.}
\label{condensedVdot}
\end{center}
\end{figure}

\begin{figure}
\begin{center}
\includegraphics[clip,scale=0.3]{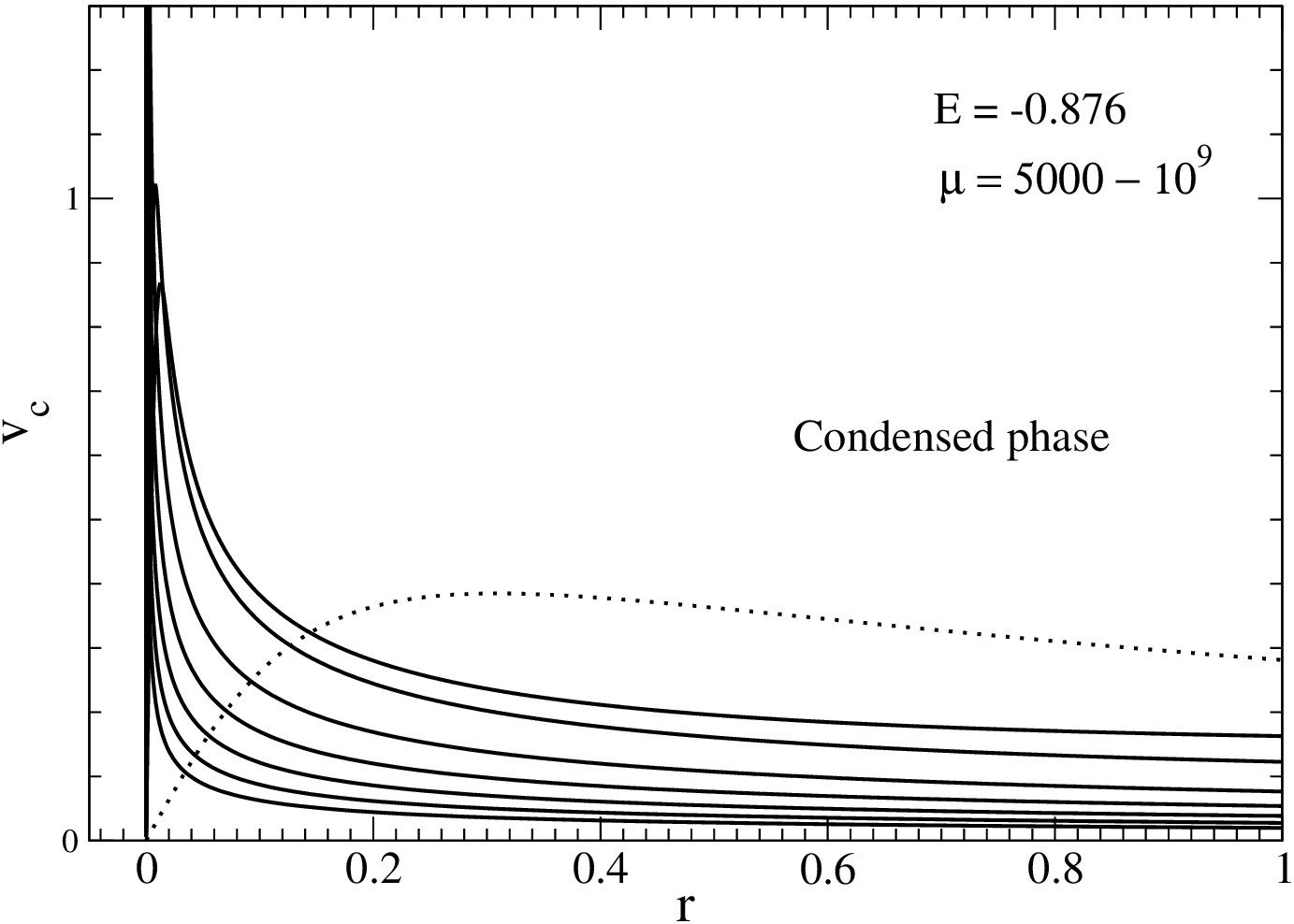}
\caption{Circular velocity profile of the condensed phase  (solution C) for different values of $\mu$ in linear
scales.}
\label{condensedVlindot}
\end{center}
\end{figure}

The solution A (gaseous phase) does not significantly change with $\mu$ and tends to the
classical King distribution for $\mu\rightarrow +\infty$ (see Figs.  \ref{gaseousRHO}-\ref{gaseousVlin}). Since the classical King model close to the point of marginal microcanonical stability ($E_c=-1.54$) describes large dark matter halos relatively well (see Paper I), and since the chosen energy $E=-0.876$ is relatively close to $E_c$, we shall take the asymptotic profile of  Figs. \ref{gaseousRHO}-\ref{gaseousVlin} as a reference in our discussion (see the dotted lines in Figs. \ref{embryonRHOdot}-\ref{condensedVlindot}).

The solution B (embryonic phase) is  similar to the solution A (gaseous phase) except that it contains a
small embryonic nucleus with high density.  This is a completely degenerate compact object equivalent to a fermion ball at $T=0$.
Therefore, the solution B has a core-halo
structure. The mass, the size and the absolute value of the potential energy of the nucleus
decrease as $\mu$ increases. As a result, for $\mu\gg 1$, the solutions A and B have
almost the same temperature ($\beta_A\simeq \beta_B$) and  the profiles A and B coincide
outside of the nucleus (see Figs. \ref{embryonRHOdot}-\ref{embryonVlindot}). This is why the branches A and B in the series of  equilibria superimpose
for $\mu\rightarrow +\infty$ (see Fig. \ref{mutresgrandprolongehenon}).  Still, the two solutions A and B are physically distinct. In particular, solution B is unstable as further discussed in Sec. \ref{sec_harbor}.

The solution C (condensed phase) is very different from the solution A (gaseous phase) and from the
solution B (embryonic phase). Like solution B, it has a core-halo structure. It contains
a small degenerate nucleus with high density (condensate) that has a small mass and a small radius. However, unlike solution B,
the nucleus  has a very negative potential
energy. Since energy is conserved in MCE, this implies that the  halo must be very hot. This is why $\beta_C$ is small
(see Fig. \ref{mutresgrandprolongehenon}). Since the halo
is hot, it expands at very large distances (see Figs. \ref{condensedRHOdot}-\ref{condensedVlindot}).
The mass and the size  of the nucleus
decrease as $\mu$ increases while the absolute value of its potential energy increases.
The mass, radius, and temperature of the halo increase as $\mu$ increases. For $\mu\rightarrow
+\infty$, the mass and the radius of the
nucleus tend to zero but its potential energy tends to $-\infty$. The temperature of the halo
tends to $+\infty$ in order to conserve the energy. The radius of the halo  also tends to $+\infty$.
Therefore, for $\mu\rightarrow +\infty$, we recover the  singular
 ``binary $+$ hot halo'' structure with infinite entropy corresponding to the strict equilibrium state
 of classical self-gravitating systems in MCE (see footnote 3). For finite values of
 $\mu$, quantum mechanics  provides a regularization of this singular structure: the ``tight binary'' is
replaced by a ``fermion ball'' whose size is fixed by quantum mechanics. The resulting structure has
a finite entropy.

We now describe the form of the physical caloric curve when $\mu\rightarrow
+\infty$. For $\mu>\mu_{MCP}$ the physical caloric curve always looks like Fig.
\ref{ETmicro5meta} (the spiral that develops for large values of $\mu$ does not
play any role since it is made of unstable states). The upper branch (gaseous
phase) does not change much with $\mu$. For $\mu>\mu_{MCP}$ it almost coincides
with the classical King model ($\mu\rightarrow +\infty$). The collapse energy
$E_c(\mu)$ is close  to $-1.54$. The lower branch (condensed phase) depends
sensibly on $\mu$. For $\mu\rightarrow +\infty$, the explosion energy $E_*(\mu)$
tends to zero and the minimum energy $E_{min}(\mu)$ tends to $-\infty$. The
transition energy $E_t(\mu)$ also tends to zero. This implies that the gaseous
phase corresponds to metastable states (LEM) while the condensed phase
corresponds to fully stable states (GEM). For  $\mu\rightarrow +\infty$ the
condensed states are singular and have an infinite entropy. They are made of a
``tight binary'' (a degenerate core with a small mass but a huge potential
energy) surrounded by a hot halo with $T\rightarrow +\infty$. As a result, the
branch of condensed states coincides with the $x$-axis at $\beta=0$. Therefore,
in the $\mu\rightarrow +\infty$ limit, the physical caloric curve is formed by
the metastable gaseous branch of Fig. \ref{ckm} up to MCE  plus a singular stable
condensed branch at $\beta=0$ coinciding with the $x$-axis (tight binary  $+$
hot halo). On the 
other hand, the saddle points are superposed to the spiral and to the branch of
gaseous states although they have a very different structure presenting a germ.

\subsection{The effect of decreasing $E$  for  fixed $\mu>\mu_{MCP}$}
\label{sec_imp}

We consider a value of $\mu$ larger than $\mu_{MCP}=1980$ for which a microcanonical phase transition (collapse) takes place at the
critical energy $E_c\sim -1.54$. Specifically, we choose $\mu=10^4$ (large halo) corresponding to the caloric curve represented
in Fig. \ref{ETmicro5summary}. We start from the gaseous phase (solution A) and progressively decrease the
energy. This is a natural evolution since the concentration parameter $k(t)$ increases with time as the system slowly evaporates until an instability takes place (see Appendix \ref{sec_dts}). As a result, the system follows the series of equilibria from high energies to low energies.
For $E_c<E<0$ (corresponding to $0\le k\le k_{MCE}$), the system is non degenerate. It can be described by the classical King model.
When $E\rightarrow 0$ (corresponding to $k\rightarrow 0$), the system is equivalent to a polytrope of index $n=5/2$. When $E$ is close
to $E_c$ (corresponding to $k\sim k_{MCE}$), the classical King profile can be approximated by the modified Hubble profile.  This profile
 provides a good description of large dark matter halos (see Paper I). When $E<E_c$, the gaseous phase
(solution A) disappears and the  system undergoes a gravitational
collapse towards the condensed phase (solution C). This corresponds to a saddle-node bifurcation.
According to the discussion of Sec. \ref{sec_mumce},
the gravitational collapse in MCE (gravothermal catastrophe) results in the formation of  a compact degenerate object
(fermion ball at $T=0$) of  much smaller mass and size than the initial cluster. This is accompanied  by the expulsion of  a hot and massive envelope
at very large distances.  Since the envelope is dispersed at large distances, only the degenerate object remains at the end.
Therefore, when $E<E_c$,  the system forms a compact object and ejects an envelope. This could be a mechanism leading to the formation of dwarf dark matter halos that are completely degenerate and whose dimensions are much smaller than the dimensions of large dark matter halos.\footnote{We can see in Fig. \ref{condensedRHOdot} that, for large values of $\mu$, the size of the degenerate object that forms after collapse is smaller than the size of the initial halo by about three orders of magnitude (or more). This is consistent with the difference of size between dwarf and large dark matter halos (see Table I of \cite{vega2}).}

This  evolution is reminiscent of the red-giant
phase where a star, having exhausted its nuclear fuel, collapses into a white dwarf star and ejects its outer layers by forming a planetary nebula. This is also reminiscent of the supernovae
explosion phenomenon leading to a degenerate compact object
such as a neutron star or a black hole and to the expulsion
of a massive envelope. We may wonder whether a similar scenario can take place
(or has already taken place!) at the galactic scale. We may speculate
that large dark matter halos are described by stable classical King models with
$k<k_{MCE}$ but that some halos can reach the critical value $k=k_{MCE}$ and
collapse to give birth to degenerate dwarf dark matter halos of much smaller
mass and size, with the expulsion of a massive envelope. Note, however, that
this phenomenon takes considerably much more time (of the order of the
Hubble time) than the supernova phenomenon (a few seconds) since the
gravothermal catastrophe is a rather slow process.

\subsection{The effect of decreasing $E$ for  fixed $\mu<\mu_{MCP}$}
\label{sec_finmce}

We consider a value of $\mu$ smaller than $\mu_{MCP}=1980$ for which there is no microcanonical phase transition (no collapse).
Specifically, we choose $\mu=100$ (small halo) corresponding  to the caloric curve represented
in Fig. \ref{ETmicro3}.  We start from $E=0$ and  progressively decrease the
energy.
At high energies, the system is non degenerate. It can be described by the classical King model.  When $E\rightarrow 0$, the solution is equivalent to a polytrope of index $n=5/2$. As $E$ decreases, the solutions become partially degenerate. They have a core-halo structure but the distinction between the core and the  halo is not clear-cut (see Figs. \ref{densityMu100} and \ref{velocityMu100}).  The core can be approximated
by  a polytrope of index  $n=3/2$ and the halo  can be approximated by a polytrope of index  $n=5/2$. These solutions lie in the region of negative specific heats between $E_{gas}$ and $E_{cond}$. When $E\rightarrow E_{min}$, the solutions are completely degenerate (ground state). They coincide with a polytrope of index  $n=3/2$.  The size of the cluster decreases as the energy decreases.

\subsection{The effect of increasing $\mu$ for fixed $T>T_{C}$}
\label{sec_muce}

We consider the series of equilibria of Fig. \ref{ETcano3summary} corresponding to $\mu=100$. We take a temperature $T=0.971$ larger than the critical temperature $T_c=0.613$ of gravitational collapse (isothermal collapse).  At that temperature, the system can be found in three different states: a gaseous phase (solution A'), an embryonic phase (solution B'), and a condensed phase (solution C').  We study the evolution of the  solutions A', B' and C' as $\mu$ increases.

For large values of $\mu$,  the branches A' and B' approach each other while the
branch C' moves away. For $\mu\rightarrow +\infty$, the 
branches A' and B' superimpose  while the branch C' is rejected to $E\rightarrow
-\infty$ (see Fig. \ref{mutresgrandprolongehenon}). In this limit, we recover
the spiral corresponding to the classical King model (see Fig. \ref{ckm}).

The description of solutions A' (gaseous phase) and B' (embryonic phase) is similar to the description of solutions A and B  in MCE
(see Sec. \ref{sec_mumce}). However, the description of solution C'  (condensed
phase) is different from that of solution C in MCE. Like solution C, it has a
core-halo structure. It contains a small degenerate nucleus with high density
(condensate) surrounded by a non degenerate halo. However, unlike solution C,
the nucleus contains almost all of the mass while the halo is very tenuous.
Actually, the halo is almost absent from the density profile C' in Fig.
\ref{densityMu100}. This is in sharp contrast with the density profile C in Fig.
\ref{profilesREAL} in MCE. The fact that almost all of the mass is in the
nucleus explains why $E_{C'}$ is very negative (see Fig.
\ref{mutresgrandprolongehenon}).
The size  of the nucleus  decreases as $\mu$ increases while its mass increases.  For $\mu\rightarrow
+\infty$, the radius of the nucleus tend to zero but its mass tends to $M$.
Therefore, for $\mu\rightarrow +\infty$, we recover the
 ``Dirac peak''  with infinite free energy corresponding to the strict equilibrium state
 of classical self-gravitating systems in CE (see footnote 3). For finite values of
 $\mu$, quantum mechanics  provides a regularization of this singular structure: the ``Dirac peak'' is
replaced by a ``fermion ball'' whose size is fixed by quantum mechanics. The resulting structure has
a finite free energy.

We now describe the form of the physical caloric curve when $\mu\rightarrow
+\infty$. For $\mu>\mu_{CCP}$ the physical caloric curve always looks like Fig.
\ref{ETcano3meta} (the spiral that develops for large values of $\mu$ does not
play any role since it is made of unstable states). The left branch (gaseous
phase) does not change much with $\mu$. For $\mu>\mu_{CCP}$ it almost coincides
with the classical King model ($\mu\rightarrow +\infty$). The collapse
temperature $T_c(\mu)$ is close  to $0.613$. The right branch (condensed phase)
depends sensibly on $\mu$. For $\mu\rightarrow +\infty$, the explosion
temperature $T_*(\mu)$ tends to $+\infty$. The transition temperature $T_t(\mu)$
also tends to $+\infty$. This implies that the gaseous phase corresponds to
metastable states (LFEM) while the condensed phase corresponds to fully stable
states (GFEM). For  $\mu\rightarrow +\infty$ the condensed states are singular
and have an infinite free energy. They are made of a ``Dirac peak'' containing
all the mass. As a result, the branch of condensed states (vertical line) is
rejected to $E\rightarrow -\infty$. Therefore, in the $\mu\rightarrow +\infty$
limit, the physical caloric curve is formed by
the metastable gaseous branch of Fig. \ref{ckm} up to CE  plus a singular stable
condensed branch at $E=-\infty$  (Dirac peak). On the other hand, the saddle
points are 
superposed to the spiral and to the branch of gaseous states although they have
a very different structure presenting a germ.

\subsection{The effect of decreasing $T$ below $T_c$ for fixed $\mu>\mu_{CCP}$}
\label{sec_imp2}

We consider a value of $\mu$ larger than $\mu_{CCP}=10.1$ for which a canonical phase transition (collapse) takes place at the
critical temperature $T_c=0.613$. We start from the gaseous phase and progressively decrease the temperature. This is a natural evolution since the concentration parameter $k(t)$ increases with time as the system slowly evaporates  until an instability takes place (see Appendix \ref{sec_dts}). As a result, the system follows the series of equilibria from high temperatures  to low temperatures. For $T>T_c$ (corresponding to $0\le k\le k_{CE}$), the system is non degenerate. It can be described by the classical King model. When $T\rightarrow +\infty$ (corresponding to $k\rightarrow 0$), the system is equivalent to a polytrope of index $n=5/2$. When $T$ is close to $T_c$ (corresponding to $k\sim k_{CE}$) the classical King profile can still be approximated by a polytrope $n=5/2$.  Such a profile does not account for the observations of large dark matter halos (see Paper I). This suggests that CE is not relevant to describe dark matter halos (see Appendix \ref{sec_vs}).   When $T<T_c$, the gaseous phase
(solution A') disappears and the  system undergoes a gravitational
collapse towards the condensed phase (solution C'). This corresponds to a saddle-node bifurcation. According to the discussion of Sec. \ref{sec_muce},  the gravitational collapse results in the formation of a compact degenerate object (fermion ball at $T=0$) of small size and high density that contains almost all the mass of the initial cluster. This object has only a very tenuous atmosphere with a small mass that is hardly visible. The mass of the nucleus increases as the temperature decreases and, at $T=0$, all the mass is in the nucleus. Therefore, when $T<T_c$, the system forms a compact object containing almost all the mass. There is almost no atmosphere. This is very different from the gravitational collapse in MCE that leads to a degenerate object with a small mass and the expulsion of a massive atmosphere. Therefore, the collapse in CE cannot account for the formation of dwarf dark matter halos because their observed mass is much smaller than the mass of large dark matter halos. This is another argument that CE is not relevant to describe dark matter halos (see Appendix \ref{sec_vs}).

\subsection{The effect of decreasing $T$ for fixed $\mu<\mu_{CCP}$}
\label{sec_fince}

We consider a value of $\mu$ smaller than $\mu_{CCP}=10.1$ for which there is no canonical phase transition (no collapse). We start from $T\rightarrow +\infty$ and progressively decrease the temperature. At high temperatures, the system is non degenerate. It can be described by the classical King model.  When $T\rightarrow +\infty$, the solution is equivalent to a polytrope of index $n=5/2$. As $T$ decreases, the solutions become partially degenerate. They have a core-halo structure but the distinction between the core and the  halo is not clear-cut. The core can be approximated by  a polytrope of index  $n=3/2$ and the halo can be approximated by a polytrope of index  $n=5/2$.  When $T\rightarrow 0$, the solutions are completely degenerate. They correspond to a polytrope of index  $n=3/2$. The size of the cluster decreases as the temperature decreases.

\section{Can large dark matter halos harbor a fermion ball?}
\label{sec_harbor}

Many observations have revealed that galaxies and dark matter halos contain a
very massive object at the center. This compact object is usually interpreted as
a black hole. Alternatively, some authors have suggested that this object could
actually be a fermion ball made of the same matter as the rest of the halo.
Indeed, some configurations of the self-gravitating Fermi gas at finite
temperature have a nucleus-halo structure resembling a large dark matter halo
with a small compact object at the center. This nucleus-halo structure is
particularly clear in the embryonic phase (solution B). These solutions are
similar to the gaseous phase (solution A) except that they contain a small
degenerate nucleus. The halo has the form of a truncated classical isothermal
gas consistent with the observations of large dark matter halos (Burkert
profile) and the nucleus has the form of a degenerate fermion ball. When $\mu$
is large, the fermion ball is very small so it does not affect the structure of
the halo. The corresponding density
profiles and rotation curves are represented in Figs. \ref{embryonRHOdot}-\ref{embryonVlindot}. The nucleus creates  a secondary peak and a dip in  the rotation curve at very small radii that may not be resolved observationally.  This type of nucleus-halo
configurations has been obtained by several authors \cite{gao,csmnras,viollier,bilic4,ijmpb}. Some of them \cite{viollier} made the interesting suggestion that the fermion ball
could mimic the effect of a central black hole. However, these authors
\cite{viollier} did not investigate the stability of such configurations. Our
study (see also \cite{csmnras,pt,ijmpb}) shows that these structures (solution
B) are thermodynamically unstable (i.e. unreachable)  because they are saddle
points of entropy at fixed mass and energy. Therefore, large dark matter halos
should not contain a degenerate nucleus (fermion ball). This is an important
prediction of
our study.\footnote{Some caution should be made. We have shown that
the solutions B are thermodynamically unstable. This means that they are
unstable with respect to a ``collisional'' evolution. However, as discussed in
Appendix \ref{sec_dts}, they are Vlasov dynamically stable. This means that they
are stable with respect to a ``collisionless'' evolution. On the other hand,
even if we consider their thermodynamical stability, we note that these
structures are saddle points of entropy. Therefore, they are unstable only for
some particular perturbations. As a result, provided that they appear
spontaneously from a collisionless relaxation (which is very unlikely because
they are saddle
points of
Lynden-Bell's entropy hence ``least probable''), they may persist
for a long time, as long as the system
does not spontaneously generate the dangerous perturbations that destabilize
them. Therefore, it may be possible to observe a fermion ball at the center of a
dark matter halo as a transient structure. Recalling that the fermion ball in
solution B corresponds to a  ``germ'' triggering a gravitational collapse (see
Sec. \ref{sec_large}), their observation would be the signal of a phase
transition to come. Finally, we should recall that our stability analysis
assumes that the parameter $A$ is fixed. It is not known whether other
assumptions can change the results of the stability analysis and make the
solutions B stable.} The fact that fermion balls are not observed at the center
of galaxies (a central black hole is indeed observationally favored over a
fermion ball \cite{nature,reid}) is in agreement with our result.

We note that the solutions of the condensed phase (solution C) also have a
core-halo structure with a degenerate nucleus and a non degenerate envelope.
These solutions are stable. However, in that case, the nucleus formed by
gravitational collapse releases an enormous energy that heats the envelope and
disperse it at very large distances. As a result, only the degenerate object
remains at the end. These solutions do not resemble a large dark matter halo
with a central nucleus because the atmosphere is too hot (compare solutions B
and C in Figs. \ref{embryonRHOdot}-\ref{condensedVlindot}). However, the nucleus
alone resembles a dwarf halo that is a completely degenerate object without
atmosphere.

In conclusion, dark matter halos cannot harbor a fermion ball,
unlike the proposition that has been made in the past \cite{viollier}, because
the nucleus-halo structures that have been considered by these authors are
unreachable: they correspond to saddle points of entropy at fixed mass and
energy. As a result, dark matter halos should be either everywhere non
degenerate (solution A) or everywhere completely degenerate with an atmosphere
dispersed at large distances (solution C).  They cannot be made of a completely
degenerate nucleus (fermion ball) surrounded by a non degenerate halo equivalent
to the halo of the gaseous phase because these intermediate structures (solution
B) are unstable. Therefore, it should not be possible to observe a dark matter
halo with a fermion ball.\footnote{If we observe a dark matter halo, it should
not contain a fermion ball. Inversely, if we observe a fermion ball, it  should
not be surrounded by a dark matter halo (the atmosphere has been expelled far
away).  It should not be possible to observe {\it simultaneously} a dark matter
halo and a fermion ball.} This may
explain why black holes at the center of
galaxies are observationally favored over fermion balls \cite{nature,reid}.

We note that similar results are obtained in the case where dark
matter is made of bosons instead of fermions. Slepian and Goodman \cite{slepian}
have calculated equilibrium states of a self-gravitating gas of self-interacting
bosons at finite temperature. They obtained nucleus-halo configurations made of
a classical isothermal halo and a nucleus equivalent to a BEC at zero
temperature. The BEC is the counterpart of the
fermion ball. They determined the density profiles and the rotation curves
of these configurations and
obtained results very similar to those obtained in Figs.
\ref{embryonRHOdot}-\ref{embryonVlindot} for fermions.\footnote{This confirms
the claim made in Paper I that it is not possible at present to distinguish
between fermionic and bosonic models of dark matter because they lead to very
similar results. Therefore, the bosonic models cannot be rejected a priori. This
claim depends, however, if the bosons are self-interacting or not as discussed
in Appendices \ref{sec_fbdm} and \ref{sec_degclass}.}  They argued that
these nucleus-halo structures are not consistent with observations because the
rotation curves do not show the secondary peak and the dip corresponding to the
presence of the nucleus. Actually, when the nucleus is very small, it is not
clear whether the secondary peak can be resolved observationally. Therefore,
their argument should be considered with caution. Anyway, it can be shown that
these nucleus-halo structures are thermodynamically unstable \cite{prep},
similarly to solutions B in the case of fermions. Therefore, they should not be
observed in nature. Note, however, that non condensed configurations of
self-gravitating bosons at sufficiently high temperatures may describe large
dark matter halos, totally condensed configurations of self-gravitating
bosons (BEC) at low
temperatures may describe dwarf halos, and partially condensed configurations
may describe intermediate size halos.

\section{Can large dark matter halos harbor a black hole?}
\label{sec_harborbh}

We have seen in the previous section that the presence of a fermion ball at the
center of large dark matter halos is unlikely because these nucleus-halo
structures are unreachable: they are saddle points of entropy. The presence of
a central black hole is more likely \cite{nature,reid}. These
black holes could be formed by the mechanism discussed by Balberg {\it et al.}
\cite{balberg} if dark matter is collisional \cite{spergel}. In that case, large
dark matter halos may undergo a gravothermal catastrophe when $E<E_c$. The
increase of the density and temperature of the core during the collapse can
trigger a dynamical (Vlasov) instability of general relativistic origin leading
to the formation of a black hole. During this process, only the core collapses.
This can form a black hole of large mass without affecting the structure of the
halo. Therefore, this process leads to large halos compatible with the Burkert
profile for $r>0$ but harboring a central black hole at $r=0$.

In this scenario, the presence of black holes at the center of dark matter halos
is conditioned by the possibility that dark matter halos may undergo a
gravothermal
catastrophe. Now, when quantum mechanics is taken into account, as in the
fermionic King model,
an important result of our study is the existence of a microcanonical critical
point
$\mu_{MCP}$ below which the microcanonical phase transition (gravothermal
catastrophe) is suppressed. Roughly speaking, this result implies that ``large''
dark matter halos ($\mu>\mu_{MCP}$) that are non degenerate can undergo a
gravothermal catastrophe (although this is not
compulsory\footnote{It is possible that a proportion of large dark matter halos
have a concentration parameter $k<k_{MCE}$ and have not undergone core collapse
(these halos do not contain a black hole) while some halos have reached the
critical threshold $k=k_{MCE}$ and have undergone core collapse (these halos
contain a black hole).}) and contain a central black hole  while ``small'' dark
matter halos ($\mu<\mu_{MCP}$) that are quantum objects stabilized by the Pauli
exclusion principle
cannot contain a central black hole because they do not experience a
gravothermal 
catastrophe. This result seems to qualitatively agree with the observations.
Therefore, the presence (or absence) of black holes at the center of galaxies
may be connected to the existence of a microcanonical critical point
($\mu_{MCP}=1980$) in the
fermionic King model. We provide a more quantitative criterion for
the presence of a black hole at the center of dark matter halos in Appendix
\ref{sec_bh}.

\section{Differences between dwarf and large halos}
\label{sec_diffsl}

The structure of dark matter halos crucially depends on their size through the
value of the degeneracy parameter $\mu$. 

For large halos with $\mu>\mu_{MCP}=1980$, the
series of equilibria is represented in Fig. \ref{ETmicro5summary}. It
displays an instability (gravothermal catastrophe) when $E<E_c$. When $\mu\gg
\mu_{MCP}$, two
possibilities can occur: (i) The system collapses into a fermion ball and expels
a halo at very large distances so that only the degenerate object remains at the
end (see Fig. \ref{condensedRHOdot}); \footnote{This is the
equilibrium state of the fermionic King model for $E<E_c$. Note, however, that
the collapse process
can take a very long time in practice so that, on intermediate times, one should
observe a contracting fermion ball surrounded by a halo similar to the
halo before collapse (see Fig. 5 of \cite{ribot} for 
a preliminary numerical simulation). We stress that this
nucleus-halo structure is an {\it out-of-equilibrium} structure.} (ii) A
general relativistic instability
develops before the system reaches equilibrium, and the system
forms a central
black hole surrounded by a halo not affected by the collapse. In that case, we
get a halo compatible with the Burkert profile but containing a central black
hole. This may explain the presence of black holes in large dark matter halos.
When $\mu>\mu_{MCP}$ is not too large, the system may be stabilized by quantum
mechanics before the relativistic instability leading to a black hole  takes
place. In that case, one obtains a core-halo configuration with a fermion ball
surrounded by a halo that is not too much dispersed (see Fig.
\ref{profilesREAL}). However, the structure of the halo is affected by the
collapse of the core so that it is different from the Burkert profile. In
particular, the density decreases as $r^{-\alpha}$ with an exponent $\alpha$
much smaller than $3$.

For  dwarf and intermediate size halos with $\mu<\mu_{MCP}$ the
series of equilibria is represented in Fig. \ref{ETmicro3}.  There is no
instability (no gravothermal catastrophe) because the collapse is prevented by
quantum mechanics. In that case, there is no possibility to form
black holes. This may explain why dwarf and intermediate size halos do not
contain black holes. These halos are partially or completely degenerate quantum
objects.  When $\mu<\mu_{MCP}$, all the configurations of the fermionic King
model are stable.  The solutions in the region of negative specific heat have a
core-halo structure with a partially degenerate nucleus surrounded by a non
degenerate atmosphere. However, in  that case, the distinction between the
nucleus and the  halo is not clear cut. In particular, the profile of these
solutions (see Fig. \ref{densityMu100}) is  very different
from the nucleus-halo configurations that have been considered in the literature
(see Fig. \ref{profilesREAL}).

Obviously, several configurations of dark matter halos  are possible within the
fermionic King model making the study of this model  very rich. The system can
be non degenerate (large halos), partially degenerate (intermediate size halos),
or completely degenerate (dwarf halos). We can obtain core-halo configurations
with a wide diversity of nuclear concentration depending on $\mu$ (i.e. the size
of the system) and $E$. This may account for the diversity of dark matter halos
observed in the universe. Large dark matter halos are non degenerate
classical objects. They may contain a black hole. Small halos are degenerate
quantum objects. They should not contain a black
hole. Our approach is the first attempt to determine the caloric curves of
dark matter halos. This allows us to study the thermodynamical stability of the
different configurations and to reject those that are unstable. In particular,
we have shown that the nucleus-halo configurations considered in the past (as in
Fig. \ref{embryonRHOdot}) are unstable. More work is needed to relate our
theoretical results to the observations.

\section{Conclusion}

In this paper, we have studied the  thermodynamical properties of the fermionic King
model. The fermionic King model is interesting from the viewpoint of  statistical mechanics
for the following reasons: (i) it takes into account the evaporation of high energy particles. As a result, the system  has a
finite mass without having to introduce an artificial  box; (ii) it
takes into account the Pauli exclusion principle for fermions.\footnote{The
Pauli exclusion principle is justified by quantum mechanics. As explained in
Paper I, if the evolution of the  particles is collisionless, an exclusion
principle similar to the Pauli exclusion principle arises because of dynamical
constraints brought by the  Vlasov equation.  After a phase of violent
relaxation, the system is expected to reach a quasi stationary state (QSS)
described by the Lynden-Bell distribution function that is similar to the
Fermi-Dirac distribution \cite{lb,csr,csmnras}. A fermionic King model can also
be introduced in this context in order to make the mass of the configurations
finite \cite{mnras,dubrovnik}.}  As a result, the system is stabilized against
gravitational collapse
and there  exist a non singular equilibrium  state (with a finite entropy and a finite free energy) for all accessible energies $E_{min}\le E\le 0$ and for all
temperatures $T\ge 0$; (iii) it exhibits interesting phase  transitions between gaseous states and
condensed states similar to those described in \cite{ijmpb} for a gas of self-gravitating
fermions enclosed within a box. Of course, the form of the caloric curves and the values of the critical parameters differ quantitatively from the box model  since the equilibrium states are different but the phenomenology of the phase transitions  is the same.

We have  studied the nature of phase transitions in the fermionic King model as a function of the degeneracy parameter $\mu$. For $\mu\rightarrow +\infty$, we recover the classical King model (Paper I). For finite values of $\mu$, phase transitions can take place between a ``gaseous'' phase  unaffected by quantum mechanics and  a ``condensed'' phase dominated by quantum mechanics. The phase diagram exhibits two critical points, one in each ensemble. The microcanonical critical point corresponds to $\mu_{MCP}=1980$ and the canonical critical point corresponds to $\mu_{CCP}=10.1$. For $\mu>\mu_{MCP}$, there exist microcanonical and canonical first order phase transitions. For $\mu_{CCP}<\mu<\mu_{MCP}$, only canonical first order phase transitions are present. For $\mu<\mu_{CCP}$, there is no phase transition at all. There also exist a region of negative specific heats and a situation of ensemble inequivalence when $\mu>\mu_{CCP}$. We have mentioned that metastable states have considerable lifetimes. As a result, the first order phase transitions do not take place in practice. The phase transitions listed above correspond to zeroth order phase transitions associated with spinodal points and saddle-node bifurcations.

The fermionic King model also provides a realistic model of dark matter halos and, as such, is interesting from the viewpoint of astrophysics
and cosmology. Large dark matter halos are non-degenerate so
the classical King model may be used. We have shown in Paper I that the marginally stable King model in MCE provides a good
description of large dark matter halos. Its density profile is flat in the core and decreases at large distances as $r^{-3}$, similarly to the
Burkert profile \cite{observations} that fits a large number of galactic rotation curves.\footnote{This $r^{-3}$ decay at large distances is also consistent with the NFW
profile \cite{nfw}. However, unlike the NFW profile, the marginally stable King model has a flat core density (in agreement with the observations and with the Burkert profile)  instead of a cuspy
profile.} We note that, for large dark matter halos, the cusp problem of the
CDM model is solved by finite
temperature effects, without the need to
invoke quantum mechanics. Therefore warm dark matter (WDM) may account for the
observations of large dark matter halos. By contrast, quantum mechanics must be
taken into account in smaller dark matter halos. Dwarf dark matter halos are
completely degenerate and they are equivalent to polytropes of index $n=3/2$. In
that case, the cusp problem is solved by quantum mechanics,
not by thermal effects.
Intermediate size halos are partially degenerate and they may be 
described by the fermionic King model at finite temperature.

In order to summarize our results, it is relevant to follow the series of equilibria from low values of the concentration parameter $k\rightarrow 0$ (high energies $E\rightarrow 0$ and high temperatures $T\rightarrow +\infty$) to high values of the concentration parameter $k\rightarrow +\infty$ (low energies $E\rightarrow E_{min}$ and low temperatures $T\rightarrow 0$). This evolution is natural because the concentration parameter $k(t)$ of dark matter halos increases monotonically with time due to collisions and evaporation until an instability takes place (see Appendix \ref{sec_dts}). Different  evolutions are possible depending on the value of $\mu$ and according to whether we work in MCE or CE.

We first summarize our results in MCE:

(i) If $\mu>\mu_{MCP}$ (large halos), the series of equilibria is represented in
Fig. \ref{ETmicro5instable}. For $E_c\le E\le 0$ (corresponding to $0\le k\le
k_{MCE}$), the system is in the gaseous phase (upper branch). The solutions of
this branch (solution A) are non degenerate. They correspond to the classical
King model. They may describe large dark matter halos for which quantum
mechanics is negligible.  The marginal King profile at $E=E_c$ (corresponding to
$k=k_{MCE}$) can be approximated by the  modified Hubble profile. This is the
last stable state on the gaseous branch. It accounts relatively well for the
observation of large dark matter halos (see Paper I).\footnote{Large dark matter
halos that are observed at present are expected to have a concentration
parameter close to $k_{MCE}$. The concentration parameter cannot be much smaller
than $k_{MCE}$ because $k(t)$ increases with time and the halos are relatively
old. The concentration parameter cannot be larger than $k_{MCE}$ because above
$k_{MCE}$ the gaseous branch disappears and the system collapses. These
arguments may explain why large dark matter halos are relatively well described
by the marginal King profile with $k=k_{MCE}$.} For $E<E_c$, the gaseous branch
disappears and the system collapses. Gravitational collapse in MCE results in
the formation of a completely degenerate object with a much smaller mass and
radius than the original halo accompanied by the expulsion of a hot massive
envelope (solution C). Since the envelope is expelled at large distances, only
the completely degenerate object remains at the end. These objects may
correspond
to dwarf dark matter halos that are completely degenerate and that have a much
smaller mass than large dark matter halos. Nucleus-halo solutions (solution B)
are attractive because they are similar to large dark matter halos (solution A)
with  a fermion ball at the center that could mimic a central black hole.
However, we have shown that these structures are thermodynamically unstable.
Therefore, when $\mu>\mu_{MCP}$, only two types of equilibrium structures are
stable: the
non degenerate solutions (solution A) corresponding to large dark matter halos
and the completely degenerate solutions with a dispersed atmosphere (solution C)
corresponding to dwarf dark matter halos. Nucleus-halo configurations (solution
B) are unreachable. Instead of forming a completely
degenerate dwarf halo (solution C), the gravitational collapse (gravothermal
catastrophe) can trigger a dynamical instability of general relativistic origin
and lead to the formation of a black hole \cite{balberg}. In that case, we
obtain a non degenerate large halo (solution A) harboring a central black hole.

(ii) If  $\mu_{CCP}<\mu<\mu_{MCP}$ (intermediate size halos),  the series of equilibria is represented in Fig. \ref{ETmicro3}. In that case, the evolution of the system along the series of equilibria is more progressive, without a sudden jump corresponding to a phase transition (collapse). Since there is no gravothermal catastrophe, there is no possibility to form black holes. At high energies, the solutions  are non degenerate.  At intermediate energies,  in the region of negative specific heats,  the solutions are partially degenerate. They have a core-halo structure but the distinction between the degenerate core and the non degenerate halo is not clear cut. At low energies, the solutions  are completely degenerate.

(iii) If  $\mu<\mu_{CCP}$ (dwarf halos), the evolution is similar to the previous case except that there is no region of negative specific heat.  In that case, quantum effects are relatively strong along the whole series of equilibria.

We now summarize our results in CE:

(i) If $\mu>\mu_{CCP}$, the series of equilibria is represented in Fig.
\ref{ETcano3instable}. For  $T\ge T_c$ (corresponding to $0\le k\le k_{CE}$),
the system is in the gaseous phase (left branch). The solutions of this branch
(solution A') are non degenerate. They correspond 
to the classical King model. Because of ensemble inequivalence, the value of the
concentration parameter corresponding to the marginal King profile in CE  is
smaller than in MCE (see Paper I).  As a result, the marginal King profile at
$T_c$ (corresponding to $k=k_{CE}$) is very different from the modified Hubble
profile (corresponding to $k=k_{MCE}$).  It almost coincides with a polytrope
$n=5/2$ that is the exact solution of the King model for $T\rightarrow +\infty$.
This profile  does not correspond to the observations of large dark matter
halos. This is an observational evidence that CE may not be appropriate to
describe dark matter halos. For $T<T_c$, the gaseous branch disappears and the
system collapses. Gravitational collapse in CE results in the formation of a
completely degenerate object with a small radius but a large mass, of the same
order as the mass of the original halo. This compact object is surrounded by a
very tenuous envelope with a small mass that is hardly visible (solution C').
Therefore, the isothermal collapse leads to a small degenerate object of the
same mass as the initial halo. Since dwarf dark matter halos have a much smaller
mass than large dark matter halos, this result is not consistent with
observations. This is another observational evidence that CE may not be
appropriate to describe dark matter halos.

(ii) If  $\mu<\mu_{CCP}$,  the evolution of the system along the series of equilibria is more progressive, without a sudden jump corresponding to a phase transition (collapse). In that case, quantum effects are relatively strong along the whole series of equilibria.

From these results, we conclude that MCE is more appropriate to describe dark
matter halos than CE: (i) the marginally stable King profile in MCE is
consistent with the observations of large dark matter halos while the marginally
stable King profile in CE is not; (ii) the gravitational collapse in MCE leads
to a small completely degenerate compact object with a much smaller mass than
the initial halo (and the expulsion of a hot massive envelope) while the
gravitational collapse in CE leads to a small completely degenerate compact
object with the same mass as the initial halo (with almost no atmosphere).
Therefore the gravitational collapse in MCE can account for the difference of
mass between large and dwarf halos (a factor $1000$ or more according to
Table 1 of \cite{vega2}) while the gravitational collapse in CE
does not. That MCE provides a better description than CE is consistent with the
fact that dark matter halos are rather isolated objects. Therefore a
microcanonical description is more adapted than a canonical description which
assumes that the system is dissipative and coupled to a thermal bath.

We note that the idea that dark matter halos contain  a fermion ball mimicking a
central black hole \cite{viollier} is very attractive but, unfortunately, our
study shows that these nucleus-halo structures (solution B) are unreachable
because they are saddle points of entropy. Therefore, fermion balls should not
be observed at the center of large dark matter halos (see, however, footnotes 16
and 20). It is more likely that the system develops a central black hole
\cite{balberg}.

In future works, we will take general relativity into account.  We will
also relate the caloric curves that we have obtained to the observations
of dark matter halos in order to show that the fermionic King model can account
for the diversity of dark matter halos observed in the universe. Finally, we
will explore other models of dark matter such as the bosonic model. To our
opinion, we cannot favor one model over the other for the moment.

\appendix

\section{Dynamical versus thermodynamical stability}
\label{sec_dts}

In this Appendix, we discuss subtle issues 
concerning the  dynamical and thermodynamical stability of self-gravitating
systems.

\subsection{Thermodynamical stability}
\label{sec_thermo}

For $t\rightarrow +\infty$, a self-gravitating system of fermions is expected to
reach a statistical equilibrium state described by the Fermi-Dirac distribution
(I-1). This distribution function is the solution of the maximization
problems (I-10) and (I-11) with the Fermi-Dirac entropy
\begin{eqnarray}
S=-\int \left\lbrack f\ln f+(\eta_0-f)\ln(\eta_0-f)\right \rbrack\, d{\bf r}d{\bf v}.
\label{thermo2}
\end{eqnarray}
These maximization problems determine the most probable distribution of particles at statistical equilibrium, i.e. the macrostate that is represented by the largest number of microstates. In the classical (non-degenerate) limit, the Fermi-Dirac distribution (I-1)  reduces to the Boltzmann distribution (I-2) and the Fermi-Dirac entropy (\ref{thermo2}) reduces to the Boltzmann entropy
\begin{eqnarray}
S=-\int \left\lbrack f\ln \left (\frac{f}{\eta_0}\right )-f\right \rbrack\, d{\bf r}d{\bf v}.
\label{thermo4}
\end{eqnarray}
If the system is isolated, the evolution of the distribution function is governed by the fermionic Landau equation
\begin{eqnarray}
\frac{\partial f}{\partial t}+{\bf v}\cdot \frac{\partial f}{\partial {\bf r}}-\nabla\Phi \cdot  \frac{\partial f}{\partial {\bf v}}=\frac{\partial}{\partial v^{\mu}}\int d{\bf v}_1 K^{\mu\nu}\nonumber\\
\times\left\lbrace f_1 (1-f_1/\eta_0)\frac{\partial f}{\partial v^{\nu}}-f (1-f/\eta_0)\frac{\partial f_1}{\partial v_1^{\nu}}\right\rbrace,
\label{landau1}
\end{eqnarray}
\begin{eqnarray}
K^{\mu\nu}=2\pi G^2 m\ln N \frac{u^2\delta^{\mu\nu}-u^{\mu}u^{\nu}}{u^3},
\label{landau2}
\end{eqnarray}
where $f=f({\bf r},{\bf v},t)$, $f_1=f({\bf r},{\bf v}_1,t)$, and ${\bf u}={\bf
v}_1-{\bf v}$. The collision term on the right hand side of Eq. (\ref{landau1})
models the effect of two-body encounters between particles.\footnote{Depending
on the nature of collisions, different kinetic equations can be considered. The
fermionic Landau equation can be
used to model systems with long-range interactions and the Boltzmann equation
can be used
to model systems with short-range interactions.} The
fermionic Landau equation conserves mass and 
energy and monotonically increases the Fermi-Dirac entropy ($H$-theorem).  This
corresponds to MCE. Alternatively, if the system is in contact with a thermal
bath fixing the temperature, the evolution of the distribution function is
governed by the fermionic Kramers equation
\begin{eqnarray}
\frac{\partial f}{\partial t}+{\bf v}\cdot \frac{\partial f}{\partial {\bf r}}-\nabla\Phi \cdot  \frac{\partial f}{\partial {\bf v}}=\nonumber\\
\frac{\partial}{\partial {\bf v}}\cdot \left \lbrace D \left \lbrack \frac{\partial f}{\partial {\bf v}}+\beta f (1-f/\eta_0) {\bf v}\right\rbrack\right\rbrace.
\label{fp1}
\end{eqnarray}
The term on the right hand side of Eq. (\ref{fp1}) models the interaction with
the thermal bath. The fermionic Kramers equation conserves mass and
monotonically increases the Fermi-Dirac free energy  ($H$-theorem). This
corresponds to CE.  The classical Landau equation and the classical Kramers
equation are recovered for $f\ll \eta_0$. A derivation of these kinetic
equations is given in \cite{dark,genkin,nfp}.

As recalled in the Introduction, there is no statistical equilibrium state 
for classical or quantum self-gravitating systems  in an unbounded domain
because the maximization problems (I-10) and (I-11) with the Boltzmann entropy
or with the Fermi-Dirac entropy have no solution. When coupled to the Poisson
equation, the Boltzmann distribution and the  Fermi-Dirac distribution have
infinite mass. The absence of statistical equilibrium state simply reflects the
fact that a system of particles has the tendency to evaporate. As a result, the
kinetic equations (\ref{landau1}) and (\ref{fp1}) do not relax towards a steady
state but  display a permanent evolution driven by evaporation. In practice,
evaporation is a slow process and it may be relevant to consider some form of
confinement in order to describe the structure of the system on intermediate
timescales.

A first possibility to avoid evaporation is to enclose the system within a box \cite{antonov,lbw}. The maximization  problems (I-10) and (I-11) have been studied for classical particles described by  the Boltzmann entropy in \cite{antonov,lbw,katz,paddy,aa,metastable,ijmpb} and for fermions described by  the Fermi-Dirac entropy in \cite{pt,dark,ispolatov,rieutord,ptd,ijmpb,epjb14}.  For classical particles, there exists equilibrium states only above a critical energy $E_c$ in MCE and only above a critical temperature $T_c$ in CE. These equilibrium states are metastable but they are long-lived. Below $E_c$ in MCE, the system undergoes a gravothermal catastrophe leading to a tight binary surrounded by a hot halo (at the collapse time, the singular density profile has infinite central density but zero central mass). Below $T_c$ in CE, the system undergoes an isothermal collapse leading to a Dirac peak containing all the particles. If the particles are fermions, these singular structures (tight binary and Dirac peak) are regularized by quantum mechanics. In that case, the collapse stops when the system becomes degenerate because  of the Pauli exclusion principle. As a result, there exist equilibrium states at all accessible energies and at all temperatures. This gives rise to phase transitions between a gaseous phase and a condensed phase.

Instead of working within an artificial box, we can use the classical and fermionic King models. The study of Katz \cite{katzking} and our study show that we get the same  phenomenology as when we work in a box. One difference is the absence of states with positive energy since the system is self-bound. There is also, of course, a qualitative change in the form of the series of equilibria and in the values of the critical points. Apart from that, the results are very similar. In this analogy, the generalized entropies defined by Eqs. (I-9) and (\ref{fkm8}) for the classical King model and by Eqs. (I-9) and (\ref{fkm7}) for the fermionic  King model play the same role as the Boltzmann entropy  (\ref{thermo4}) and the Fermi-Dirac entropy (\ref{thermo2})  in the case of box-confined configurations. As a result, it is natural to interpret  the maximization problems  (I-10) and (I-11) as conditions of thermodynamical stability for tidally truncated isothermal distributions (classical or quantum). As explained in Paper I, in order to relate  the turning points of energy and temperature to a change of microcanonical and canonical stability (Poincar\'e theorem), we must keep $A$ fixed along the series of equilibria. Therefore, the parameter $A$ is the counterpart of the radius $R$ of box-confined systems.

\begin{figure}[!h]
\begin{center}
\includegraphics[clip,scale=0.3]{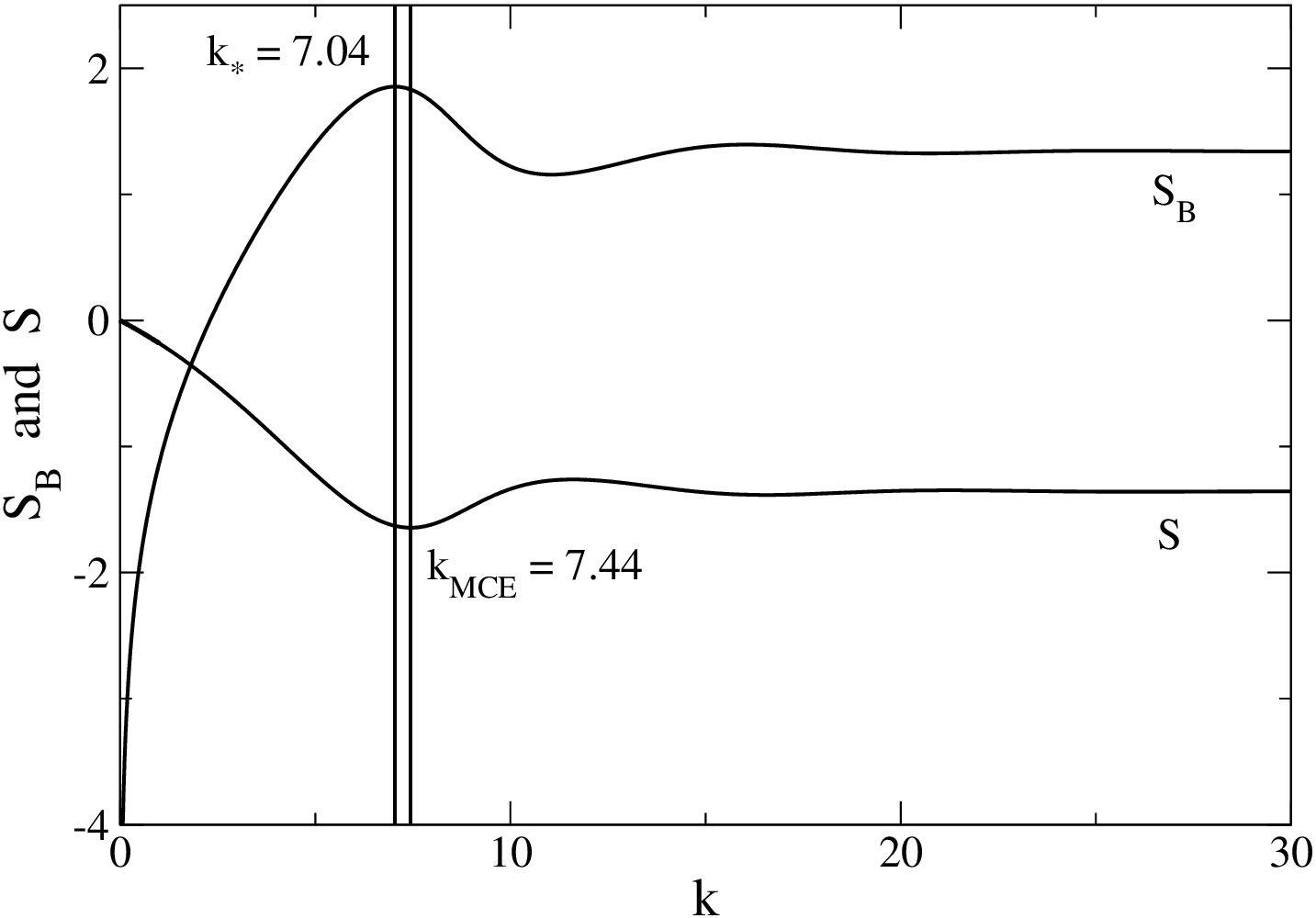}
\caption{Evolution of the Boltzmann entropy $S_B(k)$ calculated with the King
distribution along the series of equilibria. It increases and reaches a maximum at $k_*=7.04$
(corresponding to $K_*=7.84$). By contrast, the entropy of the King model $S(k)$ decreases and reaches
a minimum at $k_{MCE}=7.44$ (corresponding to $K_{MCE}=8.13$). In the two cases, the value of
$A$ is fixed so the energy changes as $k$ increases.}
\label{kSbS}
\end{center}
\end{figure}

We note, however, that the analogy with thermodynamics remains heuristic (and
maybe incorrect) because the King model is an out-of-equilibrium model. Indeed,
the system keeps evolving by losing mass and energy as a result of evaporation.
Therefore, it is not quite clear whether one can use arguments of equilibrium
thermodynamics to study the stability of the King model. Another possibility is
to use kinetic theory.\footnote{We restrict our discussion to MCE for the
reasons explained in Appendix \ref{sec_vs}.} The King distribution is a quasi
stationary solution of the Landau equation  with coefficients that slowly change
with time because of evaporation. During the collisional evolution, the system
becomes more and more isothermal so the concentration parameter  $k(t)$
increases monotonically with time. If we consider the non degenerate limit
(gaseous phase) and plot the Boltzmann entropy $S_B$ calculated with the King
distribution as a function of $k$, we obtain the curve reported in Fig.
\ref{kSbS}. We see that $S_B$ increases monotonically for $k<k_*=7.04$, then
decreases. Since $k(t)$ increases with time, a decrease of $S_B(k)$ is not
physically possible because it would  violate the $H$-theorem associated with
the Landau equation. This implies that the King distribution is unstable for
$k>k_*$. At that point, the system undergoes a gravothermal catastrophe and
experiences core collapse. This  instability takes it away from the King
sequence. This argument was first put forward by Lynden-Bell and Wood
\cite{lbw}. Cohn \cite{cohn} confirmed numerically that the actual path of
evolution departs from the King sequence at some critical $k_*$ corresponding
to the maximum of $S_B(k)$. The value of $k_*$ obtained by Cohn \cite{cohn} is
slightly different from ours. We obtain  $K_*=7.84$ ($k_*=7.04$) by fixing $A$ while Cohn
obtains $K_*=9.3$ ($k_*=8.82$) by scaling the King distribution to the mass and  energy of
the evolutionary model.  We have also represented in Fig. \ref{kSbS} the
evolution of the generalized entropy defined by Eqs. (I-9) and (I-69) as a
function of $k$. As we have seen in Paper I, it reaches its extremal value at
$k_{MCE}=7.44$. It is a little bit disturbing to see that this extremum value is
actually a minimum and that $S(k)$ decreases for $k<k_{MCE}$. There is no
paradox, however, since the Landau equation satisfies an H-theorem for $S_B$,
not
for $S$.
Therefore, $S$ may decrease with time. However, these considerations  indicate
that, for open systems, the Boltzmann entropy $S_B$ and the generalized entropy
$S$ (both calculated with the King distribution) have a different physical
meaning. The Boltzmann entropy is appropriate to interpret the dynamical
evolution of the system in relation to the $H$-theorem (kinetic theory), and the
generalized entropy is appropriate to deduce stability limits from the series of
equilibria  of the King model by using the Poincar\'e theory.

We can make an additional comment. The value of the critical concentration  $k_*=8.82$
obtained by Cohn \cite{cohn} is slightly larger than the value $k_{MCE}=7.44$
obtained from the Poincar\'e theory by fixing $A$. This suggests that the
gravothermal catastrophe occurs slightly later than predicted by Katz
\cite{katzking} and in Paper I. This allows the slope $\alpha$ of the density
profile to be substantially smaller than $3$ (the slope corresponding to
$k_*=8.82$ is $\alpha=2.52$). A larger value of the critical concentration
slightly improves the agreement between the marginal King model and the Burkert 
profile. On the other hand, we remark
that the critical concentration  $k_*=8.82$ obtained by Cohn \cite{cohn} is
relatively close to the  value $k'_{MCE}=8.50$ obtained from the Poincar\'e
theory by fixing $R$ (see Sec. VI. of Paper I). It is not clear whether this is
a coincidence or if it bears a deeper meaning than is apparent at first sight.

\subsection{Dynamical stability}
\label{sec_lb}

In the previous section, the system was assumed to have reached a statistical equilibrium state described by the fermionic King model as a result of collisions. In that case, the Fermi-Dirac distribution arises from the quantum properties of the particles (fermions). However, collisions generally take a very long time to establish a statistical equilibrium state (except possibly in the dense core). For self-gravitating systems with a large number of particles, the collisional relaxation time is in general much larger than the age of the universe because it scales with the number of particles as $(N/\ln N)t_D$, where $t_D$ is the dynamical time \cite{bt}. As a result, most self-gravitating systems are collisionless and their evolution is governed by the Vlasov equation
\begin{eqnarray}
\frac{\partial f}{\partial t}+{\bf v}\cdot \frac{\partial f}{\partial {\bf r}}-\nabla\Phi \cdot  \frac{\partial f}{\partial {\bf v}}=0.
\label{vlasov}
\end{eqnarray}

It can be shown that any steady state of the Vlasov equation of the form $f=f(\epsilon)$ with $f'(\epsilon)<0$ is linearly \cite{bt}, and even nonlinearly \cite{lmr}, dynamically stable. Therefore, all the King distributions are dynamically stable, whatever their degree of concentration $k$. The maximization problems (I-10) and (I-11) provide only sufficient, but not necessary, conditions of dynamical stability. That a stronger dynamical stability criterion exists for the Vlasov equation is due to the fact that this equation conserves an infinite number of integrals, beyond mass and energy, the so-called Casimirs integrals \cite{lmr,cc}.

However, stability analysis does not explain {\it how} a collisionless
self-gravitating system 
reaches a steady state. Collisionless relaxation is actually a very non-trivial
concept related to mechanisms known as violent relaxation, phase mixing, and
nonlinear Landau damping \cite{bt}. This form of relaxation  takes place on a
very short timescale of the order of a few dynamical times $t_D$. It is
therefore very relevant in astrophysics. Assuming ergodicity, the quasi
stationary state (QSS) that results from violent relaxation can be predicted
from the statistical theory of Lynden-Bell \cite{lb,csr,csmnras}. In that
approach, the QSS reached by the system is the solution of the maximization
problem (I-10) where $S$ is the Lynden-Bell entropy defined by Eq.
(\ref{thermo2}) with a bar on $f$ (it represents the coarse-grained
distribution).  The  Lynden-Bell distribution, given by Eq. (I-1) with a bar on
$f$,  is similar to the Fermi-Dirac distribution but  the reason has nothing to
do with quantum mechanics. To avoid the infinite mass problem arising in
Lynden-Bell's theory, we can consider the fermionic King model defined by Eq.
(\ref{fkm1}) with a bar on $f$ \cite{mnras,dubrovnik}.  The maximization problem
(I-10) with the entropy defined by Eqs. (I-9) and (\ref{fkm7}) determines the
most probable coarse-grained distribution $\overline{f}$ resulting from the
intertwinement of the fine-grained distribution function $f$. While all the
distribution functions of the form $\overline{f}=\overline{f}(\epsilon)$ with
$\overline{f}'(\epsilon)<0$ are dynamically stable, some of them are more
probable than others. This is the difference between Vlasov dynamical stability
and Lynden-Bell thermodynamical stability. For example, in the dilute limit, all
the King distributions (interpreted as tidally truncated Lynden-Bell's
distributions) are Vlasov dynamically stable but only the King distributions
with $k<k_{MCE}$ are Lynden-Bell thermodynamically stable, i.e. ``most
probable'' (local maxima of entropy at fixed mass and energy). On the other
hand, while the fine-grained distribution function $f({\bf r},{\bf v},t)$ is the
solution of the Vlasov equation (\ref{vlasov}), the coarse-grained distribution
function $\overline{f}({\bf r},{\bf v},t)$ is not. It is the solution of a
kinetic equation that has the form of the fermionic Landau equation
(\ref{landau1}) with a bar on $f$  \cite{mnras,dubrovnik}. This equation
conserves mass and energy and satisfies an $H$-theorem for the Lynden-Bell 
entropy. Therefore, the problematic of the collisionless relaxation is, in some
sense, similar to the problematic of the collisional relaxation. Because of this
analogy, the same arguments apply except that the distribution function $f$ must
be viewed as the {\it coarse-grained} distribution function $\overline{f}$. The
main differences between collisionless and collisional relaxation are the
following: (i)  the timescale of collisionless relaxation is much shorter than
the timescale of collisional relaxation; (ii)  the equilibrium distribution of
collisionless systems undergoing violent relaxation is similar to the
Fermi-Dirac distribution even if the particles are classical while the 
equilibrium distribution of collisional systems is the Boltzmann distribution
for classical particles and the Fermi-Dirac distribution for fermions; (iii) the
evolution of collisionless systems stops when the system has reached a
virialized state while collisional systems undergo a permanent evaporation.

\section{Microcanonical versus canonical ensemble}
\label{sec_vs}

For the sake of completeness, we have considered both microcanonical and canonical ensembles in the thermodynamical analysis of the King model. However, we give below some arguments that MCE may be more appropriate than CE to describe dark matter halos:

(i) Most of self-gravitating systems such 
as galaxies and globular clusters are isolated rather than being coupled to a
thermal bath \cite{bt}. This may be the case for dark matter halos also. 
Therefore, the Boltzmann and the Landau equations  (MCE) may be more appropriate
than the Kramers equation (CE) to describe the dynamics of dark matter halos.

(ii) The King model can be derived from the Landau equation in which the
diffusion coefficient $D(v)$ depends on the velocity of the particles and
decreases as $1/v^3$ for $v\rightarrow +\infty$.  This decay  is crucial in
order to obtain the King model \cite{king,lb,mnras,dubrovnik,clevap}. Since the
diffusion coefficient appearing in the Kramers equation  is constant, the King
model cannot be rigorously derived from that equation. 

(iii) The Burkert profile that fits a wide diversity of rotation curves 
of galaxies is relatively close to the King profile at the limit of
microcanonical stability $k_{MCE}=7.44$. At that point, the King profile can be
approximated by the modified Hubble profile that is reasonably close to the
Burkert profile.  By contrast, the Burkert profile  is relatively different from
the  King profile at the limit of canonical stability $k_{CE}=1.34$. In CE, only
the King distributions that are close to a polytrope $n=5/2$ are stable, and
these distributions are very different from the Burkert profile. Since
$k_{CE}<k_{MCE}$ as a result of ensemble inequivalence, the modified Hubble
profile is accessible in MCE but not in CE. Therefore, from the observation
viewpoint, MCE gives better results than CE. This is because it contains a
larger sequence of stable King models than CE as a result of ensemble
inequivalence.

(iv) The consideration of MCE solves the apparent contradiction faced by
Merafina and Ruffini \cite{mr}. These authors study the dynamical stability of
the classical King model and find, in the non-relativistic regime, that it
becomes unstable for $k>1.3654$ (in our notations). They conclude therefore that
``the configurations integrated by King are not in the stable branch''. However,
their stability criterion corresponds to the criterion of {\it canonical}
stability $k_{CE}=1.34$ (the small difference between $1.3654$ and $1.34$ is
probably a numerical effect). Since globular clusters are isolated, they should
be treated in MCE where the limit of stability is $k_{MCE}=7.44$. In this proper
ensemble, most of the configurations integrated by King are stable. Therefore,
there is no contradiction. The same remark applies to dark matter halos.

(v) The gravitational collapse of a large dark matter halo of mass $M\sim 10^{12}\, M_{\odot}$  in MCE leads to a small compact degenerate object containing a much smaller mass $M\sim 10^{6}\, M_{\odot}$ that the initial halo, accompanied by the expulsion of a hot and massive envelope (see Sec. \ref{sec_imp}). The compact object may correspond to a dwarf dark matter halo (see Table 1 of \cite{vega2}). Gravitational collapse of a large dark matter halo in CE leads to a small compact degenerate object containing almost all the mass of the initial halo, surrounded by a very tenuous atmosphere (see Sec. \ref{sec_imp2}). The compact object cannot correspond to a dwarf dark matter halo because it has a too big mass.  Therefore, from the observation viewpoint, MCE gives better results than CE.

(vi)  If the dark matter halos were coupled to a thermal bath (CE), they would 
all have the same temperature while observations reveal that their temperature
(or their velocity dispersion) depends on their size (see Table 1 of
\cite{vega2}). By contrast, isolated systems (MCE) have different
energies, hence different temperatures.

(vii) That the temperature depends on the size of the halos may be in favor of Lynden-Bell's  theory of violent relaxation \cite{lb} where $T$ is an out-of-equilibrium temperature that has no reason to be the same for all clusters. In the Lynden-Bell theory, only MCE makes sense since this theory is based on the Vlasov equation that applies to an isolated collisionless system (fixed $E$).

(viii) Violent relaxation leads to core-halo structures with a density slope
$\alpha=4$ \cite{henonVR,albada,roy,joyce}. These configurations are consistent
with a King model of index $k\sim 5$. They are stable in MCE but unstable in CE.
This is another argument that MCE is more relevant than CE.

\section{The functions ${\cal R}$, $F$, $G$, and $H$}
\label{sec_fgh}

In Paper I, we have introduced the functions
\begin{equation}
\frac{R}{r_h}=\frac{\zeta_1(k)}{\zeta_h(k)}\equiv {\cal R}(k),
\label{hr2}
\end{equation}
\begin{equation}
\frac{M_h}{\rho_0r_h^3}=-4\pi\frac{\chi'[\zeta_h(k)]}{\zeta_h(k)}\equiv F(k),
\label{dim1}
\end{equation}
\begin{equation}
\frac{\sigma_0^2}{G\rho_0r_h^2}=\frac{8\pi}{3}\frac{1}{\zeta_h^2(k)}\frac{I_2(k)}{I_1(k)}\equiv G(k),
\label{dim3}
\end{equation}
where $r_h$ is the halo radius such that $\rho(r_h)/\rho_0=1/4$ \cite{vega3}. These functions have been computed for the classical King model. For the fermionic model, we introduce the additional function
\begin{equation}
\frac{\eta_0\sigma_0^3}{\rho_0}=\frac{\mu}{3\sqrt{3}}\frac{I_2(k)^{3/2}}{I_1(k)^{5/2}}\equiv H(k),
\label{dim4}
\end{equation}
which gives the ratio between the maximum value of the distribution function
$\eta_0=gm^4/h^3$ fixed by the Pauli exclusion principle and the typical central
phase space density $\rho_0/\sigma_0^3$. To obtain the right hand side of Eq.
(\ref{dim4}), we have used Eqs. (I-28), (I-99), and (I-100). We also note that
the exact central distribution function is $f_0=A{\cal F}_s(-k)$ so that
\begin{equation}
\frac{\eta_0}{f_0}=\frac{e^k+\mu}{e^k-1}.
\label{dim5}
\end{equation}

For the fermionic King model, we have the following asymptotic results. For $k\rightarrow 0$, we get ${\cal R}(0)=2.74$, $F(0)=1.89$, $G(0)=0.944$ (see Paper I), $H(k)\sim 15(1+\mu)/(56\pi\sqrt{7}k)\sim 3.22\, 10^{-2}\, (1+\mu)k^{-1}$, and $\eta_0/f_0\sim (1+\mu)/k$. For $k\rightarrow +\infty$, we get ${\cal R}(+\infty)=1.61$, $F(+\infty)=-4\pi\theta'_h/\xi_h=1.99$, $G(+\infty)=8\pi/(5\xi_h^2)=0.975$, $H(+\infty)=3/(20\pi\sqrt{5})=2.14\, 10^{-2}$, and $\eta_0/f_0\rightarrow 1$ where we have used $\xi_h=2.27$ and $\theta'_h=-0.360$ for a polytrope $n=3/2$.

For a given value of $\mu$, Eq. (\ref{dim5}) can be used to determine the value
of the concentration parameter $k$ above which the system is degenerate. If we
consider that the system is degenerate when $f_0>\eta_0/\nu$, where $\nu$ is a
number that depends on our degree of precision (e.g. $\nu=10$), we find that the
system is degenerate when $k>\ln[(\mu+\nu)/(\nu-1)]$. Inversely, for a given
value of $k$ we find that the system is degenerate when $\mu<(\nu-1)e^k-\nu$. As
an illustration, taking
$k=k_{MCE}=7.44$, we find that the system is degenerate at that point when
$\mu<15300$.

We can use the functions $F$, $G$, $H$ to relate the theoretical results of the fermionic King model to the observations. This is beyond the scope of the present paper but this will be considered in a future work.

\section{Fermionic versus bosonic dark matter}
\label{sec_fbdm}

In this Appendix, we determine the mass of the particles that compose dark matter halos according to whether they correspond to fermions, bosons without self-interaction, or bosons with self-interaction.

The smallest known dark matter halo is Willman 1 that has $r_h=33\, {\rm pc}$, $\rho_0=6.8\, M_{\odot}/{\rm pc}^3$, $M_h=0.39\, 10^6\, M_{\odot}$, and $\sigma_0=4\, {\rm km/s}$ \cite{vega2,vega3}. To determine the mass of the particles that compose dark matter halos, we consider that this most compact halo is completely degenerate (for fermions) or completely condensed (for bosons), i.e. that it corresponds to the ground state of a self-gravitating Fermi or Bose gas.  We compare the result with the one obtained by considering  that large dark matter halos such as the Medium Spiral (with $r_h=1.9\, 10^4 \, {\rm pc}$, $\rho_0=7.6\, 10^{-3}\, M_{\odot}/{\rm pc}^3$, $M_h=1.01\, 10^{11}\, M_{\odot}$, and $\sigma_0=76.2\, {\rm km/s}$ \cite{vega2}) are completely degenerate or completely condensed. This is incorrect but this estimate has been made in the past so it is interesting to do the comparison.

A completely degenerate system of self-gravitating fermions at $T=0$ has the mass-radius relation $MR^3=1.49\, 10^{-3}\, h^6/(G^3 m^8)$ \cite{chandra}. This gives
\begin{equation}
\frac{m}{{\rm eV}/c^2}=2.27\, 10^4 \left (\frac{\rm pc}{R}\right )^{3/8}\left (\frac{M_{\odot}}{M}\right )^{1/8}.
\label{fb1}
\end{equation}
Using the values of $M$ and $R$ corresponding to Willman 1, we obtain a fermion
mass $m=1.23\, {\rm keV}/c^2$.  This is the typical
mass\footnote{This result assumes that (i) Willman 1 is completely degenerate,
and that (ii) the observational values of $r_h$ and $M_h$ are accurate. More
precise observational data may change the value of the particle mass $m$ but its
order of magnitude should remain the same. On the other hand, if the
distribution function of the halos corresponds to the Lynden-Bell distribution
function instead of the Fermi-Dirac distribution function, we must divide
$\eta_0=2m^4/h^3$ by $2$ since $\eta_0^{LB}=\eta_0^{Pauli}/2$ (see footnote 8 in
Paper I).  This implies that the particle mass $m$ must be multiplied by
$2^{1/4}$ giving $m=1.46\, {\rm keV}/c^2$.} obtained by de Vega and Sanchez
\cite{vega,vega2}. This particle could be a sterile neutrino. In the past, some
authors \cite{gr,stella,mr,gao,mr2,merafina} have determined the fermion mass by
considering that large dark matter halos are completely degenerate. Using the
values of $M$ and $R$ corresponding to the Medium Spiral, this leads to a
fermion mass $m=28.8\, {\rm eV}/c^2$. This is the typical mass obtained in
\cite{gr,stella,mr,gao,mr2,merafina}. However, if large dark matter halos were
completely degenerate there would not be smaller halos such as Willman 1.
Therefore, this prediction is not correct. Furthermore, the density profiles and
the rotation curves of large dark matter halos do not correspond to that of a
completely degenerate self-gravitating Fermi gas. Indeed, large dark matter
halos are closer to a classical self-gravitating isothermal gas (see Paper I).

A completely condensed system of self-gravitating bosons (BEC) without self-interaction at $T=0$  has the mass-radius relation $MR=9.95\, \hbar^2/(G m^2)$ \cite{rb,membrado,prd2}. This gives
\begin{equation}
\frac{m}{{\rm eV}/c^2}=9.22\, 10^{-17} \left (\frac{\rm pc}{R}\right )^{1/2}\left (\frac{M_{\odot}}{M}\right )^{1/2}.
\label{fb2}
\end{equation}
Using the values of $M$ and $R$ corresponding to Willman 1, we obtain a boson
mass $m=2.57\, 10^{-20}\, {\rm eV}/c^2$. This is a new prediction. In the past,
some authors \cite{baldeschi} have determined the
boson mass by considering that large dark matter halos are completely condensed.
Using the values of $M$ and $R$ corresponding to the Medium Spiral, this leads
to a boson mass  $m=2.10\, 10^{-24}\, {\rm eV}/c^2$.  This is the typical mass
obtained in \cite{baldeschi}. However, the Medium Spiral cannot be completely
condensed otherwise there would not be smaller halos such as Willman 1.
Therefore, this prediction is not correct. In large dark matter halos, we
must take into account the envelope made of scalar radiation that surrounds the
solitonic core (see footnote 4 of
Paper I). This envelope fixes the size of large dark matter halos.

A completely condensed system of self-gravitating bosons (BEC)  with self-interaction at $T=0$ in the Thomas-Fermi limit has a unique radius $R=\pi(a\hbar^2/Gm^3)^{1/2}$ ($a$ is the scattering length) \cite{goodman,arbey,bohmer,prd1}. This gives
\begin{equation}
\left (\frac{{\rm fm}}{a}\right )^{1/3} \left (\frac{m}{{\rm eV}/c^2}\right )=6.73 \left (\frac{\rm pc}{R}\right )^{2/3}.
\label{fb3}
\end{equation}
Using the value of $R$ corresponding to Willman 1, we obtain $({\rm
fm}/a)^{1/3}(mc^2/{\rm eV})=0.654$. This is a new prediction. If we take
$m=1.23\, {\rm keV}/c^2$ we
obtain $a=6.65\, 10^9\, {\rm fm}$. If we take  $m=2.57\, 10^{-20}\, {\rm
eV}/c^2$ we obtain $a=6.07\, 10^{-59}\, {\rm fm}$. If we take
$a=10^6\, {\rm fm}$ which corresponds to the typical value of the scattering
length observed in laboratory BEC experiments, we obtain $m=65.4\, {\rm
eV}/c^2$. We
can show that the Thomas-Fermi approximation is valid when
\cite{prd2,bookspringer}:
\begin{equation}
\frac{M}{M_{\odot}}\gg 1.54\, 10^{-34} \left (\frac{{\rm fm}}{a}\right )^{1/2} \left (\frac{{\rm eV}/c^2}{m}\right )^{1/2}.
\label{fb4}
\end{equation}
Therefore, when $M=0.39\, 10^6\, M_{\odot}$,  $m=2.57\, 10^{-20}\, {\rm eV}/c^2$
and $a=6.07\, 10^{-59}\, {\rm fm}$, the Thomas-Fermi approximation is not fully
valid and one must take into account {\it both} the pressure due to the
self-interaction and the quantum pressure. The corresponding mass-radius
relation has been obtained in \cite{prd1,prd2,bookspringer}. In the past, some
authors \cite{bohmer} have determined the ratio $m/a^{1/3}$ by considering that
large dark matter halos are completely condensed. Using the value of $R$
corresponding to the Medium Spiral, we obtain $({\rm fm}/a)^{1/3}(mc^2/{\rm
eV})=9.45\, 10^{-3}$. However, if large dark matter halos were completely
condensed there would not be smaller halos such as Willman 1. Therefore, this
prediction is not correct. Finite temperature effects must be taken into account
to describe large dark matter halos.

We note that, for a given dark matter particle mass $m$, the 
ground state of self-interacting bosons determine the radius of the most compact
dwarf halos while the ground state of fermions and non-interacting bosons
determine their mass-radius relation.

Finally, we can make the following remark. In the fermionic model, large halos are classical (see Appendix \ref{sec_degclass}) so we have the relation $\sigma_0^2=k_B T/m$. Introducing relevant scales, this can be rewritten as
\begin{equation}
\frac{m}{{\rm eV}/c^2}=7.74\, 10^{12} \left (\frac{{\rm m/s}}{\sigma_0}\right )^{2}\frac{T}{{\rm K}}.
\label{fb5}
\end{equation}
For the medium spiral, $\sigma_0=76.2\, {\rm km/s}$. If we argue that the temperature is in the Kelvin range, as is the case for the radiation background, and take $T\sim 1\, {\rm K}$, we obtain $m=1.33\, {\rm keV}/c^2$. Thus, we find that the fermion mass is in the ${\rm keV}/c^2$ range. This is a completely different argument than the one given previously (relying on the ground state of the self-gravitating Fermi gas) but we may find confidence in the fact that  these two arguments lead to similar results. On the other hand, in the bosonic scenario, a boson mass $m=2.57\, 10^{-20}\, {\rm eV}/c^2$ leads to an effective temperature $T=1.93\, 10^{-23}\, {\rm K}$. It is not clear which meaning can be attached to such a small temperature.

\section{Quantum versus classical halos}
\label{sec_degclass}

Once we know the mass of the dark matter particle (see Appendix
\ref{sec_fbdm}), we can determine if a given halo is quantum (Fermi degenerate
or Bose condensed) or classical. Let us first consider 
the case of fermions. The parameter 
\begin{equation}
H=\frac{2m^4\sigma_0^3}{\rho_0 h^3} 
\label{fund1}
\end{equation}
measures the
degree of degeneracy of the core of dark matter halos (see Appendix
\ref{sec_fgh}). Introducing scaled variables, this parameter may be written as
\begin{equation}
H=1.03\, 10^{-24} \left (\frac{m}{{\rm eV}/c^2}\right )^4 \left (\frac{\sigma_0}{{\rm m/s}}\right )^3\left (\frac{M_{\odot}/{\rm pc}^3}{\rho_0}\right ).
\label{degclass1}
\end{equation}
Using the results of Sec. \ref{sec_compdeg}, we find that a completely
degenerate  system of self-gravitating fermions at $T=0$ has $H_0=0.0214$. On
the other hand, one can show that a self-gravitating Fermi gas is non degenerate
(classical) when $H>1$ \cite{prep}. 

According to the discussion of Appendix
\ref{sec_fbdm}, we take a fermion mass $m=1.23\, {\rm keV}/c^2$. For Willman 1,
we find $H=0.0221\simeq H_0$, a value expected for a completely degenerate
system (this is how the mass $m=1.23\, {\rm keV}/c^2$ has been obtained). For
the Medium Spiral, we find $H=1.35\, 10^5\gg 1$ indicating that this system is
non
degenerate.\footnote{We have to be careful that the observed central density
$\rho_0=7.6\, 10^{-3}\, M_{\odot}/{\rm pc}^3$ reported in Table 1 of
\cite{vega2} may be an apparent one. Large dark matter halos may contain a
central nucleus (black hole) of very small size and huge density $\rho'_0\gg
\rho_0$ that may not be resolved observationally (see Appendix \ref{sec_bh}).}
We can also compute $H$ for a globular cluster with  $r_h=10 \, {\rm pc}$,
$\rho_0=8\, 10^{3}\, M_{\odot}/{\rm pc}^3$, $M_h=6\, 10^{5}\, M_{\odot}$,
$\sigma_0=7\, {\rm km/s}$, and $m_*=M_{\odot}=1.12\, 10^{66} \, {\rm eV}/c^2$
\cite{bt}. In that case, we get the huge number $H=6.80\, 10^{247}\gg 1$.

We now evaluate the parameter $H$ in the case where dark matter is made of
bosons. It can be shown that the core of the halo is condensed (BEC) when
$H<4.86\, 10^{-2}$ (see Appendix \ref{sec_temp}) and non condensed (thermal
bosons) when $H>4.86\, 10^{-2}$. If we
take a boson mass $m=2.57\, 10^{-20}\, {\rm eV}/c^2$
appropriate to non-interacting bosons (see Appendix \ref{sec_fbdm}), we find
that $H\ll 4.86\, 10^{-2}$ even for large dark matter halos such as the Medium
Spiral (or
even larger halos) indicating that their core is always condensed.
If we
take a boson mass $m=1.23\, {\rm keV}/c^2$, which is allowed for
self-interacting bosons, we find that dwarf halos such
as Willman 1 are Bose condensed ($H=0.0221$)\footnote{Actually, for this value
of $m$ we see that Willman 1 is only marginally condensed. This suggests that
the boson mass should be smaller than $m=1.23\, {\rm keV}/c^2$. We recall that,
for self-interacting bosons, observations only
determine the ratio $({\rm
fm}/a)^{1/3}(mc^2/{\rm eV})=0.654$, not $m$ and $a$ individually. Therefore,
the mass $m$ is un-determined and the value $m=1.23\, {\rm keV}/c^2$ was chosen
as an example.} while large
halos such as the Medium Spiral are
non-condensed ($H=1.35\, 10^5$).

These results are potentially very important because they can help determining 
if dark matter is made of fermions or bosons. Indeed, in the fermion
case ($m=1.23\, {\rm keV}/c^2$), small
dark matter halos are quantum objects while large halos are classical objects.
By contrast, in the non-interacting boson case ($m=2.57\, 10^{-20}\, {\rm
eV}/c^2$), small and large dark matter halos are quantum objects. They have a
core-halo structure made
of a solitonic core surrounded by a halo of scalar radiation (see footnote 4 of
Paper I). In the self-interacting boson case, the halos can be classical or
quantum depending on the boson mass.

Since the surface density $\Sigma_0=\rho_0 r_h$ is approximately the same for
all the halos \cite{vega3}, it is relevant to express $H$ in terms of this
quantity. Using the results of Appendix \ref{sec_fgh}, we obtain
\begin{equation}
H=\frac{2 m^4 G^{3/2}M_h^{5/4}}{\Sigma_0^{3/4}h^3} \frac{G(k)^{3/2}}{F(k)^{5/4}}.
\label{degclass2}
\end{equation}
As shown in Paper I, the function $F(k)$ and $G(k)$ do not sensibly change with
$k$. Taking $F(k)\sim 1.8$ and $G(k)\sim 0.95$, and introducing relevant scales,
we obtain
\begin{equation}
H=1.29\, 10^{-19} \left (\frac{m}{{\rm eV}/c^2}\right )^4 \left (\frac{M_h}{M_{\odot}}\right )^{5/4}\left (\frac{M_{\odot}/{\rm pc}^2}{\Sigma_0}\right )^{3/4}.
\label{degclass3}
\end{equation}

Considering that $\Sigma_0=120 M_{\odot}/{\rm pc}^2$ is the same for all
the halos,
and assuming that dark matter is made of fermions with mass $m=1.23\, {\rm
keV}/c^2 $, we find that the halos are degenerate ($H<1$) for $M_h<2.97\, 10^6
\, M_{\odot}$ and classical ($H>1$) for $M_h>2.97\, 10^6 \, M_{\odot}$.
Therefore, the majority of the observed halos reported in Table 1 of
\cite{vega2} are classical. Still, the dwarf halos that correspond to the ground
state of the self-gravitating Fermi gas are crucially important for determining
the mass of the dark matter particle \cite{vega2}. 

Assuming
that dark matter is made of interacting or non-interacting bosons of
mass $m=2.57\,
10^{-20}\, {\rm eV}/c^2$, we find that all the observed halos are Bose
condensed. The bosonic halos would become  classical for  $M_h>1.11\, 10^{79} \,
M_{\odot}$. 

Assuming that dark matter is made of self-interacting bosons of
mass $m=1.23\, {\rm keV}/c^2$, we find that the halos are condensed
($H<4.86\, 10^{-2}$) for $M_h<2.63\, 10^5
\, M_{\odot}$ and non condensed ($H>4.86\, 10^{-2}$) for $M_h>2.63\, 10^5 \,
M_{\odot}$.

Finally, assuming that dark matter is made of particles with mass
$m\sim \, {\rm GeV}/c^2$, corresponding to the CDM model, we find that all the
observed halos are classical ($H\gg 1$). These halos would become quantum for 
$M_h<1.44\, 10^{-3} \, M_{\odot}$. This bound is so small that quantum mechanics
can be neglected in the CDM model. Therefore, if the CDM model were valid, we
should observe halos of any
size. The fact that we do not observe halos below
the size of Willman 1 (missing satellite problem) shows that there exists  a
minimum scale in the universe (ground state) that is fixed by quantum mechanics
(fermions or bosons).

\section{The temperature of the halos}
\label{sec_temp}

The results of the previous section can be expressed in terms of the
temperature of the halos. We have seen that the halos are classical if
$H=2m^4\sigma_0^3/\rho_0 h^3>1$. Since $\sigma_0^2=k_B T/m$ for a classical
isothermal equation of state, the classical limit corresponds to $T>T_F$ where
\begin{equation}
k_B T_F=\frac{2^{4/3}\pi^2\hbar^2\rho_0^{2/3}}{m^{5/3}}
\label{temp1}
\end{equation}
is the Fermi temperature. The parameter $H$ may be rewritten as
$H=(T/T_F)^{3/2}$. Introducing scaled variables, we obtain
\begin{equation}
\frac{T_F}{{\rm K}}=1.27\, 10^3 \left (\frac{\rho_0}{M_{\odot}/{\rm pc}^3}\right
)^{2/3}\left (\frac{{\rm eV}/c^2}{m}\right )^{5/3}.
\label{temp2}
\end{equation}
Using the results of Sec. VII.E of Paper I, the temperature of classical halos 
is given by $k_B T=G(k) m G\rho_0 r_h^2$, or by $k_B T=G(k) m G\Sigma_0 r_h$, or
by $k_B T=\lbrack G(k)/F(k)\rbrack m G M_h/r_h$, or by $k_B
T=\lbrack G(k)/F(k)^{1/2}\rbrack m G \Sigma_0^{1/2}M_h^{1/2}$ with $F(k)\sim
1.8$ and $G(k)\sim 0.95$. Introducing scaled variables, we obtain
\begin{equation}
\frac{T}{{\rm K}}=5.28\, 10^{-10} \frac{m}{{\rm eV}/c^2}\frac{\rho_0}{M_{\odot}/{\rm pc}^3}\left (\frac{r_h}{{\rm pc}}\right )^{2},
\label{temp3}
\end{equation}
\begin{equation}
\frac{T}{{\rm K}}=5.28\, 10^{-10}  \frac{m}{{\rm eV}/c^2}\frac{\Sigma_0}{M_{\odot}/{\rm pc}^2}\frac{r_h}{{\rm pc}},
\label{temp4}
\end{equation}
\begin{equation}
\frac{T}{{\rm K}}=2.93\, 10^{-10} \frac{m}{{\rm
eV}/c^2}\frac{M_h}{M_{\odot}}\frac{{\rm pc}}{r_h},
\label{temp5}
\end{equation}
\begin{equation}
\frac{T}{{\rm K}}=3.94\, 10^{-10} \frac{m}{{\rm
eV}/c^2}\left (\frac{\Sigma_0}{M_{\odot}/{\rm pc}^2}\right
)^{1/2}\left (\frac{M_h}{M_{\odot}}\right )^{1/2}.
\label{temp5c}
\end{equation}
We can also obtain the temperature directly from Eq. (\ref{fb5}). Let us make a
numerical application. We take a
fermion mass $m=1.23\, {\rm keV}/c^2 $. For large halos such as the Medium
Spiral, we find that $T=1.78\, {\rm K}$ and $T_F=3.48\, 10^{-4}\, {\rm K}$ so
these halos are classical. For dwarf halos such as Willman 1, we find that
$T=4.81\, 10^{-3}\, {\rm K}$ and $T_F=2.23\, 10^{-2}\, {\rm K}$ so these halos
are quantum. This is of course equivalent to the results of Appendix
\ref{sec_degclass}.

The condensation temperature of bosons is
\begin{equation}
T_c=\frac{2\pi\hbar^2\rho_0^{2/3}}{k_B\zeta^{2/3}m^{5/3}},
\label{temp5b}
\end{equation}
with $\zeta=2.61$. We note that the condensation temperature
(\ref{temp5b})  has the same scaling as the Fermi
temperature (\ref{temp1}). They differ by a factor
$T_F/T_c=2^{1/3}\pi\zeta^{2/3}=7.51$. Writing $\sigma_0^2=k_B T/m$, the
parameter $H$ defined by Eq. (\ref{fund1})  may be written as
$H=4.86\, 10^{-2}(T/T_c)^{3/2}$. The bosons are condensed when $H<4.86\,
10^{-2}$ and non-condensed when $H>4.86\, 10^{-2}$. Let us make a numerical
application. We take a boson
mass $m=2.57\, 10^{-20}\, {\rm eV}/c^2 $ appropriate to non-interacting bosons.
For large halos such as the Medium
Spiral, we find that $T=3.72\, 10^{-23}\, {\rm K}$ and
$T_c=2.92\, 10^{33}\,
{\rm K}$. For dwarf halos such as Willman 1, we find that $T=1.00\, 10^{-25}\,
{\rm K}$ and $T_c=2.72\, 10^{35}\, {\rm K}$. Therefore, all the
halos are quantum objects.  We now take a boson
mass $m=1.23\, {\rm keV}/c^2 $ which is allowed for self-interacting bosons.
For large
halos such as the Medium
Spiral, we find that $T=1.78\, {\rm K}$ and
$T_c=4.63\, 10^{-5}\,
{\rm K}$ so these halos are classical. For dwarf halos such as
Willman 1, we find that $T=4.81\, 10^{-3}\,
{\rm K}$ and $T_c=2.97\, 10^{-3}\, {\rm K}$ showing that they are marginally
condensed (see footnote 28 and recall that our estimate of $T$ is approximate in
the case of cold systems).

\section{Maximum mass of relativistic compact objects}
\label{sec_relat}

Once we know the mass of the dark matter particle (see Appendix
\ref{sec_fbdm}), we can determine the maximum mass and the minimum radius (fixed
by general relativity) of a
completely degenerate or completely condensed compact object at $T=0$. While we
have argued that large dark matter halos should not contain such objects at
their center (at least in the form of the solutions B in Figs.
\ref{profilesREAL} and \ref{embryonRHOdot} because these solutions are saddle
points of entropy), it is nevertheless interesting to make the numerical
application.

If dark matter is made of fermions, the maximum mass is $M_{max}=0.376\, (\hbar
c/G)^{3/2}/m^2$ and the minimum radius is $R_{min}=9.36 \, GM_{max}/c^2$
\cite{ov}. Introducing scaled variables, we get
\begin{equation}
\frac{M_{max}}{M_{\odot}}=6.13\, 10^{17}\left (\frac{{\rm eV}/c^2}{m}\right
)^2,\quad \frac{R_{min}}{{\rm km}}=13.8 \frac{M_{max}}{M_{\odot}}.
\label{relat1}
\end{equation}
For $m=1.23\, {\rm keV}/c^2$, we obtain $M_{max}=4.05\, 10^{11}\, M_{\odot}$
and $R_{min}=0.181\, {\rm pc}$.

If dark matter is made of non interacting bosons, the maximum mass is
$M_{max}=0.633\, \hbar
c/Gm$ and the minimum radius is $R_{min}=9.53 \, GM_{max}/c^2$
\cite{kaup}. Introducing scaled variables, we get
\begin{equation}
\frac{M_{max}}{M_{\odot}}=8.48\, 10^{-11}\frac{{\rm
eV}/c^2}{m},\quad \frac{R_{min}}{{\rm km}}=14.1 \frac{M_{max}}{M_{\odot}}.
\label{relat2}
\end{equation}
For $m=2.57\, 10^{-20} {\rm eV}/c^2$, we obtain $M_{max}=3.30\, 10^{9}\,
M_{\odot}$
and $R_{min}=1.51\, 10^{-3}\, {\rm pc}$. 

If dark matter is made of self-interacting bosons, the maximum mass is
$M_{max}=0.307\, \hbar c^2\sqrt{a}/(Gm)^{3/2}$ and the minimum radius is
$R_{min}=6.25 \, GM_{max}/c^2$
\cite{chavharko}. Introducing scaled variables, we get
\begin{equation}
\frac{M_{max}}{M_{\odot}}=1.12\, \left (\frac{a}{{\rm fm}}\right
)^{1/2}\left (\frac{{\rm
GeV}/c^2}{m}\right )^{3/2},
\label{relat3}
\end{equation}
\begin{equation}
\frac{R_{min}}{{\rm km}}=9.27
\frac{M_{max}}{M_{\odot}}.
\label{relat4}
\end{equation}
For $({\rm fm}/a)^{1/3}(mc^2/{\rm eV})=0.654$, we obtain $M_{max}=6.70\,
10^{13}\, M_{\odot}$ and $R_{min}=20.2\, {\rm pc}$.

\section{A criterion for the possible existence of a black hole at the center of dark matter halos}
\label{sec_bh}

It is known that certain dark matter halos contain a central black hole. In Sec.
\ref{sec_harborbh}, we have argued that ``large'' halos may contain a black hole
because they can experience a gravothermal catastrophe while ``small'' halos
should not contain a black hole because the  gravothermal catastrophe is
prevented by quantum mechanics. In this Appendix, we determine a more precise
criterion for the possible existence of a black hole at the center of dark
matter halos. 

To
that purpose, we use the parameters of the box model \cite{ijmpb} that are more
convenient for numerical applications. In this model, the degeneracy parameter
writes $\mu_{box}=\eta_0\sqrt{512\pi^4G^3MR^3}$. If we identify $R$ with the
halo radius $r_h$ and $M$ with the halo mass $M_h$, and introduce relevant
scales, we obtain
\begin{equation}
\mu_{box}=6.41\, 10^{-17} \left (\frac{m}{{\rm eV}/c^2}\right )^4 \left (\frac{M_h}{M_{\odot}}\right )^{1/2}\left (\frac{r_h}{\rm pc}\right )^{3/2}.
\label{bh1}
\end{equation}
Our criterion for the possible existence of a black hole at the center of dark
matter halos is that $\mu>\mu^{box}_{MCP}$ where $\mu^{box}_{MCP}=2670$
\cite{ijmpb}.\footnote{Actually, $\mu$ must be substantially
larger than $\mu^{box}_{MCP}$ so that the gravothermal catastrophe is
sufficiently efficient to allow the system to enter in the relativistic
regime and trigger the dynamical instability to a black hole \cite{balberg}.}
Now, $\mu_{box}$ can be related to the parameter $H$. Using the
results of Appendix \ref{sec_fgh}, we get
\begin{equation}
\mu_{box}=\sqrt{512\pi^4}\frac{F(k)^{1/2}H(k)}{G(k)^{3/2}}.
\label{bh2}
\end{equation}
As we have seen in Paper I, the function $F(k)$ and $G(k)$ do not sensibly
change with $k$. Taking $F(k)\sim 1.8$ and $G(k)\sim 0.95$, we obtain
$\mu_{box}\simeq
324 H$. Therefore, our criterion for the existence of a black hole at the center
of a dark matter halo can be written as $H>8.24$. Taking $m=1.23\, {\rm
keV}/c^2$ and using Eq. (\ref{degclass3}) this
corresponds to a halo mass $M_h>1.60\, 10^7\, M_{\odot}$.

In conclusion, large dark matter halos of mass
$M_h>1.60\, 10^7\,
M_{\odot}$ can experience a gravothermal catastrophe and may contain a central
black hole. Small dark matter halos of mass  $0.39\, 10^6\,
M_{\odot}<M_h<1.60\, 10^7\, M_{\odot}$ should not
contain a black hole because they cannot experience a gravothermal catastrophe.

As we have seen in Appendix \ref{sec_degclass}, dark matter halos made of
non-interacting bosons of mass $m=2.57\, 10^{-20}\,
{\rm eV}/c^2$ are quantum objects. As a result, they cannot experience a
gravothermal catastrophe and should not contain a black hole. They instead
contain a central solitonic object (BEC) surrounded by a halo of scalar
radiation. Therefore, the nature of the object
that lies at the center of dark matter halos (black hole or BEC) may tell
the nature of dark matter (fermions or non-interacting bosons).
Self-interacting bosons with a mass $m=1.23\,
{\rm keV}/c^2$ behave as fermions except
that they are stabilized by the pressure arising from the
scattering instead of the pressure arising from the Pauli exclusion principle.

\section{Scenarios of formation of dark matter halos}
\label{sec_scen}

In this Appendix, we discuss different scenarios of formation of  dark matter
halos depending on the nature of the dark matter particle.

We first assume that dark matter is made of classical particles (i.e. heavy
particles of mass  $m\sim \, {\rm GeV}/c^2$) as in the CDM model. Initially,
dark matter can be considered as a spatially homogeneous gas at $T=0$. This
distribution is unstable and undergoes gravitational collapse  (Jeans
instability). Since the classical Jeans wavenumber $k_J=\sqrt{4\pi
G\rho}/c_s\rightarrow +\infty$ for $c_s\rightarrow 0$, the gas
is unstable at
all wavelengths and, consequently, structures form at all scales. There is no
ground state so we expect to observe dark matter halos of all sizes.
Furthermore, small halos are cuspy because nothing prevents the divergence of
the density resulting from gravitational collapse. Cusps are preserved during
successive mergings so that large halos are also cuspy. As we know, these
results do not agree with observations: halos are cored (cusp problem) and they
are not observed below a certain scale (missing satellite problem). This
suggests that quantum mechanics has to be taken into account.\footnote{Another
possibility is to consider warm dark matter (WDM) with $T\neq 0$. In that case,
the Jeans wavenumber $k_J$ and the maximum value of the distribution function
$f_0$ are determined by thermal effects (i.e. by the velocity dispersion of the
particles).}

We now assume that dark matter is made of fermions.  Initially, dark matter can
be considered as  a spatially homogeneous gas described by the  relativistic
Fermi distribution $f=\eta_0^{Pauli}/(1+e^{p c/k_B T})$  where $\eta_0^{Pauli}=g
m^4/h^3$ is the Pauli bound \cite{tg}. The maximum value of the distribution
function is $f_0\sim (1/2)\eta_0^{Pauli}=(g/2)m^4/h^3$. Since this gas is
collisionless, it is described by the Vlasov-Poisson system. A spatially
homogeneous distribution is unstable and undergoes gravitational collapse 
(Jeans instability). The fermionic Jeans wavenumber $k_J=\sqrt{12\pi
G}(8\pi/3)^{1/3}m^{4/3}\rho^{1/6}/h$, obtained from Eq. (\ref{eos2}), is finite
so that quantum mechanics prevents the formation of small-scale structures and
fixes a ground state. This produces a sharp cut-off in the power spectrum. In
the linear regime, some regions of over-density form. When the density has
sufficiently grown, these regions collapse under their own gravity at first in
free fall. Then, as nonlinear gravitational effects become important at higher
densities, these regions undergo damped oscillations (due to an exchange of
kinetic and potential energy) and finally settle into a quasi stationary state
(QSS) on a coarse-grained scale. This corresponds to the process of violent
relaxation first reported by Lynden-Bell \cite{lb} for stellar systems like
elliptical galaxies. This process is related to phase mixing and nonlinear
Landau damping. It is applied here to dark matter. In this context, the QSSs
represent dark
matter halos. Because of violent relaxation, the halos are almost isothermal and
have a core-halo structure.  The density of the core is relatively large and can
reach values at which quantum effects or Lynden-Bell's type of degeneracy are
important.\footnote{In the case of dark matter, the  Lynden-Bell bound and the
Pauli bound are of the same order, differing by a factor two, since
$\eta_0^{LB}=f_0= \eta_0^{Pauli}/2=(g/2)m^4/h^3$.} On the other hand, the
halo is relatively hot  and behaves more or less as a classical isothermal gas.
Actually, it cannot be exactly isothermal otherwise it would have an infinite
mass. The finite extension of the halo may be due to incomplete violent
relaxation \cite{lb}. The extension of the halo may also be limited by tidal
effects. In that case, the complete configuration of the system can be described
by the fermionic King model \cite{mnras,dubrovnik}. As we
have demonstrated, the
fermionic King model can show a wide diversity of configurations with different
degrees of nuclear concentration. The system can be everywhere non degenerate,
everywhere completely degenerate, or have a core-halo structure with a
degenerate core and a non degenerate halo. Small halos, that are compact, are
degenerate. Their flat core is due to quantum mechanics. Assuming that the
smallest and most compact observed dark matter halo of mass $M_h=0.39\, 10^6\,
M_{\odot}$ and radius $r_h=33\, {\rm pc}$ (Willman 1) is completely
degenerate ($T=0$) leads to a fermion mass of the order of $1.23\, {\rm
keV}/c^2$ \cite{vega,vega2}. 
These particles may be sterile neutrinos. Small halos can merge with each other
to form larger halos. This is called hierarchical clustering.
The merging
of the halos also corresponds to a process of collisionless violent relaxation.
Large halos, that are dilute, are non degenerate. Their flat core is due to
thermal effects.\footnote{Here, the temperature is effective and it must be
understood in the sense of Lynden-Bell.} Knowing the mass of the fermions, we
can deduce from observations that halos of mass $0.39\, 10^6\,
M_{\odot}<M_h<2.97\, 10^6 \, M_{\odot}$  are quantum (degenerate) objects while
halos of mass $M_h>2.97\, 10^6 \, M_{\odot}$  are classical (non degenerate)
objects \cite{vega,vega2}. In the classical limit, numerical simulations of
violent relaxation generically  lead to configurations presenting an isothermal
core and a halo whose density decreases as $r^{-\alpha}$ with  $\alpha=4$
\cite{henonVR,albada,roy,joyce}. These configurations are relatively close to
H\'enon's isochrone profile. They can be explained by models of incomplete
violent relaxation \cite{bertin1,bertin2,hjorth}. A density slope $\alpha=4$ in
the halo is also consistent with a King profile of concentration $k\sim 5$
(see Paper I). If the
halos were truly collisionless, they would remain in a virialized configuration.
However, if the core is dense enough, collisional effects can come into play and
induce an evolution of the system on a long timescale (driven
by the gradient of temperature - velocity dispersion - between the core and the
halo) during which the
concentration parameter $k(t)$ increases while the slope $\alpha(t)$ of the
density profile decreases much like in globular clusters \cite{cohn}. We now
have to distinguish between small halos of mass $M_h<1.60\, 10^7\, M_{\odot}$
(i.e. $\mu<\mu_{MCP}=1980$) and large halos of mass $M_h>1.60\, 10^7\,
M_{\odot}$ (i.e. $\mu>\mu_{MCP}$). For small halos, the series of equilibria
(see Fig. \ref{ETmicro3}) does not present any instability so that $k(t)$
increases and $\alpha(t)$ decreases regularly due to collisions and evaporation.
These halos are degenerate. They are stabilized against gravitational collapse
by quantum mechanics. As a result, they do not experience the gravothermal
catastrophe so they should not contain black holes. For large halos, the series
of equilibria (see Fig. \ref{ETmicro5summary}) presents an instability at
$k_{MCE}=7.44$.  Because of collisions and evaporation, the concentration
parameter increases from $k\sim 5$ corresponding to a density slope  $\alpha=4$
(a typical outcome of violent relaxation) up to the critical value
$k_{MCE}=7.44$ 
corresponding to a density slope $\alpha\sim 3$. Less steep halos ($\alpha<3$)
are
unstable ($k>k_{MCE}$). Large halos are expected to be close to the point of
marginal
stability (see solution A in Figs. \ref{profilesREAL} and \ref{gaseousRHO}). At
that point, the King profile can be approximated by the modified Hubble profile
that is relatively close to the Burkert profile fitting observational halos.
Some halos may be stable ($k<k_{MCE}$) but some halo may have undergone a
gravothermal catastrophe ($k>k_{MCE}$). In that case, they experience core
collapse. The evolution is self-similar. The system develops an isothermal core
surrounded by a halo with a density slope $\alpha=2.2$ \cite{cohn,balberg}. The
core radius decreases with time while the central density and
the central temperature increase. The halo does not change. The
specific heat of the core is negative. Therefore, by loosing heat to the profit
of the halo, the core grows hotter and enhances the gradient of temperature
with the halo so the collapse continues. This is the origin of the gravothermal
catastrophe. For weakly collisional classical systems (globular
clusters), core collapse leads to a
finite time singularity with a profile $\rho\propto r^{-2.2}$ at $t=t_{coll}$.
The singularity has infinite density but contains no mass. It corresponds to a
tight binary surrounded by a hot halo \cite{cohn}. However, for
collisional dark matter halos, the situation is different. If the particles
are fermions, and if the mass of the halo is not
too large ($\mu>\mu_{MCP}$ not too large), the gravothermal catastrophe stops
when the core of the system becomes degenerate. This leads to a configuration
with a small degenerate nucleus (condensed state) surrounded by an extended
atmosphere that is relatively different from the structure of the halo before
collapse (see solution C in Figs. \ref{profilesREAL} and \ref{condensedRHOdot}).
However, the formation of this equilibrium
 structure can be very long (of the order of the Hubble
time) so that, on an intermediate timescale, the system is made of a
contracting fermion
ball surrounded by an atmosphere that is not too much affected
by the collapse of the nucleus. Alternatively, if the halo mass is large
($\mu>\mu_{MCP}$ large), during the gravothermal catastrophe the system can
develop a (Vlasov) dynamical instability of general relativistic origin and form
a central black hole without affecting the structure of the halo \cite{balberg}.
In this way, the system is similar to the halo before collapse (Burkert profile)
except that it contains a central black hole.\footnote{More
precisely, the core collapse of fermionic dark matter halos is a two-stages
process.
In a first stage \cite{balberg}, the core collapses while the halo does not
change. Only the
density, the radius and the temperature of the core change. This creates strong
gradients of temperature between the core and the halo. At sufficiently high
temperatures (achievable  if $\mu$ is large) the system becomes relativistic and
triggers a dynamical instability leading to a black hole with a large mass.
Alternatively, if $\mu$ is small, quantum mechanics can stop the increase of the
central density and central temperature before the system enters in the
relativistic regime. In that case, core collapse stops. Then, in a second stage
(never studied until now because it requires quantum simulations),
the temperature uniformizes between the core and the halo. Therefore, the halo
heats up and extends at large distances until an equilibrium state with a
uniform temperature is reached (solution C in Figs. \ref{profilesREAL} and
\ref{condensedRHOdot}).} Large halos should not
contain a fermion ball because these nucleus-halos structures
(solution B in
Figs. \ref{profilesREAL} and \ref{embryonRHOdot}) are unreachable (saddle point
of entropy).

We now assume that dark matter is made of  bosons without self-interaction. Since the temperature of the universe is very low, they form a BEC so  they are described by the Schr\"odinger-Poisson system. These equations are equivalent to fluid equations with a quantum pressure. Initially, the distribution of dark matter is spatially homogeneous. This distribution is unstable and undergoes gravitational collapse (Jeans instability). The Jeans wavenumber $k_J=(16\pi Gm^2\rho/\hbar^2)^{1/4}$ \cite{hu,prd1} is finite so that quantum mechanics prevents the formation of small-scale structures and fixes a ground state.
In the linear regime, some regions of over-density form. In the nonlinear
regime, these over-density regions oscillate and settle into a compact bosonic
object through the radiation of a complex scalar field.   This corresponds to
the process of gravitational cooling 
first reported by Seidel and Suen \cite{seidel} in the context of boson stars.
It is extended here to dark matter. This process is similar to violent
relaxation. The resulting structures correspond to dark matter halos. Because of
gravitational cooling, the halos have a core-halo configuration. The core is
equivalent to a self-gravitating BEC at $T=0$ (soliton) stabilized against
gravitational collapse by the Heisenberg uncertainty principle. The halo
corresponds to quantum fluctuations and scalar radiation. It behaves
similarly to a thermal halo. Assuming that the smallest and most compact
observed dark matter halo (Willman 1) is completely condensed without halo (no
quantum fluctuation) implies that the boson mass is  $m=2.57\, 10^{-20}\, {\rm
eV}/c^2$. Knowing the mass of the bosons, we can deduce from observations that
all the halos are Bose condensed. They have a condensed core (soliton) and are
surrounded by a halo of scalar waves that gives them their proper size.

Self-interacting bosons are described by the Gross-Pitaevskii-Poisson system. In
the TF approximation, the Jeans wavenumber is given by
$k_J=(Gm^3/a\hbar^2)^{1/2}$ \cite{prd1} and its general expression (valid beyond
the TF approximation) is given by Eq. (139) of \cite{prd1}. Self-interacting
bosons are more complicated to study because we can just predict the ratio
$({\rm
fm}/a)^{1/3}(mc^2/{\rm eV})=0.654$ between the mass of the bosons and their
scattering length (see
Appendix \ref{sec_fbdm}), but we cannot have direct information on the mass $m$
of the bosons. This leaves open many possibilities leading to potentially very
rich studies. If $m\sim 1.23 \,{\rm keV}/c^2$, self-interacting bosons behave as
fermions except that the fermion ball (stabilized by the Pauli exclusion
principle) is replaced by a BEC (stabilized by the self-interaction of the
particles). If $m\sim 2.57\, 10^{-20}\,  {\rm eV}/c^2$, we must take into
account both the quantum pressure
and the pressure arising from the self-interaction of the particles as in
\cite{prd1,prd2}.

In conclusion, dark matter halos made of fermions of mass  $m=1.23\, {\rm
keV}/c^2$ are quantum object for $M_h<2.97\, 10^6 \, M_{\odot}$ and classical
objects for $M_h>2.97\, 10^6 \, M_{\odot}$. Small halos are degenerate. Large
halos are non degenerate. They may contain a central black hole but not a
central fermion ball (see Sec. \ref{sec_harbor}). Dark matter halos made of non
interacting bosons of mass  $m=2.57\, 10^{-20}\, {\rm eV}/c^2$ are quantum
objects for all relevant sizes. They have a solitonic core surrounded by a halo
of scalar radiation. Since fermionic and non-interacting bosonic dark matter
halos
present different characteristics, it should be possible to determine which of
them better agrees with observations. This may allow to determine the nature
of the dark matter particle. It is also important to perform
cosmological simulations taking quantum mechanics into account. This was claimed
by Feynmann long ago (see Appendix A of \cite{vega2}). Recent simulations have
been performed in \cite{ch1,ch2,ch3} for non-interacting bosons. The
case of fermions and self-interacting bosons should
also be considered. We guess that future years will see the development of
quantum
cosmological simulations.

\end{document}